\documentclass{USC-Thesis}
\usepackage{epsfig,amsmath,amsfonts,amssymb,array,calc,rotating,psfrag}
\usepackage[dvips]{feynmp}

\newcommand{\nc}{\newcommand}
\nc{\rnc}{\renewcommand}
\nc{\N}{{\cal N}}
\nc{\A}{{\cal A}}
\nc{\F}{{\cal F}}
\nc{\Lag}{{\cal L}}
\nc{\G}{{\cal G}}
\nc{\Veff}{{V_{\mbox{\it \!eff}}}}
\nc{\Weff}{{W_{\mbox{\it \!eff}}}}
\nc{\Wtree}{{W_{\mbox{\it \!tree}}}}
\nc{\Wlow}{{W_{\mbox{\it \!low}}}}
\nc{\diag}{{\mbox{diag}}}
\nc{\SU}{{\mbox{\it SU}}}
\nc{\SO}{{\mbox{\it SO}}}
\nc{\Sp}{{\mbox{\it Sp}}}
\nc{\U}{{\mbox{\it U}}}
\nc{\SUSY}{supersymmetry}
\nc{\susy}{supersymmetry}
\nc{\tr}{\mbox{Tr}}
\nc{\str}{\mbox{\scriptsize Tr}}
\nc{\MM}{{\bf M}}
\nc{\dslash}{\slash\!\!\!}
\nc{\im}{{\mbox{Im}}}
\nc{\CY}{Calabi-Yau}
\nc{\CYM}{Calabi-Yau manifold}
\nc{\CYMs}{Calabi-Yau manifolds}
\nc{\DB}{D-Brane}
\nc{\DBs}{D-Branes}
\nc{\Kah}{K\"ahler}
\nc{\cs}{complex structure}
\nc{\beq}{\begin{equation}}
\nc{\eeq}{\end{equation}}
\nc{\ntwo}{${\cal N}=2$}
\nc{\nOne}{${\cal N}=1$}
\nc{\hs}{\hspace{0.2in}}
\nc{\Z}{{\mathbb Z}}
\rnc{\P}{{\mathbb P}}
\nc{\RP}{{\mathbb {RP}}}
\nc{\WP}{\mathbb{WP}}
\nc{\slag}{special Lagrangian}
\nc{\cn}{\C^n}
\nc{\rn}{\R^n}

\nc{\del}{\partial}
\def\pder#1#2{{\frac{\partial{#1}}{\partial{#2}}}}
\def\oder#1#2{{\frac{d{#1}}{d{#2}}}}

\def\CO{{\cal O}}

\def\BZ{{\mathbb {Z}}}
\def\BP{{\mathbb {P}}}
\def\BC{{\mathbb {C}}}
\def\half{{\frac12}}

\def\odd{{\rm odd}}

\nc{\Cos}{\hbox{cos}}
\nc{\Sin}{\hbox{sin}}
\nc{\WLax}{W_{\hbox{\it lax}}}
\nc{\comb}[2]{\left(\begin{array}{c}#1\\#2\end{array}\right)}
\nc{\tLambda}{\tilde{\Lambda}}

\def\omit#1{}

\begin{document}

\title{String Theory and the Vacuum Structure \break
of Confining Gauge Theories}

\author{Kristian David Kennaway}
\major{Physics}
\month{August}
\year{2004}
\maketitle

\dedication
    \vspace{2.5in}
    \begin{center}
        {\em To Deb, for her love and support.}
    \end{center}
    
\epigraph
    \vspace{1.8 in}
    {\Huge REALITY, n.}
{\Large The dream of a mad philosopher. That which would remain in the cupel if one should assay a phantom. The nucleus of a vacuum.
    \begin{flushright}
    -- {\it from ``The Devil's Dictionary'', by Ambrose Bierce (1911)}
    \end{flushright}}

\tableofcontents 	

\listoftables	

\listoffigures	

\abstract
We discuss recent progress in the understanding of the vacuum structure (effective superpotentials) of confining gauge theories with $\N=1$ supersymmetry, in particular theories with softly broken $\N=2$ supersymmetry.  We show how the new techniques improve upon older calculations in non-supersymmetric quantum field theories.  A common feature of both approaches is that appropriate perturbative field theory calculations (e.g.~using the background field method) give non-perturbative information about the vacuum structure of the theory.  However, in supersymmetric theories, these results are often exact.

The geometric engineering of supersymmetric gauge theories in string theory provides powerful tools for studying gauge theories.  Central to the analysis is a particular class of hyperelliptic curve, which emerges from the Calabi-Yau geometry of the string theory background and encodes the gauge theory effective superpotential.  These curves may be rederived using other techniques based on zero-dimensional matrix integrals, the dynamics of integrable systems and the factorization of Seiberg-Witten curves, and we describe in detail how each technique highlights complementary aspects of the gauge theory.

We find that the use of the spectral curve requires the introduction of additional fundamental matter fields, which act as regulators for the UV divergences of the calculation by embedding the gauge theory in a UV-finite theory.  Theories with $0 \le N_f < 2N_c$ fundamental multiplets may thus be treated uniformly.  We focus in detail on maximally-confining vacua of $\N=1$ gauge theories with fundamental matter, and of gauge theories with $\SO$ and $\Sp$ gauge groups.  Both cases require refinements to the basic techniques used for $\SU$ gauge theory without fundamental matter.

We derive explicit general formulae for the effective superpotentials of $\N=1$ theories with fundamental matter and arbitrary tree-level superpotential, which reproduce known results in special cases.  The problem of factorizing the Seiberg-Witten curve for $\N=2$ gauge theories with fundamental matter is also solved and used to rederive the corresponding $\N=1$ effective superpotential.

\topmatter{Preface}  

I am thankful to Sujay Ashok, Richard Corrado, Nick Halmagyi, Christian R\"omelsberger, and Nick Warner for fruitful collaboration, and for Nick Warner's advice and guidance during my PhD.  I thank the students, postdocs and faculty of the USC string theory group for providing a stimulating learning environment, especially during my first few years at USC, and my colleagues here and at other universities from whom I have learned.  Most of all, I thank God that I am finally finished writing this thesis.

This work was supported in part by DOE grant number DE-FG03-84ER-40168, and by a USC Final Year Dissertation Fellowship.

\mainmatter 		

\chapter{Introduction}

A central problem in theoretical particle physics is to understand the nature of the strong nuclear interactions at low energies.  A quantitative theory of the strong nuclear force (quantum chromodynamics, or QCD) has been known for over 30 years, but computational difficulties prevent accurate analytical calculations at low energies or long (nuclear-scale) distances.  Specifically, the effective coupling constant of perturbative QCD increases at low energies, becoming of order 1 at energies $\sim 200$ MeV (conversely, the coupling constant approaches 0 at short distances or high energies, a property called ``asymptotic freedom'').  Therefore the main analytical tool used to study quantum field theory -- perturbation theory -- breaks down as this energy scale is approached from above.

Qualitatively, we expect QCD below this energy scale to ``confine'', or tightly bind quarks into color-neutral bound states, which are the familiar hadrons of particle physics (such as protons, neutrons, pions and other particles).  The analogous theory without quarks (non-Abelian gauge theory, also called Yang-Mills theory) is also asymptotically free and is expected to manifest similar behavior at low energies: the massless gluons of Yang-Mills perturbation theory, which mediate the strong nuclear force, bind together into hadronic ``glueball'' bound states and become massive.  Approximate numerical results in QCD and Yang-Mills theory (such as the value of the hadron masses) can be obtained by simulating the theory on a discrete spacetime lattice, and various qualitative proposals have been made for the mechanism of confinement, but a solid theoretical understanding of confinement is still lacking\footnote{See \cite{Apollodorus} for a classical problem of comparable difficulty.}.

In the absence of analytical tools for studying non-perturbative phenomena in QCD such as confinement, one alternative is to turn to related models in the hope of finding a more tractable problem that may nonetheless provide insight into the theory of interest.  A profitable tool in this regard is supersymmetry, a symmetry that relates bosonic and fermionic degrees of freedom.  The extra symmetry constraints present in the supersymmetric version of Yang-Mills theory and QCD are surprisingly tight and allow for greater depth of analytic computation; at the same time, the supersymmetric versions of Yang-Mills and QCD are expected to share many of the same qualitative features, in particular confinement at low energies.

In fact, there are several theoretical and experimental indications -- and widespread anticipation among high energy particle theorists -- that supersymmetry may be realized in nature at suitably high energies.  Thus, the study of supersymmetric gauge theories may be directly relevant for describing the nature of fundamental interactions at sufficiently high energies.

It has long been suspected that four-dimensional gauge theories such as QCD are related to string theories.  The tube of confined gauge field flux that extends between two quarks has string-like properties, and in fact, modern string theory emerged from an attempt to model the strong interactions.  However, despite over three decades of intensive study there is still no known consistent quantum theory of strings propagating in four dimensions; for example, worldsheet anomaly cancellation of the supersymmetric string requires the (suitably generalized) dimension of spacetime  to be 10.

The resolution to this dichotomy is that four-dimensional gauge theories may be equivalent to (limits of) string theories in {\it higher dimensions}; the dynamics of strings propagating in the extra dimensions can give rise to gauge dynamics in four dimensions.  These ``gauge/string dualities'' have provided many fascinating and unexpected results, some of which are the subject of this thesis.  

As in the heuristic example of QCD, open strings carry matter degrees of freedom (``quarks'') at their endpoints, and their excitation spectrum contains a massless spin-1 particle.  Thus, open strings give rise to matter coupled to gauge fields.  Taking into account string interactions, the endpoints of open strings may join together to form a closed string.  Closed strings include a massless spin-2 particle in their excitation spectrum; this particle must couple to the stress-energy tensor of the theory, and the spacetime theory is required by consistency to have diffeomorphism invariance.  The spin-2 particle is therefore identified with the graviton, and quantum theories including closed strings are theories of quantum gravity that reduce in the classical limit to classical general relativity coupled to additional fields.  Thus, string theory has the potential to unify the interactions of matter with all four fundamental forces in a consistent quantum theory; this is a long-standing theoretical problem that has resisted many previous attempts at solution.

The link between these two aspects of string theory, and the main string-theoretical tool for studying gauge theories in the modern context, are D-branes.  These are extended ``membrane'' objects, of various dimensions, which are required by non-perturbative consistency to be present in the spectrum: when the theory contains open strings, these strings may only end on a D-brane.  Therefore, the matter fields at the endpoints of the strings are confined to live on the D-brane, and open strings with both endpoints on the brane give rise to gauge fields propagating along it.  Thus, the study of D-branes and strings propagating in appropriate 10 dimensional geometries can teach us about four-dimensional gauge theories. 

This is one of the great advantages of string theory, fully realized only since the mid 1990s; it can be used to translate certain problems of quantum field theory into geometrical language.  This allows the application of powerful geometrical tools to study the corresponding quantum field theories. In cases where theoretical control of the calculation is presently available, the corresponding field theories are typically supersymmetric, and the more supersymmetries that are present, the greater the constraints on the mathematical structures that underly the theory.

String theory is now known to possess many remarkable properties, and while there remain many difficult problems to solve before it can be quantitatively applied to study the physics of our observed universe, it has nonetheless provided deep insights into many aspects of theoretical physics and mathematics. In this thesis, we will describe a set of tools that have emerged from string theory over the past few years, which allow the computation of exact results in a class of confining supersymmetric gauge theories at low energies.  These string theoretical tools have provided some unexpected insights into the structure of quantum field theory.

Central to the analysis is a particular class of hyperelliptic curves related to a string theory background geometry, the periods of which encode the superpotential of the gauge theory and define its vacuum structure.  These ``spectral curves'' also emerge from the study of a number of mathematical systems that appear at first sight to be unrelated to the gauge theory (such as matrix integrals, and integrable systems), and understanding this connection provides new insights into the structure of the vacua of the quantum field theory\footnote{Conversely, this relationship provided a link between previously unrelated areas of mathematics, for example that the combinatorics of planar diagrams is related to the special geometry of Calabi-Yau manifolds.}.

To provide context for the later results on supersymmetric gauge theories, we will begin by reviewing some known techniques and results on the vacuum structure of non-supersymmetric gauge theories.  We will explain the limitations of these calculations, and describe how they are avoided in supersymmetric theories.  The remainder of the thesis will discuss various techniques that have emerged from string theory and allow the computation of exact results about the low energy structure of supersymmetric gauge theories.

This thesis is based on material previously published in the original collaborative works \cite{Ashok:2002bi,Kennaway:2003jt}, and on the review article \cite{Kennaway:2004}, although some details and aspects of the composition are new.

\chapter{Effective Potentials in Quantum Field Theories}


When a quantum field theory possesses continuous symmetries, the form of the effective potential (the non-derivative terms in the effective Lagrangian) is constrained by the corresponding (anomalous) Ward identities, which give rise to partial differential equations that must be satisfied by the quantum corrected effective potential.  For example, as we will discuss in section \ref{sec:anom.sym} the differential equation associated to the anomalously broken scaling symmetry is the Callan-Symanzik equation.

The {\it background field method} can be used to derive the one-loop effective action from the path integral of the theory; in theories with non-trivial vacua, such as asymptotically free theories, this gives an approximation to the vacuum state.  Evaluating the 1-loop effective action is equivalent to the summation of an infinite class of Feynman diagrams where one includes the couplings of a set of fluctuating fields to a classical background field, but ignore the self-interactions of the fluctuating fields.

We begin by studying the Gross-Neveu model, a two-dimensional theory of chiral fermions which is asymptotically free.  This model exhibits several of the features of more interesting four-dimensional theories such as Yang-Mills theory and QCD, including asymptotic freedom and spontaneous chiral symmetry breaking.  We will solve for the 1-loop effective potential of this model, as a warm-up exercise for studying four-dimensional gauge theories.

Due to the Landau pole (divergence of the perturbative gauge coupling at low energies), the one-loop approximation to the Yang-Mills effective potential cannot be extrapolated to the vacuum of the theory, but it gives a qualitative picture of some of the features of the vacuum.  When the theory has supersymmetry, the constraints on the effective (super)potential become much more powerful, and the one-loop perturbative gauge theory computations can be extrapolated all the way to low energies to obtain exact, non-perturbative information about the vacuum.


\section{A toy model: the Gross-Neveu model}
\label{sec:gn}

The Gross-Neveu model \cite{Gross:1974jv} is a simple model that exhibits spontaneous symmetry breaking through a quantum-mechanical symmetry-violation.  It is a two-dimensional, asymptotically-free theory of $N$ massless interacting fermions, with Lagrangian:

\begin{equation}
\label{eq:lag.gn}
{\cal L}_{GN} = \overline \psi_i \imath \dslash{\partial} \psi_i + \frac{g^2}{2} (\overline \psi_i \psi_i)^2
\end{equation}
The classical Lagrangian has a discrete chiral symmetry

\begin{equation}
\label{eq:gn.chisym}
\psi_i \rightarrow \gamma_5 \psi_i \quad
\overline \psi_i \rightarrow - \overline \psi_i \gamma_5
\end{equation}
By summing the contribution of Feynman diagrams with vanishing external momenta, we will derive the effective potential of the Gross-Neveu model, and find that the chiral symmetry is spontaneously broken in the quantum theory.  This perturbative 1-loop computation provides exact non-perturbative results about the vacuum of the theory at large $N$.

A useful technique for studying the response of quantum field theories to non-trivial field backgrounds is the {\it background field method}.  One splits the external field into a classical, background field, and a fluctuating quantum field, and then evaluates the path integral perturbatively in the fluctuations around the given background.  
We will make full use of the background field method when we study non-Abelian Yang-Mills theory in section \ref{sec:ym}.  Because this technique is non-perturbative in the background field, it can be used to probe for phenomena that are invisible in perturbation theory around the usual zero-field background.

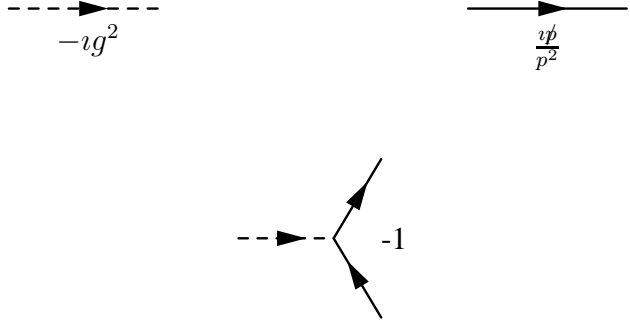
\begin{figure}[t]
\begin{center}
\begin{fmffile}{gnfeyn}
\parbox{60mm}{\begin{center}
	\begin{fmfgraph*}(60,60)
		\fmfleft{i}
		\fmfright{o}
		\fmf{scalar,label=$-\imath g^{2}$,label.side=right}{i,o}
	\end{fmfgraph*}\end{center}
}
\parbox{60mm}{\begin{center}
	\begin{fmfgraph*}(60,60)
		\fmfleft{i}
		\fmfright{o}
		\fmf{fermion,label=$\frac{\imath \dslash p}{p^{2}}$,label.side=right}{i,o}
	\end{fmfgraph*}\end{center}
}
\parbox{60mm}{\begin{center}
	\begin{fmfgraph*}(60,60)
		\fmfleft{i}
		\fmfright{o2,o1}
		\fmf{scalar}{i,v}
		\fmf{fermion}{v,o1}
		\fmf{fermion}{o2,v}
		\fmflabel{\quad-1}{v}
	\end{fmfgraph*}\end{center}
}
\end{fmffile}
\end{center}
\caption{The Feynman rules for the Gross-Neveu Lagrangian (\ref{eq:lag.gn2})}
\label{fig:gnfeyn}
\end{figure}

Fermionic (Grassman-valued) fields are not usually considered as classical field theories, for example as possible background fields for a quantum field theory calculation.  However, fermionic quantum fields can pair up and form a composite bosonic field $\sigma \sim \overline \psi  \psi$ which can attain a vacuum expectation value.  The Gross-Neveu Lagrangian can be rewritten as

\begin{equation}
\label{eq:lag.gn2}
\tilde{\cal L} = \overline \psi_i \imath \dslash \partial \psi_i - \frac{1}{2 g^2} \sigma^2 - \sigma \overline\psi_i \psi_i
\end{equation}
which re-expresses it in terms of a coupling to the composite bosonic operator $\sigma$.  This field is treated as a non-dynamical, external background field since it has no kinetic term.  It is easily verified that integrating over this auxiliary $\sigma$ field recovers the original form of the Lagrangian (\ref{eq:lag.gn}).
The Feynman rules for (\ref{eq:lag.gn2}) are shown in figure \ref{fig:gnfeyn}.
 
We will analyze this theory in two ways:  by performing a path integral computation that amounts to summing the Feynman diagrams that can contribute to the effective potential of the theory due to the interaction with the external $\sigma$ field, and by using the anomalously broken scale invariance to constrain the form of quantum corrections to the potential.

\subsection{Path-integral computation of the effective potential}

We can probe the response of the Gross-Neveu model to the formation of a non-zero fermion condensate by introducing an external source $J$ for the field $\sigma$ into the path integral, finding the minimum-energy field configurations in the presence of the source, and then turning off the source.  Define

\begin{equation}
e^{-\imath E[J]} = \int {\cal D} \sigma \prod {\cal D} \psi_{i} \prod {\cal D} \overline \psi_{i}  \exp\left(\imath \left(\Lag(\sigma,\psi_{i},\overline \psi_{i}) + J \sigma\right) \right)
\end{equation}
where $-E[J]$ is the generating functional of connected correlation functions of $\sigma$.  Define the classical field

\begin{equation}
\sigma_{cl}(x) = -\frac{\delta E}{\delta J} = \langle 0|\sigma(x) |0\rangle_{J}
\end{equation}
the vacuum expectation value of $\sigma(x)$ in the presence of the source $J$.  Then the Legendre transform of the energy functional $-E[J]$ defines the effective action $\Gamma(\sigma_{cl})$

\begin{equation}
\Gamma(\sigma_{cl}) = -E[J] - \int d^{4} x~\sigma_{cl}(x) J(x)
\end{equation} 
subject to the constraint
\begin{equation}
\frac{\delta \Gamma(\sigma_{cl})}{\delta \sigma_{cl}(x)} = - J(x)
\end{equation}
Thus, turning off the source $J$ we obtain that the stable configurations for the external field $\sigma_{cl}$ are those for which

\begin{equation}
\frac{\delta \Gamma(\sigma_{cl})}{\delta \sigma_{cl}(x)} = 0
\end{equation}
In the translation-invariant vacuum states of the theory, $\sigma_{cl}(x)$ is constant, and the effective action can be written as

\begin{equation}
\Gamma[\sigma_{cl}] = -(V T) \Veff(\sigma_{cl})
\end{equation}
where $V$ is the 3-dimensional volume, $T$ is the time interval of the integration region, and we defined $\Veff(\sigma_{cl})$ the effective potential for the classical field $\sigma_{cl}$.  The vacua of the theory satisfy

\begin{equation}
\frac{\partial \Veff(\sigma_{cl})}{\partial \sigma_{cl}} = 0
\end{equation}

The effective action is the generating functional of 1-particle irreducible (1PI) correlation functions of the $\sigma$ field.  Therefore in the background of $\sigma_{cl}$

\begin{equation}
\Veff(\sigma_{cl}) = \sum \frac{1}{n\!} \sigma_{cl}^{n} \Gamma_{n}(0,0,\ldots,0)
\end{equation}
where the 1PI diagrams that contribute to $\Gamma_{n}$ carry 0 external momenta on all legs, and each leg comes with a coupling to the background field.  To 1-loop order, the diagrams contributing to the effective potential are shown in figure \ref{fig:gndiags}.  Since they all involve a single fermion loop, we can evaluate the 1PI generating functional to 1-loop order by integrating over the fermions, which appear quadratically in the path integral of the original Lagrangian:

\begin{figure}[t]
\begin{center}
\begin{fmffile}{feyn}
\parbox{20mm}{\begin{center}
	\begin{fmfgraph}(40,40)
		\fmfleft{i}
		\fmfright{o}
		\fmf{dashes}{i,o}
	\end{fmfgraph}\end{center}
} \quad + \quad
\parbox{20mm}{\begin{center}
	\begin{fmfgraph}(40,40)
		\fmfleft{i}
		\fmfright{o}
		\fmf{dashes}{i,v1}
		\fmf{dashes}{v2,o}
		\fmf{fermion,left,tension=0.3}{v1,v2,v1}
	\end{fmfgraph}\end{center}
} \quad + \quad
\def \gnloop#1{%
	\begin{fmfgraph}(40,40)
		\fmfsurroundn{e}{#1}
		\begin{fmffor}{n}{1}{1}{#1}
			\fmf{dashes}{e[n],i[n]}
		\end{fmffor}
		\fmfcyclen{fermion,right=0.2,tension=#1/8}{i}{#1}
	\end{fmfgraph}}
\parbox{20mm}{\begin{center}
	\gnloop{4}
	\end{center}
} \quad + \quad
\parbox{20mm}{\begin{center}
	\gnloop{6}
	\end{center}
} \quad + \quad \ldots
\end{fmffile}
\end{center}
\caption{Diagrams contributing to the 1-loop effective potential for the background field $\sigma$}
\label{fig:gndiags}
\end{figure}
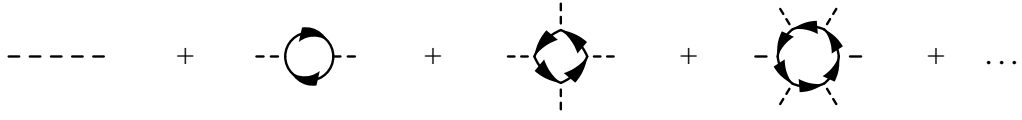

\begin{eqnarray}
Z = \int \prod_i  {\cal D} \psi_i {\cal D} \overline \psi_i {\cal D} \sigma e^{\imath S(\sigma, \psi_i, \overline \psi_i)}
&=& \int \prod_i {\cal D} \psi_i {\cal D} \overline \psi_i  {\cal D} \sigma e^{\imath \int d^2 x \overline \psi_i ( i \dslash{\partial} + \sigma) \psi_i - \frac{\sigma^2}{2 g^2}} \nonumber \\
&=& \int {\cal D} \sigma e^{\imath \int d^2 x \frac{-\sigma^2}{2 g^2}} \det(\imath \dslash{\partial} + \sigma)^N \nonumber \\
&=& \int {\cal D} \sigma e^{\imath \int d^2 x {\cal L}(\sigma)}
\end{eqnarray} 
with
\begin{equation}
\label{eq:gn.1l}
{\cal L}(\sigma) = - \frac{\sigma^2}{2 g^2} + \imath N \log \det(\imath \dslash \partial + \sigma)
\end{equation}
 
Using the two-dimensional gamma matrices $\gamma^0 = \sigma^2$, $\gamma^1 = \imath \sigma^1$ and performing a Fourier transform, we can evaluate the determinant in (\ref{eq:gn.1l}):

\begin{eqnarray}
\log \det(\imath \dslash \partial + \sigma) &=& \int \frac{d^2 p}{(2 \pi)^2} \log \det(\dslash p + \sigma) \nonumber \\
&=& \int \frac{d^2 p}{(2 \pi)^2} \log \det\left(\begin{array}{cc} \sigma & -\imath p_0 + \imath p_1 \\
\imath p_0 + \imath p_1 & \sigma \end{array} \right) \nonumber \\
&=& \int \frac{d^2 p}{(2 \pi)^2} \log(\sigma^2-p^2)
\end{eqnarray}
Therefore

\begin{equation}
\label{eq:gn1loop}
\Lag(\sigma) = - \frac{\sigma^2}{2 g^2}  + \imath N \int \frac{d^2 p}{(2 \pi)^2} \log( \sigma^2 - p^2)
\end{equation}
The 1-loop 1PI correlation functions of (\ref{eq:lag.gn2}) may be derived from ${\cal L}(\sigma)$, and to this order we can identify the Lagrangian $\Lag(\sigma)$ with the effective Lagrangian associated to the effective action $\Gamma(\sigma)=\int d^{d}x~\Lag_{\mbox{\it \!eff}}(\sigma)$, or in other words

\begin{equation}
\Veff(\sigma_{cl}) = - {\cal L}(\sigma_{cl})
\end{equation}
We can recover the diagram sum explicitly by writing
\begin{eqnarray}
\log (\sigma^{2} - p^{2}) &=& \log(-p^{2}(1-\frac{(\imath \sigma)^{2}}{p^{2}})) \nonumber \\
&=& \log(1-\frac{(\imath \sigma)^{2}}{p^{2}}) + \log(-p^{2})\nonumber \\
&\sim& \log(1-\frac{(\imath \sigma)^{2}}{p^{2}}) \nonumber \\
&=& - \sum_{n=1}^{\infty} \frac{1}{n} (\frac{\imath \sigma}{p})^{2n} \nonumber \\
&=& - \tr \sum_{n=1}^{\infty} \frac{1}{n} (- (\imath \sigma) \frac{\imath \dslash p}{p^{2}})^{2n}
\end{eqnarray}
where in the third line we dropped the second term since it just gives rise to an infinite constant upon Wick rotation and integration over $p$.  Comparing to the Feynman rules in figure \ref{fig:gnfeyn}, each term in the series corresponds to a 1-loop diagram of the form shown in figure \ref{fig:gndiags}; therefore, integrating over the fermions to quadratic order is equivalent to computing the 1-loop diagram sum to all orders.  

Returning to the 1-loop effective Lagrangian, the integral (\ref{eq:gn1loop}) is divergent and needs to be regularized.  Wick rotating to Euclidean space and using dimensional regularization we obtain

\begin{eqnarray}
\Lag(\sigma) &=& -\frac{\sigma^2}{2 g^2} - N \int \frac{d^2 p_E}{(2 \pi)^2} \log(p_E^2 + \sigma^2) \nonumber \\
&=& -\frac{\sigma^2}{2 g^2} + N\int \frac{d^d p_E}{(2 \pi)^d} \frac{\partial}{\partial \alpha} \left.\left( \frac{1}{p_E^2 + \sigma^2} \right)^{-\alpha}\right|_{\alpha=0} \nonumber \\
&=& -\frac{\sigma^2}{2 g^2} + N\int \frac{d^d p_E}{(2 \pi)^d} \frac{\partial}{\partial \alpha} \left( \frac{(-1)^{-\alpha} \imath}{(4 \pi)^{\frac{d}{2}}} \frac{\Gamma(-\alpha-\frac{d}{2})}{\Gamma(-\alpha)} \left.\left( \frac{-1}{\Delta} \right)^{-\alpha-\frac{d}{2}}\right)\right|_{\alpha=0}
\end{eqnarray}
where $\Delta = \sigma^2$.  Using the expansion of $\Gamma(x)$ near its poles, $\Gamma(x) \sim \frac{(-1)^n}{n! (x+n)} - \gamma + 1 + \ldots + \frac{1}{n} + {\cal O}(x+n)$ and $\Gamma(x+1) = x \Gamma(x)$
 we expand $\Gamma(-\alpha-\frac{d}{2})$ and write the singular terms in the form suitable for the modified minimal subtraction scheme (adapted to 2 dimensions):
 
\begin{eqnarray}
\frac{\Gamma(1-\frac{d}{2})}{(4 \pi)^{d/2}} \left( \frac{-1}{\Delta} \right)^{1-d/2} &=&\frac{1}{4 \pi} \left( \frac{1}{\epsilon} - \gamma + \log 4 \pi - \log \Delta  + {\cal O}(\epsilon) \right) \nonumber \\
&\longmapsto& -\frac{1}{4 \pi} \log(\frac{\Delta}{\mu^2})
\end{eqnarray}

\begin{figure}[t]
\begin{center}
\epsfig{file=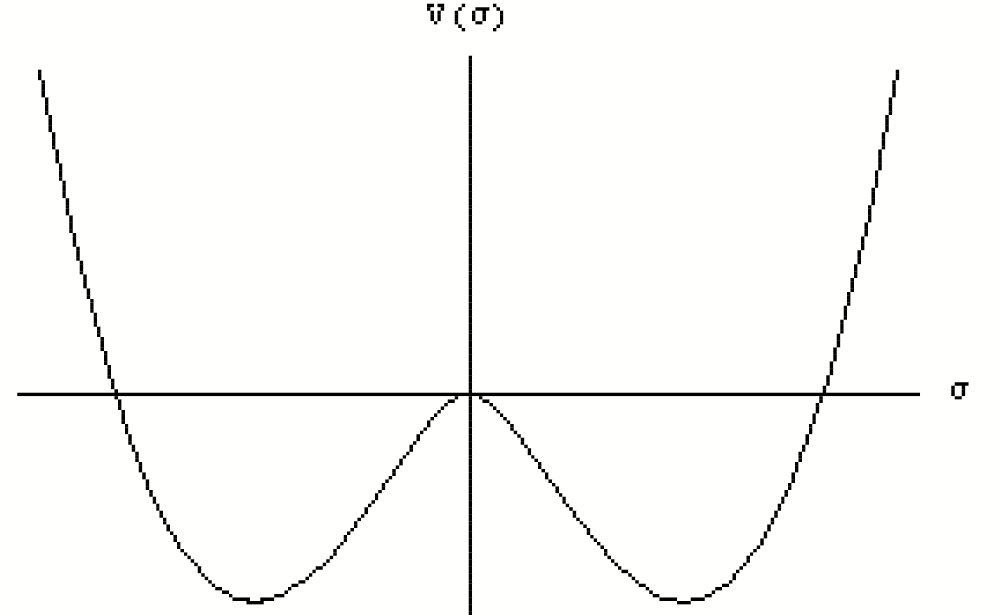}
\parbox{5.5in}{
\caption{1-loop effective potential of the Gross-Neveu model.\label{fig:gn.pot}}}
\end{center}
\end{figure}

We obtain for the effective potential
 
\begin{eqnarray}
\label{eq:gnpot}
\Veff(\sigma_{cl}) &=& \frac{\sigma_{cl}^2}{2 g^2} + \frac{N}{4 \pi} \sigma_{cl}^2 \left( \log \frac{\sigma_{cl}^2}{\mu^2} - 1 \right) \nonumber \\
 &=& \frac{N \sigma_{cl}^2}{4 \pi} \left( \log \frac{\sigma_{cl}^2}{\Lambda^2}  - 1 \right)
\end{eqnarray}
where in the second line we defined the dynamical scale $\Lambda^2 = \mu^2 \exp(\frac{-2 \pi}{N g^2})$.  The potential (\ref{eq:gnpot}) is of Coleman-Weinberg type \cite{Coleman:1973jx} and has the form shown in figure \ref{fig:gn.pot}.

Extremizing (\ref{eq:gnpot}), we find that what was the classical minimum $\langle \sigma_{cl} \rangle = 0$ is now a local maximum, and there are degenerate global minima at $\langle \sigma_{cl} \rangle = \pm \Lambda$.  Thus, the original ``perturbative'' vacuum can minimize its energy by spontaneously generating a background of paired fermions,

\begin{equation}
\langle \overline \psi \psi \rangle = \frac{1}{g^2} \langle \sigma \rangle = \pm \frac{\mu}{g^2} \exp(\frac{-\pi}{N g^2})
\end{equation}
and since this fermion bilinear does not respect the chiral symmetry (\ref{eq:gn.chisym}), the Gross-Neveu model exhibits spontaneous chiral symmetry breaking.  

Higher loop corrections to the effective potential necessarily involve $\sigma$ propagators and are therefore suppressed by powers of $g$; in fact all higher loop corrections vanish in the 't Hooft limit $N \rightarrow \infty$, $g \rightarrow 0$, $g^{2} N = \mbox{\it const.}$ \cite{Gross:1974jv}.  Therefore, in this limit the 1-loop result is exact.  Unfortunately, for most interesting non-supersymmetric theories (such as Yang-Mills or QCD) the higher-loop corrections do not vanish in this limit, and the infinite diagram series cannot be summed explicitly even at large $N$ \footnote{Although a generating function that enumerates the infinite series of Feynman diagrams is known for QCD \cite{'tHooft:1998mh}}; the complication comes from performing the loop momentum integrals at higher orders.  However we can obtain partial results by organizing the diagrams as a loop expansion: in section \ref{sec:ym} we will show how summing the one-loop diagrams for Yang-Mills theory in a covariantly constant background field strength gives a (not particularly good) approximation to the vacuum state of Yang-Mills theory.

However, simplifications even more powerful than those of the Gross-Neveu model were observed recently in certain four-dimensional $\N=1$ theories, where supersymmetry provides additional constraints on the effective potential that allows us to sum the diagram expansion to all orders.  We will come back to this in section \ref{sec:susy.matter}.

The value of the fermion condensate $\langle \overline \psi \psi \rangle = \pm \frac{\mu}{g^2} \exp(\frac{- \pi}{N g^2})$ is a non-perturbative quantity, since its Taylor expansion around $g=0$ vanishes to all orders.  Therefore the non-trivial vacuum of the Gross-Neveu model is invisible in the perturbation theory of the original Lagrangian (\ref{eq:lag.gn}), which preserves chiral symmetry to all orders.  It was only by rewriting the Lagrangian by introducing a coupling to the appropriate background field that we could probe for the existence of a chiral symmetry breaking condensate.  We have seen that by introducing an appropriate variable in which to perform a perturbative loop expansion (the composite background field $\sigma$), we can obtain non-perturbative information about the vacuum of the theory, order by order in the perturbative evaluation of a {\it different} Lagrangian.

\subsection{Anomalous symmetries and effective potentials}
\label{sec:anom.sym}

In quantum field theories, continuous symmetries of the classical Lagrangian may sometimes be violated in the quantum theory.  An example of an anomalous symmetry are scale transformations (dilatations) in massless field theories\footnote{Another anomalous symmetry is axial rotations of massless Dirac fermions in gauge theories; the corresponding effective Lagrangian including quantum corrections from the axial anomaly can be obtained by similar techniques, and has been used to study the role of the anomaly in the low-energy dynamics of mesons \cite{DiVecchia:1980ve,Witten:1980sp}.}.  The continuous dilatation symmetry is associated to a current $D_\mu = \Theta^{\mu \nu} x_\nu$, where $\Theta^{\mu \nu}$ is the stress-energy tensor of the theory, defined by

\begin{equation}
\Theta^{\mu \nu} = 2 \frac{\delta}{\delta g_{\mu \nu}(x)} \int d^d x {\cal L}
\end{equation}

Classically the dilatation current is conserved; $\partial_{\mu} D^{\mu} = \Theta^{\mu}_{\mu} = 0$.  However under a change of renormalization scale this symmetry is broken by the running of the coupling constant (see \cite{Peskin:1995ev}), and the one-loop trace anomaly is given by:

\begin{equation}
\partial_\mu D^\mu = \beta(g) \frac{\partial}{\partial g} {\cal L}
\end{equation}
The trace anomaly receives contributions from all orders in perturbation theory, as well as possible non-perturbative corrections, through the beta function.

In a quantum field theory the ``charge'' of fields under a scale transformation (their scaling dimension) may receive quantum corrections as we change the renormalization scale; operators can have anomalous dimensions.  The Callan-Symanzik equation encodes the scaling behavior of the effective potential under a change of renormalization scale (renormalization group invariance):

\begin{equation}
\label{eq:cs}
\left[ d - \sum_i (d_i + \gamma_{{\cal O}_i})  {\cal O}_i \frac{\partial}{\partial {\cal O}_i} +  \beta(g) \frac{\partial}{\partial g} + \mu \frac{\partial}{\partial \mu} \right] \Veff = 0
\end{equation}
where $d$ is the space-time dimension, $d_i$ are the classical scaling dimensions of the operators ${\cal O}_i$, $\gamma_{{\cal O}_i}$ are their anomalous dimensions, and $\mu$ is the renormalization scale.  This equation imposes that the effective potential must scale with dimension $d$, and reproduce the trace anomaly under a scale transformation.

In order to use the Callan-Symanzik to obtain predictions about the form of $\Veff$, we need to know the $\beta$ function and anomalous dimensions $\gamma$.   These are typically only known through explicit loop calculations, such as the one we did in the previous section.  However, as we will discuss in section \ref{sec:ym.anom}, once we know $\beta$ and $\gamma$ from a particular calculation, we can use the Callan-Symanzik equation to constrain the allowed form of the effective potential for an {\it arbitrary} field background.

We impose
\begin{equation}
\left[2 - (1 + \gamma_{\sigma}) \sigma \frac{\partial}{\partial \sigma} + \beta(g) \frac{\partial}{\partial g} + \mu \frac{\partial}{\partial \mu} \right] \Veff = 0
\end{equation}
and find that $\beta(g)=-\frac{N g^3}{2 \pi}$, $\gamma_\sigma = 0$ \footnote{The field $\sigma$ has vanishing anomalous dimension due to the normalization of the Lagrangian (\ref{eq:lag.gn2}).  A wavefunction renormalization of $\sigma$ cannot be balanced by a coupling-constant renormalization since the coefficient of the $\sigma$ interaction term is fixed to $1$.}. As we noted in the previous section, in the 't Hooft limit these quantities are exact.

\section{Four-dimensional gauge theories}

Before considering non-Abelian Yang-Mills theory, it is instructive to review the calculation of the effective potential for QED in external electromagnetic fields, which shares many technical features with the Yang-Mills case.  These results were first obtained by Euler and Heisenberg in 1936 \cite{Heisenberg:1936qt}, and were cast in a rigorous quantum field theory framework by Schwinger in 1951 \cite{Schwinger:1951nm}.  The presentation here includes elements from \cite{Salam:1975xe,Flyvbjerg:1980qv}.

\subsection{QED}
\label{sec:qed}

The Lagrangian of QED is

\begin{equation}
{\cal L} = -\frac{1}{4} F_{\mu \nu} F^{\mu \nu} + \overline \psi \dslash D \psi + m  \overline \psi \psi
\end{equation}
where the covariant derivative is $D_{\mu} = \partial_\mu + i e A_\mu$.  As in the previous section, the effective action for the gauge field is given to 1-loop order by

\begin{eqnarray}
e^{i \Gamma[A]} &=& \int {\cal D} \overline \psi {\cal D} \psi e^{\imath \int d^4 x {\cal L}} \nonumber \\
&=& \det(\imath \dslash D - m) e^{-\frac{\imath}{4} \int d^4 x  F^2} \nonumber \\
&=& \exp(\imath {{\int d^4 x {{\cal L}_{\mbox{\it \!eff}}}}})
\end{eqnarray}
where we defined the 1-loop effective Lagrangian

\begin{eqnarray}
{\cal L}_{\mbox{\it \!eff}} &=& -\frac{1}{4} F_{\mu \nu} F^{\mu \nu} - \imath \log \det(\imath \dslash \partial - e\dslash A - m) \nonumber \\
\end{eqnarray}

For comparison to Yang-Mills theory in the next section, we henceforth restrict to massless electrons, although the massive case can be easily treated in a similar manner.  To evaluate the fermion determinant $\det(\imath \dslash D)$ it is convenient to evaluate the determinant of $(\imath \dslash D)^2$ and take the square root.  Expanding and using the anticommutation relation $\{ \gamma^\mu, \gamma^\nu\} = 2 g^{\mu \nu}$, we find

\begin{eqnarray}
\label{eq:qed.det}
(\imath \dslash D)^2 = -D^2 - \frac{e}{2} \sigma_{\mu \nu} F^{\mu \nu}
\end{eqnarray}
where $\frac{\imath}{2} [ \gamma^\mu, \gamma^\nu ] = \sigma^{\mu \nu}$ is the generator of Lorentz transformations on the spin-$\frac{1}{2}$ representation.  Therefore

\begin{equation}
\log \det( \imath \dslash D) = \frac{1}{2}  \log \det(-D^2 - \frac{e}{2} \sigma_{\mu \nu} F^{\mu \nu})
\end{equation}

As we discussed in the previous section, the determinant corresponds to summing up the infinite series of 1-loop Feynman diagrams of the theory, where the electron runs in the loop, and we consider arbitrary insertions of the background gauge field.  The one-loop effective Lagrangian for massless QED is therefore

\begin{equation}
{\cal L}_{\mbox{\it \!eff}} = -\frac{1}{4} F_{\mu \nu} F^{\mu \nu} - \frac{\imath}{2} \tr \log((p_\mu - A_\mu)^2 - \frac{e}{2} \sigma_{\mu \nu} F^{\mu \nu})
\end{equation}
This Lagrangian exhibits the anomalous magnetic moment interaction $\frac{e}{2} \sigma_{\mu \nu} F^{\mu \nu}$ of the electron with the background electromagnetic field.  A similar magnetic moment interaction for the charged gluons of Yang-Mills theory will be vital for understanding the vacuum properties of that theory.

In diagonalizing this operator one needs the eigenvalues of the field strengths $F_{\mu \nu}$.  Defining the Lorentz scalar and pseudo-scalars

 \begin{eqnarray}
 \F &=& \frac{1}{4} F_{\mu \nu} F^{\mu \nu} = \frac{1}{2}(B^2-E^2) \nonumber \\
 \G &=& \frac{1}{4} F_{\mu \nu} {\tilde F}^{\mu \nu} = E \cdot B
 \end{eqnarray}
where $\tilde F^{\mu \nu} = \frac{1}{2} \imath \epsilon^{\mu \nu \rho \sigma} F_{\rho \sigma}$ is the dual field-strength tensor.  Using the identities

\begin{eqnarray}
F_{\mu \rho} {\tilde F}^{\rho \nu} &=& -\delta_{\mu}^{\nu} \G \\
{\tilde F}_{\mu \rho} {\tilde F}^{\rho \nu } - F_{\mu \rho} F^{\rho \nu} &=& 2 \delta_{\mu}^{\nu} \F
\end{eqnarray}
the eigenvalues $\lambda$ of $\F_{\mu \nu}$ are found to satisfy

\begin{equation}
\lambda^4 + 2 \F \lambda^2 - \G^2 = 0
\end{equation}
which has solution $\pm \lambda^{(1)}$, $\pm \lambda^{(2)}$, with
\begin{eqnarray}
\label{eq:qed.ev}
\lambda^{(1)} &=& \frac{\imath}{\sqrt{2}} ((\F + \imath \G)^{1/2} + (\F - \imath \G)^{1/2}) \\
\lambda^{(2)} &=& \frac{\imath}{\sqrt{2}} ((\F + \imath \G)^{1/2} - (\F - \imath \G)^{1/2})
\end{eqnarray}
The magnetic moment operator satisfies

\begin{equation}
(\frac{1}{2} \sigma_{\mu \nu} F^{\mu \nu})^2 =2 (\F + \gamma_5 \G)
\end{equation}
therefore using $\gamma_5^2 = -1$ and (\ref{eq:qed.ev}) the eigenvalues are

\begin{equation}
\pm (2 (\F \pm \imath \G))^{1/2}
\end{equation}

In a particular Lorentz frame, a constant magnetic field may be specified by taking $\G=0, \F>0$, and the eigenvalues $\lambda$ are real.  For a constant electric field $\G=0, \F<0$ they are purely imaginary; this difference is the cause of the vacuum instability we will find for the constant electric field.

First, we consider a constant magnetic field, which we take to be along the 3 direction, $A = (0,0, -B x_1, 0)$, $B>0$, and we have $\G=0$, $\F=\frac{1}{2}B^2$, and

\begin{equation}
\label{eq:mageigen}
\frac{e}{2} \sigma_{\mu \nu} F^{\mu \nu}  \sim \diag(eB, eB, -eB, -eB)
\end{equation}
In this gauge the d'Alembertian $D^2$ becomes

\begin{eqnarray}
D^2 = p_0^2 - p_1^2 - (p_2 + e B x_1)^2 - p_3^2
\end{eqnarray}
and becomes after a unitary transformation
\begin{eqnarray}
D^2 = & e^{\imath p_1 p_2/e B} ( p_0^2 - p_1^2 - e^2 B^2  x_1^2 - p_3^2 ) e^{-\imath p_1 p_2/e B} 
\end{eqnarray}
where we have used the commutation relations $[ x_\mu, p_\nu ] = i g_{\mu \nu}$, and in particular \\
$[x_1, e^{a p_1}]~=~i a e^{a p_1}$. 

\begin{figure}[t]
\begin{center}
\epsfig{file=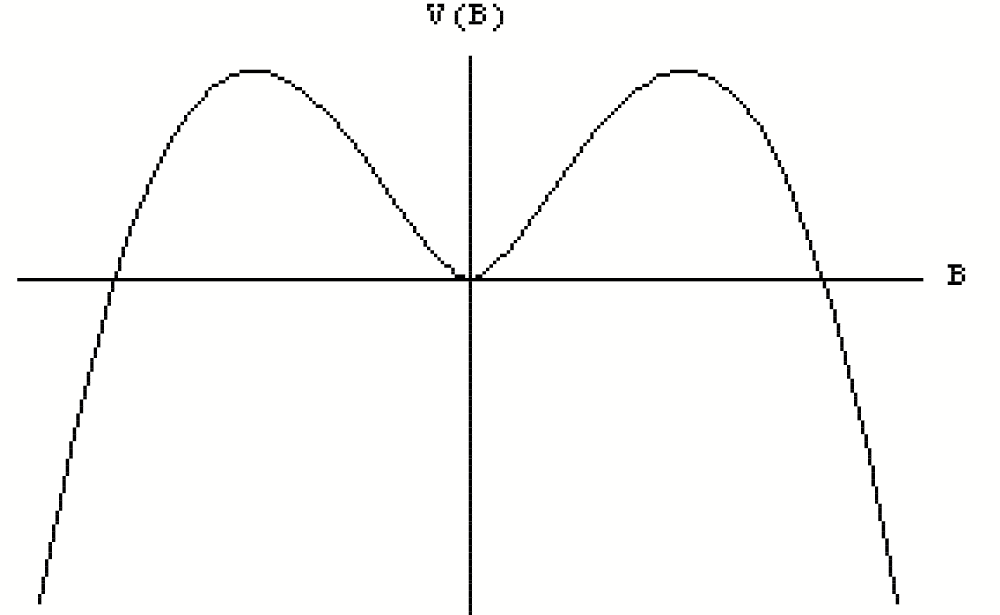}
\parbox{5.5in}{
\caption{1-loop effective potential for QED in a constant background magnetic field.  The apparent instability at large magnetic field strengths is an artifact of the 1-loop approximation.\label{fig:qed.pot}}}
\end{center}
\end{figure}

Therefore the 1-loop contribution to the effective Lagrangian is

\begin{equation}
\Lag^1 = -\frac{\imath}{2} \tr \log\left( e^{\imath p_2 p_1/e B} (p_0^2 - p_1^2 - e^2 B^2 x_1^2 - p_3^2) e^{-\imath p_2 p_1/eB} - \frac{e}{2} \sigma_{\mu \nu} F^{\mu \nu}\right)
\end{equation}
To evaluate this trace, we use the identity

\begin{equation}
\log(x) = \lim_{\epsilon \rightarrow 0} \frac{-\imath^\epsilon}{\Gamma(1+\epsilon)} \int_0^{\infty} dt\ t^{-1+\epsilon} e^{-\imath t x}
\end{equation}
This is related to the method used by Schwinger \cite{Schwinger:1951nm} (who introduced a lower cutoff into the integral instead of dimensionally continuing the argument), and amounts to rewriting the four-dimensional space-time loop momentum integral as the world-line integral of a particle moving in an external potential.  This is a close analogy of the world-sheet formalism of string theory; the world-line proper time parameter $t$ corresponds to a ``world-line modulus'' of the loop in the Feynman graph\footnote{The analogy between string theory and the Schwinger formulation of loop integrals was used in \cite{Dijkgraaf:2002xd} to calculate effective superpotentials in theories with $\N=1$ supersymmetry, by reducing a topological string theory calculation to a field theory calculation in Schwinger's formalism.  We will explain some key points of this work in section \ref{sec:susy.matter}.}.

\begin{eqnarray}
\Lag^1 &=& \frac{\imath^{1+\epsilon}}{2 \Gamma(1+\epsilon)}  \tr \int_0^{\infty} dt~t^{-1+\epsilon} e^{\imath p_2 p_1/eB} e^{-\imath t(p_0^2 - p1^2 - e^2 B^2 x_1^2 - p_3^2)} e^{-\imath p_2 p_1/eB} e^{\imath t \frac{e}{2} \sigma_{\mu \nu} F^{\mu \nu}} \nonumber \\
&=& \frac{\imath^{1+\epsilon}}{2 \Gamma(1+\epsilon)}  \int_0^{\infty} dt~t^{-1+\epsilon} \times \nonumber \\
&&\hspace{0.5in}2 \sum_{\lambda=\pm1} e^{\imath t e B \lambda} \langle x | e^{\imath p_2 p_1/eB} e^{-\imath t(p_0^2 - p_1^2 - e^2 B^2 x_1^2 - p_3^2)} e^{-\imath p_2 p_1/eB}  | x \rangle
\end{eqnarray}
For suitably large $\epsilon$ the integral converges, therefore this representation regulates the calculation.  In the second line we evaluated the trace over the anomalous magnetic moment operator using (\ref{eq:mageigen}), since the operator commutes with everything else in the trace.   The remaining trace may be evaluated as follows \cite{Itzykson:1980rh}
\begin{eqnarray}
&&\langle x | e^{\imath p_2 p_1/eB} e^{-\imath t(p_0^2 - p_1^2 - e^2 B^2 x_1^2 - p_3^2)} e^{-\imath p_2 p_1/eB}  | x \rangle \nonumber \\
&&\quad= \int d^4 p~ d^4 p'~\langle x |p \rangle \langle p |  e^{\imath p_2 p_1/eB} e^{-\imath t(p_0^2 - p_1^2 - e^2 B^2 x_1^2 - p_3^2)} e^{-\imath p_2 p_1/eB}  | p' \rangle \langle p' | x \rangle \nonumber \\
&&\quad= \int d^4 p~ d^4 p'~ \frac{e^{\imath (p-p').x}}{(2 \pi)^4}\langle p |  e^{\imath p_2 p_1/eB} e^{-\imath t(p_0^2 - p_1^2 - e^2 B^2 x_1^2 - p_3^2)} e^{-\imath p_2 p_1/eB}  | p' \rangle\nonumber\\
&&\quad= \int d^4 p~ d^4 p'~ \frac{e^{\imath (p-p').x}}{(2 \pi)} e^{\imath((p_2 p_1)/e B - (p_2' p_1')/eB)} e^{-\imath t (p_0^2 - p_3^2)} \nonumber \\
&&\hspace{2 in}\langle p_1 |  e^{-\imath -t( p_1^2 - e^2 B^2 x_1^2)}  | p_1' \rangle \delta^3((p-p')_{0,2,3}) \nonumber\\ 
&&\quad= \int d\omega~ d\omega'~ \frac{d^3p}{(2 \pi)^3}~ e^{- \imath t(p_0^2 - p_3^2)} e^{\imath(\omega - \omega') (x_1  + p_2/eB)} \langle \omega | e^{\imath t(p_1^2 + e^2 B^2 x_1^2)} | \omega' \rangle \nonumber \\
&&\quad= \frac{e B}{(2 \pi)^2 (\imath t)^{1/2}(-\imath t)^{1/2}} \int d\omega~ d\omega'~ \delta(\omega - \omega') \langle \omega | e^{\imath t(p_1^2 + e^2 B^2 x_1^2)} | \omega' \rangle \nonumber \\
&&\quad= \frac{e B}{(2 \pi)^2 t} \sum_{n=0}^{\infty} \exp(\imath t(n+\frac{1}{2}) 2 e B)
\end{eqnarray}
where we used the result for the energy levels of a harmonic oscillator

\begin{equation}
\tr \exp(\imath t(\frac{P^2}{2 m} + \frac{m \omega}{2} Q^2)) = \sum_{n=0}^\infty \exp(\imath t(n+\frac{1}{2}) \omega)
\end{equation}
Therefore the effective Lagrangian reduces to

\begin{eqnarray}
\Lag^1 &=& \frac{e B \imath^{1+\epsilon}}{8 \pi^2 \Gamma(1+\epsilon)} \int_0^\infty dt~t^{-2+\epsilon} \sum_{\lambda=\pm1} \exp(\imath e t B \lambda) \sum_{n=0}^\infty \exp(i e B t(2n+1)) \nonumber \\
&=& \frac{eB \imath^{1+\epsilon}}{8 \pi^{2}\Gamma(1+\epsilon)} (2 e B)^{1-\epsilon} \int_{0}^{\infty} dt~t^{-2+\epsilon} \frac{e^{-\imath t}+1}{1-e^{-\imath t}}
\end{eqnarray}
Rotating the integration contour $t \rightarrow \imath t$ we obtain
\begin{equation}
\Lag^1 = -\frac{e^{2}B^{2}}{4 \pi^{2}\Gamma(1+\epsilon)} (2 e B)^{-\epsilon} \int_{0}^{\infty} dt~t^{-2+\epsilon} \frac{e^{-t}+1}{1-e^{-t}}
\end{equation}
The integral may now be evaluated using the identity

\begin{equation}
\int_{0}^{\infty} dt\ t^{\sigma-1} \frac{e^{-\nu t} }{1-e^{-t}} = \Gamma(\sigma) \zeta(\sigma, \nu)
\end{equation}
where $\zeta(\sigma, \nu) = \sum_{n=0}^\infty (\nu + n)^{-\sigma}$ is the generalized Riemann zeta function.  Therefore

\begin{eqnarray}
\Lag^1 &=& -\frac{e^2 B^2}{4 \pi^2} \left(\frac{1}{2 e E}\right)^\epsilon \frac{-\Gamma(-1+\epsilon)}{\Gamma(1+\epsilon)} (\zeta(-1+ \epsilon, 0) + \zeta(-1+\epsilon, 1)) \nonumber \\
\end{eqnarray}
In taking the limit $\epsilon \rightarrow 0$, we renormalize the expression using a variant of the $\overline{MS}$ scheme \cite{Peskin:1995ev}\footnote{The difference is that we also subtract the $\log 2$ coming from the coefficient of $\Delta$}:

\begin{equation}
\frac{\Gamma(\epsilon)}{(4 \pi)^{2+\epsilon}} \left(\frac{1}{2 \Delta}\right)^{\epsilon} \rightarrow -\frac{1}{4 \pi^2} \log\left(\frac{\Delta}{\mu^2}\right)
\end{equation}
and use the property of the $\zeta$-function
\begin{equation}
\zeta(-m, \nu) = - \frac{B_{m+1}(\nu)}{m+1}
\end{equation}
where $m=0, 1, \ldots$, and $B_{m+1}(\nu)$ are the Bernoulli polynomials, in particular \\
$B_2(x) = x^2 -x + 1/6$.  Putting this all together, we find for the effective potential

\begin{eqnarray}
\label{eq:qed.veff}
\Veff =  \frac{B^2}{2} - \frac{e^2 B^2}{24 \pi^2} \log(e B/\mu^2)
&=& \frac{B^2}{2} - \frac{b_0 B^2}{2 e} \log(e B/\mu^2)
\end{eqnarray}
where we recognize the 1-loop QED $\beta$-function coefficient $b_0 = \frac{e^3}{12 \pi^2}$.  This potential is plotted in figure \ref{fig:qed.pot}.

For small external fields $e B < \mu^2$ the second term is positive, and the effective potential has a local minimum at $B=0$.  At larger field strengths there appears to be a local maximum and the potential eventually becomes arbitrarily negative.  However, in precisely this limit the 1-loop approximation breaks down, because the quantum correction term dominates and is no longer small compared to the classical term.  Therefore, for large enough magnetic fields one needs to also consider the higher loop corrections.

We turn now to the electric case.  Using the form of $\F = \frac{1}{2}(B^2 -E^2)$, we may obtain the effective potential for a constant background electric field $E \neq 0, B = 0$ by formally continuing $B \rightarrow \imath B \equiv E$ in (\ref{eq:qed.veff}). This introduces a factor of $\imath$ into the argument of the logarithm, and therefore the effective Lagrangian in a background electric field is complex.

Since the amplitude for a vacuum in the far past to remain in the far future is given by

\begin{equation}
\langle 0_{+} | 0_{-} \rangle = e^{\imath \Gamma}
\end{equation}
the probability of vacuum decay, per unit time and volume, is given by 
\begin{equation}
2 \im \Lag = -\im\frac{e^{2}E^{2}}{12 \pi^{2}} \log(\imath)= \frac{e^{2}E^{2}}{24 \pi}
\end{equation}
and the constant electric field background is unstable against pair production of positron/electron pairs.

The result for a non-zero electron mass can also be computed following the above steps, and one finds

\begin{equation}
2 \im \Lag = \frac{e^{2}E^{2}}{4 \pi^{3}} \sum_{N=1}^{\infty} \frac{1}{N^{2}} \exp(\frac{-N \pi m^{2}}{e E})
\end{equation}
which is non-perturbative in the RG-invariant field combination $eE$.  Again we see that the background field method produces non-perturbative information from a perturbative calculation.

In a general constant background with $\F \neq 0, \G \neq 0$, the effective Lagrangian is that of Euler and Heisenberg \cite{Heisenberg:1936qt}, which takes the form (before regularization and renormalization)

\begin{equation}
\label{eq:eh}
\Lag^{1} = \frac{1}{8 \pi^{2}} \int_{0}^{\infty} dt~t^{-1} e^{\imath s m^{2}}\left( e^{2} a b \frac{\cosh(e a t) \cos(e b t)}{\sinh(e a t) \sin(e b t)} \right)
\end{equation}
where $a^{2} - b^{2} = E^{2}-B^{2}$, $a b = E \cdot B$.  A list of references to recent work on this Lagrangian and related matters may be found in \cite{Dunne:2003tr}.

\subsection{Yang-Mills theory}
\label{sec:ym}

To calculate the 1-loop effective action for four-dimensional Yang-Mills theory we again use the background field method.  This calculation and related results were developed by a number of authors, including \cite{Duff:1975ue,Batalin:1977uv,Matinyan:1978mp,Nielsen:1978rm,Pagels:1978dd,Brown:1979bv,Yildiz:1980vv,Flyvbjerg:1980qv,Flyvbjerg:1981rs,Jezabek:1981bs}.

The Yang-Mills Lagrangian is

\begin{equation}
{\cal L} =  -\frac{1}{4} F_{\mu \nu}^a F^{\mu \nu a}
\end{equation}
We split the gauge field into a classical background field $A$ and a fluctuating quantum field $a$:

\begin{equation}
A_\mu^a(x) \rightarrow A_\mu^a(x) + a_\mu^a(x)
\end{equation}
The covariant derivative $(D_\mu)^{ac}=\partial_{\mu} \delta^{ac} + \imath g f^{a b c}A_{\mu}^{b}$ is defined with respect to the background gauge field, and we will integrate over the quantum field $a$ in the path integral.  Then the field strength becomes

\begin{equation}
F_{\mu \nu}^a \rightarrow F_{\mu \nu}^a + D_\mu a_\nu^a - D_\nu a_\mu^a + \imath g f^{a b c} a_\mu^b a_\nu^c
\end{equation}
In background gauge $D_{\mu} A^{\mu a} = 0$, the gauge-fixed Lagrangian is
 
\begin{eqnarray}
\label{eq:lag.ym}
{\cal L} &=&  -\frac{1}{4} (F_{\mu \nu}^a + D_{\mu} a_{\nu}^{a} - D_{\nu} a_{\mu}^{a} +\imath g f^{a b c} a_{\mu}^{b} a_{\nu}^{c})^{2} \nonumber \\
&&\hspace{0.5in}-\frac{1}{2} (D^\mu a^{\mu a})^2 + \overline c^a(-(D^2)^{a c}  -\imath g D^{\mu a d} f^{d b c} a_\mu^b) c^c
\end{eqnarray}
where $c, \overline c$ are the Faddeev-Popov ghosts corresponding to the gauge fixing.

As before, the effective action to 1-loop order is given by evaluating the path integral 
\begin{equation}
e^{\imath\Gamma[A]} = \int {\cal D} a {\cal D} c {\cal D} \overline c e^{\imath \int d^4x {\cal L}}
\end{equation}
to quadratic order in the fluctuations.  Expanding (\ref{eq:lag.ym}) to quadratic order, we find

\begin{equation}
{\cal L}_{\hbox{\it quad}} = -\frac{1}{2} a_\mu^a \left[ (-D^2)^{a b} g^{\mu \nu} - 2 \imath g F^{\mu \nu c} f^{c a b} \right] a_{\mu}^b + \overline c^{a} \left[ -(D^2)^{ab} \right] c^b
\end{equation}
As in QED, the new interaction term $- 2 \imath g F^{\mu \nu c} f^{c a b}a^{\mu a} a^{\nu b}$ is an anomalous magnetic moment interaction of two spin-1 gluons with the background field $F^{\mu \nu c}$.  Introducing the generator of spin-1 Lorentz transformations
\begin{equation}
(J^{\rho \sigma})_{\alpha \beta} = \imath(\delta^{\rho}_{\alpha} \delta^{\sigma}_{\beta} - \delta^{\sigma}_{\alpha} \delta^{\rho}_{\beta})
\end{equation}
the operator $- 2 \imath g F^{\mu \nu c} f^{c a b}$ can be rewritten as $-2 \imath (\frac{1}{2} F^{c}_{\rho \sigma} J^{\rho \sigma})^{\mu \nu} f^{c a b}$, emphasizing the similarity to the operator (\ref{eq:qed.det}) for spin-$\frac{1}{2}$ electrons in QED.  The spin interaction for the ghost fields vanishes since they have spin 0.

Therefore the path integral to 1-loop order is Gaussian and can be evaluated, giving the 1-loop effective Lagrangian

\begin{equation}
{\cal L}_{\mbox{\it \!eff}} = -\frac{1}{4} F_{\mu \nu}^a F^{\mu \nu a} + \frac{\imath}{2} \log \det((-D^2)^{a b} g^{mu \nu} - 2 \imath g F^{\mu \nu c} f^{c a b}) - \imath \log \det((-D^2)^{a b}))
\end{equation}
We can evaluate these determinants by restricting to covariantly constant fluctuations of the gauge fields:

\begin{eqnarray}
D_\rho F^{\mu \nu} = 0
\Leftrightarrow [ D_\rho, F^{\mu \nu} ] = 0 
\end{eqnarray}
where we write the field strength as a matrix in colour space $(F_{\mu \nu})^{a b} = f^{a b c} F_{\mu \nu}^c$, and the second form follows because the covariant derivative in the adjoint representation acts by matrix commutation.  Using the Jacobi identity $[D_\sigma, [D_\rho, F_{\mu \nu}]] + \hbox{\it perm.} = 0$ it follows that

\begin{equation}
[F_{\mu \nu},F_{\rho \sigma}] = 0
\end{equation}
i.e.~the colour matrices $F_{\mu \nu}$ form a commuting set and may be simultaneously diagonalized.  In other words, by a gauge transformation we may rotate a given gauge field configuration into the Cartan subalgebra.  Then

\begin{eqnarray}
\label{eq:ym.color}
{\cal L}^{1} &=& \frac{\imath}{2}\tr \sum_{\alpha} \log(-D^{(\alpha)2} g^{\mu \nu} - 2 \imath g F^{\mu \nu {(\alpha)}}) - \imath \sum_{\alpha} \tr \log((-D^{(\alpha)})^{2}) \nonumber \\
&=& \imath\tr \sum_{\alpha>0} \log(-D^{(\alpha)2} g^{\mu \nu} - 2 \imath g F^{\mu \nu (\alpha)}) - 2 \imath \sum_{\alpha>0} \tr \log((-D^{(\alpha)})^{2})
\end{eqnarray}
where the sum is over the positive roots $\alpha$ of ${\cal G}$.  In the second line we used that each root $\alpha$ is paired with a negative root $-\alpha$, and the zero roots do not contribute.  We also defined effective quantities

\begin{eqnarray}
D_{\mu}^{(\alpha)} &=& \partial_\mu + i g {\alpha}_j A_{\mu}^j \nonumber \\
F_{\mu \nu}^{(\alpha)}&=& \alpha_{j} F_{\mu \nu}^{j}  \nonumber \\
A_{\mu}^{(\alpha)} &=& \alpha_j A_{\mu}^j
\end{eqnarray}
in terms of the simple roots $(\alpha_1, \ldots, \alpha_r)$, $r=\hbox{rank}({\cal G})$, which span root space.

In other words, we have reduced the computation of the 1-loop effective action for a non-abelian gauge group ${\cal G}$ to that of an Abelian $\U(1)^r$ gauge theory, where the $j$'th ``photon'' carries charges $g \alpha_{j}$ with respect to the different $\U(1)$ gauge factors.  The situation is therefore quite similar to that of QED, which we studied in the previous section, except there is more than one type of ``electromagnetic field'', and the charged particles are spin-1 photons, not spin-$\frac{1}{2}$ electrons.

At this point we need to choose the orientation for the effective $\U(1)$ gauge fields in four-dimensional space; when the rank of the gauge group is larger than 1, the ``electromagnetic fields'' may point in different spatial directions.  Most of the early work on this problem either considered $\SU(2)$ \cite{Batalin:1977uv,Matinyan:1978mp}, or chose to align all the effective $\U(1)$ gauge fields parallel \cite{Ambjorn:1979ff}. However, it was shown in subsequent work that for $2 < N \le 4$ the lowest-energy configuration is to choose the fields to be mutually orthogonal \cite{Flyvbjerg:1980qv}.  For $N \ge 4$, i.e.~rank higher than 3, it is no longer possible to choose all vectors to be orthogonal in three-dimensional space, and for $N\rightarrow \infty$ the minimum energy configuration corresponds to an isotropic distribution in space \cite{Flyvbjerg:1981rs,Jezabek:1981bs}.

For simplicity, we will henceforth restrict to the $\SU(2)$ case.  The essential features are seen in this case; in particular we will see that {\it any} choice of covariantly constant field strength gives rise to a vacuum instability, and therefore the 1-loop result is at best only an approximation to the true vacuum.  This instability persists for the non-parallel gauge field orientations mentioned above.

We can now proceed as in section \ref{sec:qed}.  Again taking a constant magnetic field, the eigenvalues of $-2 \imath g F_{\mu \nu}$ are $(\pm 2 g B, 0, 0)$.  The two zero eigenvalues cancel with the contribution from the ghost determinant in (\ref{eq:ym.color}), giving
\begin{equation}
\Lag^{1} = \sum_{\lambda = \pm 1} \tr \log(-D^{2} - 2 \lambda B)
\end{equation}

After manipulations similar to QED, we find

\begin{eqnarray}
\Lag^{1} &=& - \frac{g B }{8 \pi^{2}} (g B)^{-\epsilon}\frac{\imath^{1+\epsilon}}{\Gamma(1+\epsilon)} \int_{0}^{\infty}dt~t^{-2+\epsilon} \times \sum_{\lambda=\pm 1}\exp(-\imath t(1-2 \lambda)) \nonumber \\
&&\hspace{0.5in} \times \sum_{N=0}^{\infty} \exp(-2 \imath t N)
\end{eqnarray}
Note that we can no longer unconditionally rotate the contour by taking $t \rightarrow - \imath t$, because the mode with $(\lambda, N) = (1,0)$ would diverge like $e^{t}$.  This is the {\it unstable mode} found by Nielsen and Olesen \cite{Nielsen:1978rm}, which will give rise to an imaginary part for the 1-loop effective Lagrangian even in the magnetic case.  To proceed, we subtract and add the $(\lambda, N) = (1,0)$ term:

\begin{eqnarray}
\Lag^{1} &=& - \frac{g B }{8 \pi^{2}}(g B)^{-\epsilon} \frac{1}{\Gamma(1+\epsilon)} \left\{ \imath^{1+\epsilon}\int_{0}^{\infty}dt~t^{-2+\epsilon} \frac{e^{\imath t} + e^{-3 \imath t}}{1-e^{-\imath t}} - e^{\imath t} \nonumber \right.\\
&&\hspace{0.5in}\left.+ \imath^{1+\epsilon}\int_{0}^{\infty}dt~t^{-2+\epsilon} e^{\imath t}\right\} \nonumber \\
&=& - \frac{g B}{8 \pi^{2}}(g B)^{-\epsilon} \frac{1}{\Gamma(1+\epsilon)} \left\{ \imath^{1+\epsilon}\int_{0}^{\infty}dt~t^{-2+\epsilon} \frac{e^{-\imath t} + e^{-3 \imath t}}{1-e^{-\imath t}}\right. \nonumber \\
&&\hspace{0.5in}\left.+ \imath^{1+\epsilon}\int_{0}^{\infty}dt~t^{-2+\epsilon} e^{\imath t}\right\} \nonumber \\
&=& - \frac{g B}{8 \pi^{2}}(g B)^{-\epsilon} \frac{1}{\Gamma(1+\epsilon)} \left\{ \imath^{1+\epsilon} \left(\frac{-\imath}{2}\right)^{-1+\epsilon} \int_{0}^{\infty}dt~t^{-2+\epsilon} \frac{e^{-t} + e^{-3 t}}{1-e^{-t}} \right.\nonumber \\
&&\hspace{0.5in}\left.+ \imath^{1+\epsilon} \imath^{-1+\epsilon} \int_{0}^{\infty}dt~t^{-2+\epsilon} e^{- t}\right\} \nonumber \\
&=& - \frac{g B}{8 \pi^{2}}(g B)^{-\epsilon} \frac{1}{\Gamma(1+\epsilon)} \left\{ -2\left(\frac{1}{2}\right)^{\epsilon} \int_{0}^{\infty}dt~t^{-2+\epsilon} \frac{e^{-t} + e^{-3 t}}{1-e^{-t}} \right.\nonumber \\
&&\hspace{0.5in}\left.+ (-1)^{\epsilon} \int_{0}^{\infty}dt~t^{-2+\epsilon} e^{- t}\right\}
\end{eqnarray}
where we rotated the two integration contours by $t \rightarrow -\imath t$, $t \rightarrow \imath t$ respectively.  The integrals may now be evaluated in terms of zeta functions, giving

\begin{eqnarray}
\Lag^{1} &=& -\frac{(g B)^{2}}{8 \pi^{2}} \frac{\Gamma(\epsilon)}{(-1+\epsilon) \Gamma(1+\epsilon)} \left[ \left(\frac{1}{2 g B}\right)^{\epsilon}(-2) (\zeta(-1+\epsilon, \frac{1}{2}) \right.\nonumber \\
&&\hspace{0.5in}\left.+ \zeta(-1+\epsilon, \frac{3}{2})) + \left(\frac{-1}{g B}\right)^{\epsilon}\right] \nonumber \\
&=& -\frac{(g B)^{2}}{8 \pi^{2}} \log(g B/\mu^{2}) \left[ -2(\zeta(-1,\frac{1}{2}) + \zeta(-1,\frac{3}{2})) + 1\right] \nonumber \\
&&\hspace{0.5in}- \frac{(g B)^{2}}{8 \pi^{2}} \log(-1) \nonumber \\
&=& - \frac{11}{6}\frac{(g B)^{2}}{8 \pi^{2}} \log(g B/\mu^{2}) + \imath \frac{(g B)^{2}}{8 \pi} \nonumber \\
&=& + \frac{\beta(g)}{2 g} B^{2} \log(g B/\mu^{2}) + \imath \frac{(g B)^{2}}{8 \pi}
\end{eqnarray}
where in the last line we recognized the 1-loop $\beta$-function coefficient.  As before the pure electric field result may be obtained by analytic continuation. If we consider a background with $\G \neq 0$, then the effective Lagrangian will be a generalization of the Euler-Heisenberg Lagrangian (\ref{eq:eh}) \cite{Batalin:1977uv,Matinyan:1978mp}.  In all cases the background is unstable, in contrast to QED, for which only the electric background is unstable.

Note that because of asymptotic freedom the sign of the 1-loop term is opposite to that of  QED (\ref{eq:qed.veff}); therefore the effective potential has a similar form to figure \ref{fig:gn.pot}.  The lesson we can draw from this analysis is that the ``perturbative vacuum'', where we consider excitations around the zero-field background, is an unstable field configuration.  The Yang-Mills vacuum lowers its energy by spontaneously generating a non-zero background field.  This can be seen as a vacuum anti-screening effect by the gluons, which are charged under the gauge group and can act as sources for other gluons.  Turning on a covariantly constant background field indeed lowers the vacuum energy, but this field configuration is itself unstable (not to mention violating Lorentz invariance), so the ``true'' vacuum is some other field configuration.  An ansatz for the vacuum (the ``Copenhagen vacuum'') was proposed in \cite{Nielsen:1979tr}, based on exciting the unstable mode of the constant-field vacuum.

The background field method is non-perturbative in the background field (since it is not used as an expansion parameter), which allowed us to make some progress, but excitations around this field still must be calculated perturbatively.  This means that we can only trust our 1-loop calculation when the effective coupling constant is small, however this is counteracted by the negative sign of the 1-loop $\beta$-function, which tells us that $g$ will grow towards the IR.  

Explicitly, to 1-loop order the running of the Yang-Mills coupling constant is given by

\begin{equation}
g_{\hbox{\it eff\,}}^{2}(q) = \frac{g^{2}}{1+\frac{11 g^{2}}{96 \pi^{2} N} \log(q/\mu)}
\end{equation}
which diverges at the finite energy scale

\begin{equation}
q = \mu \exp(-\frac{96 \pi^{2} N}{11 g^{2}}) \equiv \Lambda_{\hbox{\small YM}}
\end{equation}
Therefore, we can not trust our 1-loop effective potential at energies comparable to or lower than $\Lambda_{\mbox{\small YM}}$.  Nevertheless, it is expected (based on lattice simulations and other theoretical work) that the qualitative picture remains true, and the vacuum of Yang-Mills theory is associated to non-trivial gauge field backgrounds, which give rise to confinement, generation of a mass gap (the appearance of massive glueballs in the spectrum replacing the massless gluons), and other poorly-understood low-energy physics.

\subsection{Constraints on the effective potential from the trace anomaly}
\label{sec:ym.anom}

We have seen that the effective potential of quantum field theories must be consistent with the trace anomaly, in particular it satisfies the Callan-Symanzik equation.  Once we have calculated the quantities $\beta$ and $\gamma$ for a particular theory, we can use the Callan-Symanzik equation to constrain the possible form of corrections to the classical potential in an $\it arbitrary$ field background.

For $\SU(2)$ Yang-Mills theory we found that the effective potential in a covariantly constant field background with $\F = \frac{1}{4} F_{\mu \nu}^{a} F^{\mu \nu a} \neq 0$, $\G = \frac{1}{4} F_{\mu \nu}^{a} \tilde{F}^{\mu \nu a} = 0$ is (suppressing the trace over the colour indices):

\begin{equation}
\label{eq:ym.veff}
\Veff = \frac{1}{4} F^{2} + \frac{11 g^{2}}{16 \times 48 \pi^{2} N} F^{2} \log(g^{2} F^{2}/\mu^{4})
\end{equation}
Applying the Callan-Symanzik equation

\begin{equation}
\label{eq:cs2}
\left[ \mu \frac{\partial}{\partial \mu} + \beta(g) \frac{\partial}{\partial g} - \gamma F \frac{\partial}{\partial F} \right] \Veff = 0
\end{equation}
we find that $\beta = \gamma g = -\frac{11g^{3}}{3 (4 \pi)^{2} N}$.  These are properties of the Lagrangian, and do not depend on the particular background we evaluate it in; moreover to 1-loop order they are independent of the renormalization scheme.

We now look for more general functions $V$ that solve (\ref{eq:cs2}), to see what possible corrections may appear in other field backgrounds.  The equation (\ref{eq:cs2}) can be solved by a series of the form

\begin{equation}
V = \sum_{i=0}^{\infty}  a_{i}(g) F^{2} \log(g F/\mu^{2})^{i}
\end{equation}
where the $a_{i}(g)$ satisfy a set of coupled differential relations of the form

\begin{equation}
\label{eq:ym.diffrel}
\gamma g \frac{d a_{i}}{d g} - 2 \gamma a_{i} + (i+1) a_{i+1} = 0
\end{equation}
where we have used the relation $\beta = \gamma g$ that we found above.

If we assume that to 1-loop order, the correction series in a particular background terminates at some order $k$, then we can integrate the relations 
 (\ref{eq:ym.diffrel}) and impose that the function $V$ reduces to the classical potential $V = \frac{1}{4}F^{2}$ plus corrections that are higher powers of $g$.  We find
 
\begin{eqnarray} 
a_{k} &=& 0 \nonumber \\
a_{k-1} &=& C_1 g^{2}\nonumber \\ 
a_{k-2} &=& C_2 g^{2} - \frac{(k-1) C_1}{\alpha} \nonumber \\
a_{k-3} &=& \ldots
\end{eqnarray}
where we define the 1-loop $\beta$ function $\beta(g) = \alpha g^{3}$, $\alpha = -\frac{11}{3 (4 \pi)^{2} N}$.  Thus, consistency with tree level fixes $k=2$ and the value of $C_1$, and subject to the assumptions above, the general effective potential for a (not necessarily constant) background with $\F \neq 0, \G = 0$ is

\begin{eqnarray}
\label{eq:ym.univ}
V &=& \frac{1}{4} F^{2} + C_2 g^{2} F^{2} - \frac{\alpha g^{2}}{8} F^{2} \log(g^{2} F^{2}/\mu^{4}) \nonumber \\
&=& \frac{1}{4} F^{2} + C_2 g^{2} F^{2} + \frac{11 g^{2}}{8 \times3 (4 \pi)^{2} N } F^{2} \log(g^{2} F^{2}/\mu^{4})
\end{eqnarray}
The unfixed constant $C_2$ reflects the ability to shift the arbitrary renormalization scale $\mu$, as well as the possible instability of the field background if $C_2$ is complex.  Similar arguments constrain the form of $V$ in an arbitrary background with $\G \ne 0$, which gives a generalization of the Euler-Heisenberg Lagrangian \cite{Batalin:1977uv}. Note in particular that the sign of the 1-loop contribution -- and therefore the existence of the unstable perturbative vacuum -- depends on the negative sign of $\beta(g)$.

Note that this method does not rely on knowledge of the precise the form of $F_{\mu \nu}^{a}$ in 4-dimensional space-time, or in the internal (colour) space.  Non-constant field configurations may have complicated derivative terms in their effective Lagrangian, but for configurations that satisfy our assumptions, the trace anomaly constrains the non-derivative terms to reduce essentially to the form of the constant field result obtained above. 
However, as noted above this does not allow us to reliably estimate the vacuum expectation value $\langle F^2 \rangle$, because the 1-loop approximation still breaks down before we reach the dynamical scale $\Lambda$ characteristic of confinement\footnote{A more reliable estimate of $\langle F^2 \rangle$ for QCD was made by Shifman et.~al.~\cite{Shifman:1979bx} using charmonium sum rules.}.

In section \ref{sec:susy} we will turn this argument around, and use 1-loop anomalies to compute the effective superpotential of $\N=1$ supersymmetric Yang-Mills theory directly.  The 1-loop anomaly calculation is exact in supersymmetric theories, which allows us to find the exact effective superpotential without needing to perform an explicit path integral calculation around the vacuum field configuration.  Indeed, the precise nature of the $\N=1$ vacuum is unknown, although we can compute some of its properties exactly.

\section{$\N=1$ supersymmetric gauge theories}
\label{sec:susy}

In a supersymmetric theory, the Lagrangian may contain terms of the form

\begin{equation}
\int d^{2} \theta~W(\Phi_{i}) + h.c.
\end{equation}
where the integral is over half of superspace, and $W$ is the {\it superpotential} of the theory.  It has dimension 3 and is a function of the chiral superfields $\Phi_{i}$ and not of their antichiral hermitian conjugates $\overline \Phi_{i}$.  The supersymmetric vacua of the theory are determined by the ``F-term'' constraints

\begin{equation}
\frac{\partial W}{\partial \Phi_i} = 0
\end{equation}
modulo complexified gauge transformations.  In terms of the superpotential, the ordinary bosonic potential of the theory is given by

\begin{equation}
V(\phi_i) = \sum_i | \frac{\partial W}{\partial \phi_i}|^2 + \frac{g^2}{2} (D^a)^2
\end{equation}
where $\phi_i$ are the lowest components of the chiral superfields $\Phi_i$ and $D^a = \sum_i | \phi_i | ^2 t^a$, where $t^a$ are the generators of the gauge group.

There are two key results that allow us to compute the effective superpotential exactly in many supersymmetric theories: in a {\it Wilsonian} approach where we integrate over loop momenta down to a momentum cutoff, the superpotential only receives one-loop and non-perturbative corrections; and it is a holomorphic function of the chiral superfields and coupling constants.  The meaning of these statements is somewhat subtle, and bears further explaining.

Until now, we have considered the effective potential defined by the non-derivative terms in the generating functional of 1-particle irreducible (1PI) diagrams of the theory that is obtained by integrating over the fluctuating fields.  We found that in four-dimensional gauge theories this object receives contributions to all loop orders in perturbation theory, corresponding to Feynman diagrams in the background field with arbitrarily many internal loops.  This remains true in a supersymmetric theory.  Moreover, higher loop corrections will generically not be holomorphic.

The Wilsonian approach to the effective action is to integrate over all loop momenta down to some cutoff scale; the resulting functional depends on the lower-momentum modes but has no dependence on momenta higher than the cutoff.  If we integrate all the way to zero momentum we would recover the 1PI generating functional.  In supersymmetric gauge theories, Shifman and Vainshtein \cite{Shifman:1986zi} showed that the 2-loop and higher contributions are infrared effects; they only enter the Wilsonian effective action as the cutoff is taken to zero, and in computing matrix elements of Wilsonian quantities (averaging them over the external fields).  For finite cutoff, the terms appearing in the Wilsonian effective action arise only from tree-level and 1-loop contributions.

It is important to note that the parameters (fields, coupling constants) that appear in the Wilsonian effective action are not the physical quantities that would be measured in an experiment; indeed, the latter receive corrections to all orders.  It would appear that the Wilsonian approach is missing the effects of the higher-loop contributions; as we saw in non-supersymmetric Yang-Mills theory the higher loop corrections are vital for understanding the vacuum structure, because they dominate at low energies.

The resolution, emphasized by \cite{Shifman:1991dz,Dine:1994su}, is that the all-loop, non-holomorphic 1PI effective superpotential may be brought into the 1-loop, holomorphic Wilsonian form by a suitable (non-holomorphic, field- and coupling- dependent) change of variable.  In other words, the 1PI effective superpotential is {\it resummed} into the Wilsonian form by this change of variable.  This means that in supersymmetric theories the higher order corrections to the effective superpotential arising from the trace anomaly must all be related to the form of the 1-loop term, written in different variables.  For example, in $\N=1$ supersymmetric Yang-Mills theory this is intimately related to the existence of the exact NSVZ $\beta$-function \cite{Novikov:1983uc}, which has the form of a geometric series.

Therefore, for supersymmetric theories we can confidently use the 1-loop Wilsonian effective potential to study the theory beyond the range where 1-loop perturbation theory naively breaks down, because we know that written in terms of physical quantities the 1-loop calculation sums the contributions to all loop orders.  If in addition the non-perturbative corrections to the effective superpotential are calculable (by holomorphy and symmetry constraints, this is often the case), then we can obtain the exact effective superpotential, and by extension, exact results about the vacuum of the theory.  The price is that to rewrite these exact Wilsonian results in terms of physical quantities one must undo the complicated change of variables.

\subsection{$\N=1$ Yang-Mills}
\label{sec:susy.ym}

The effective superpotential for $\N=1$ Yang-Mills was constructed in \cite{Veneziano:1982ah}, by writing an effective Lagrangian whose symmetry transformations reproduced the correct 1-loop anomalies.  This is essentially the approach we used in earlier sections.

The Lagrangian for $\N=1$ Yang-Mills theory is:
\begin{eqnarray}
{\cal L} =-\frac{1}{4g^{2}} F_{\mu \nu}^{a} F^{\mu \nu a} + \theta F_{\mu \nu}^{a} \tilde{F}^{\mu \nu a}+  \frac{\imath}{2} \overline{\lambda}^{a} \dslash D_{a b} \lambda^{b} + \ldots
\label{sym}
\end{eqnarray}
where we have suppressed the gauge-fixing, ghost and auxilliary terms.  In superfield notation this can be written as

\begin{equation}
\Lag = -\int d^2 \theta~\frac{1}{4 g^{2}} \tr W_{\alpha} W^{\alpha}  + h.c. = \int d^2 \theta~\tau  S + h.c.
\end{equation}
where we define

\begin{eqnarray}
S&=& -\frac{1}{32 \pi^{2}} \tr W_{\alpha}^{2} \nonumber \\
\tau &=& \frac{8 \pi^{2}}{g^{2}} + \imath \theta
\end{eqnarray}
$S$ is the ``gaugino bilinear superfield'', whose lowest component is $\tr \lambda_{\alpha}^{2}$.  In particular, $S$ and $\tau$ are both complex.

The expansion of the composite superfield $S$ in terms of component fields includes a term $\tr (F_{\mu \nu}^{a})^2$ quadratic in the Yang-Mills field-strength tensors, which one might be tempted to identify with a scalar ``glueball'' operator of the Yang-Mills theory.  However, $S$ cannot be interpreted as a dynamical glueball superfield, because the Yang-Mills field-strengths appear as auxilliary fields in $S$ and are therefore non-dynamical \cite{Sannino:2003xe}.  The approach of studying the vacuum of $\N=1$ Yang-Mills theory by introducing a non-dynamical composite field is essentially the same approach we took in probing the Gross-Neveu model for the existence of a symmetry-breaking fermion condensate; here we are probing for a gaugino condensate, to which we associate the composite field $S$ that includes the gaugino bilinear.
In this sense, the effective superpotential $W(S)$ we will obtain is part of a ``minimal Lagrangian'' that describes the symmetries and anomalies of the theory, but is not an effective Lagrangian for physical degrees of freedom.  In particular, upon extremizing the effective superpotential $W(S)$ we will obtain the value of the gaugino condensate in the vacua of $\N=1$ Yang-Mills.

As before, the Callan-Symanzik equation constrains the form of corrections arising from the anomalous breaking of scale-invariance\footnote{In $\N=1$ Yang-Mills theory the trace anomaly is part of an anomaly multiplet that also includes the axial anomaly, and a superconformal anomaly.  By supersymmetry, the constraints from the other anomalies are equivalent to that of the trace anomaly.}:

\begin{equation}
\left[\gamma S \frac{\partial}{\partial S}- \beta(g) \frac{\partial}{\partial g} - \mu \frac{\partial}{\partial \mu} \right]  \Weff(S) = 0
\end{equation}
As we have seen in previous examples, it can be solved by a function of the form

\begin{equation}
\Weff(S) = \frac{C_1}{g^{2}} S + C_2 S + C_3 S \log(S/\mu^{3})
\end{equation}
and we find $\gamma = 0$, $C_1=8 \pi^{2}$, $C_3 = \frac{16 \pi^{2} \beta(g)}{3 g^{3}} = N$, where $\beta(g) = -\frac{3N g^{3}}{(4 \pi)^{2}}$ to 1 loop.  Therefore

\begin{eqnarray}
\Weff(S)&=& \tau S + C_2 S + N S \log(\frac{S}{\mu^{3}}) \nonumber \\
&=& C_2 S + N S\log(S/\Lambda^{3})
\end{eqnarray}
where we introduced the dynamical scale $\Lambda$ via the running coupling relation

\begin{equation}
\label{eq:runningcoupling}
\tau(\mu) - 3 N \log \mu = 3 N \log \Lambda 
\end{equation}

As in other examples, the constant $C_2$ is not fixed by symmetries and may depend on the renormalization scheme.  A value can be fixed following the approach of \cite{Cachazo:2002ry}.  Using an instanton calculation \cite{Novikov:1983uc}, the value of the gaugino condensate can be obtained directly, giving rise to the value of the superpotential in the vacuum:

\begin{equation}
\label{eq:sym.vac}
\Weff(\Lambda)= N (\Lambda^{3 N})^{1/N}
\end{equation}
The field $S$ can be introduced by performing a Legendre transformation

\begin{equation}
\Weff(\Lambda, C, S) = N C^{3} + S \log(\frac{\Lambda^{3 N}}{C^{3 N}})
\end{equation}
Integrating out $S$ recovers the previous expression (\ref{eq:sym.vac}).  If instead we integrate out $C$, then we recover the Veneziano-Yankielowicz superpotential

\begin{equation}
\Weff(S, \Lambda) = N S( \log(\frac{S}{\Lambda^{3}}) - 1)
\end{equation}
which fixes the constant $C_2 = -N$.

Since the field $S$ here is complex, the F-term constraint $\frac{\partial W}{\partial S}=0$ gives $N$ distinct vacua (related by a phase, i.e.~vacuum angle $\theta$)

\begin{equation}
\label{eq:gauginocond}
\langle S \rangle = e^{2 \pi \imath k/N_c} \Lambda^3\quad\quad k=0, \ldots, N_c-1
\end{equation}

Furthermore, as noted in the previous section, this Wilsonian effective superpotential does not receive corrections beyond one loop.  Therefore the vacuum expectation value $\langle S \rangle \propto \langle \tr W_\alpha W^\alpha \rangle$ is exact, and the $N$ vacua of $\N=1$ supersymmetric $\SU(N)$ Yang-Mills theory have a non-vanishing gaugino condensate.

Note that the Callan-Symanzik anomaly calculation does not assume a particular form of the background gauge field configuration.  A covariantly constant background field strength was considered in \cite{Kay:1983mh}, generalizing the Yang-Mills calculations reviewed in section \ref{sec:ym}.  As in the non-supersymmetric case, a constant background field strength causes the vacuum energy to decrease, but there is still an instability at the 1-loop level\footnote{This is not surprising since this field configuration is not supersymmetric.}.  A field theoretical derivation of the Veneziano-Yankielowicz superpotential is not known - this would amount to knowing the field configuration in the $\N=1$ Yang-Mills vacuum and integrating over the fluctuations around this background.

\section{$\N=1$ theories with matter}
\label{sec:susy.matter}

One of the starting-points for the recent work on $\N=1$ gauge theories with adjoint matter was the conjecture \cite{Dijkgraaf:2002fc,Dijkgraaf:2002vw,Dijkgraaf:2002dh} that the effective superpotential is computed by an associated bosonic large-$N$ matrix integral, which may be evaluated by counting planar diagrams.  This conjecture comes from string theory, and follows a chain of reasoning that is the culmination of extensive research on the relationship between string theory and gauge theories.

The steps in the conjecture can be summarized as follows: type II string theory on certain Calabi-Yau manifolds (``generalized conifolds'') is known to reduce to $\N=1$ Yang-Mills theories in a limit that decouples gravity; at low energies these geometrical spaces undergo a ``geometric transition'', where a cycle in the geometry shrinks to zero size and is replaced by a different cycle of finite size.  This is a geometrical analogue to confinement of the Yang-Mills theory at low energies.  If we instead consider B-type topological strings on these spaces, the topological string amplitudes reproduce the F-terms (superpotential) of the corresponding gauge theory.  Therefore, after the geometric transition they should give us the gauge theory effective superpotential.  However, the path integral of the topological B-model on these spaces reduces to a large $N$ matrix integral.  Following the chain of arguments, the effective superpotential of $\N=1$ Yang-Mills theories should reduce to a large $N$ matrix integral.  Thus, string theory provided an entirely unexpected computational tool for studying the effective superpotential of $\N=1$ gauge theories with matter.  

In practical terms, we can illustrate the technique as follows.  Suppose we start with a $\SU(N_c)$ gauge theory with $\N=1$ supersymmetry and a chiral superfield $\Phi$ in the adjoint representation, with a tree-level superpotential that contains a mass term and cubic self-interaction:

\begin{equation}
W = \int d^2 \theta~\left(\frac{m}{2} \Phi^{2} + \frac{g}{3} \Phi^3\right)
\end{equation}
String theory suggests that the effective superpotential of this theory $\Weff(S)$, written in terms of the gaugino bilinear $S$, receives contributions from two sources:  Veneziano-Yankielowicz terms arising from the strongly-coupled dynamics of the gauge field, and contributions from the matter field $\Phi$.  According to the conjecture, the only contributions of the matter field $\Phi$ to the effective superpotential come from the {\it planar} $\Phi$ diagrams of the theory (even at finite $N_c$) where we insert the external $S$ field once into each of the index loops of the $\Phi$ diagrams. Furthermore the effective superpotential has {\it no dependence} on the internal loop momenta of the diagrams!

The meaning of this result is that the superpotential for such theories is an essentially combinatorial object, depending only on the counting of ribbon diagrams with planar topology.  It has been known for a long time that these planar diagrams are counted by a zero-dimensional matrix integral \cite{Brezin:1978sv}, and we can often evaluate the free energy of this ``matrix model'' exactly.  

We saw in the previous section that in non-supersymmetric field theories the need to integrate over loop momenta was a serious complication for extending the computation of the effective action to higher orders.  What is the field theory process that removes the contribution of loop integrals when supersymmetry is present?  As in non-supersymmetric theories, we can understand the field theory results in two ways: using anomalies \cite{Cachazo:2002ry} and by evaluating the path integral \cite{Dijkgraaf:2002xd}.  We will summarize the results of these papers, and refer to the original papers for the details.

The technique of using anomalous symmetries to solve for the effective superpotential has been extended to a large class of $\N=1$ theories \cite{Cachazo:2002ry, Seiberg:2002jq, Brandhuber:2003va}, where the relevant anomalies are of generalized Konishi type. This approach relies on the fact that the set of chiral primary fields -- those that can enter the effective superpotential -- are closed under addition and operator product, up to terms that vanish when evaluated in a supersymmetric vacuum; in other words the chiral primary fields generate a ring structure, the {\it chiral ring}.  Moreover, elements of the chiral ring are independent of position, so the chiral ring is a global structure.

Using the properties of the chiral ring, it was shown that the (anomalous) symmetries of the theory (particularly the generalized Konishi anomalies) restrict the possible superpotential contributions to the planar diagrams with insertions of $S$~\footnote{By contrast to the trace anomaly, the generalized Konishi anomalies contribute to all orders of perturbation theory, although in a simple, and often summable way.}.  Then, the Ward identities associated to the generalized Konishi anomalies are shown to be equivalent to the loop equations of the matrix model, which are Dyson-Schwinger equations for the correlation functions, and which can be solved using matrix integral techniques to determine the effective superpotential exactly.

A complementary field theory approach \cite{Dijkgraaf:2002xd} used the background field method to study $\N=1$ gauge theories.  They showed that as a consequence of symmetries, it is again only the planar diagrams of the gauge theory that can contribute to the effective superpotential, and moreover supersymmetry implies that after the loop diagrams are summed in the Schwinger formalism, the loop momentum dependence in the diagram sum exactly cancels between bosonic and fermionic contributions.  Since there is no remaining dependence on loop momenta, the resulting effective superpotential reduces to the zero-dimensional matrix model calculation.  A key feature seen in this approach is that the individual gauge theory loop diagrams do depend on loop momenta, but after summing over all diagrams the momentum dependence exactly cancels.

There are several remarkable consequences of these results.  In many cases the associated matrix integral can be directly solved (corresponding to summing the Feynman diagram expansion to all orders).  However, in more complicated examples where the diagram series cannot easily be summed using known techniques, a perturbative expansion of the ribbon diagrams (up to some order in the number of index loops) gives a perturbative expansion of the effective superpotential $W(S)$, which upon extremization generates an expansion of the vacuum gluino condensate $\langle S \rangle \sim \langle \lambda \lambda \rangle$ as a sum of fractional instanton contributions.  As emphasized in \cite{Dijkgraaf:2002dh}, and as we have seen in other examples above, the perturbative loop expansion of the gauge theory in terms of an appropriate choice of composite operator yields non-perturbative information about the vacuum.

These results have been checked and extended in a large number of papers, and the deeper consequences for the quantum structure of gauge theories are still being explored.  In the remainder of this thesis we will discuss our contributions to this area of research.

\chapter{Effective Superpotentials from Geometry}
\label{ch:geom}

In this chapter we first review how gauge theories with $\N=1$ supersymmetry may be obtained from string theory, and how string theory provides new tools for analyzing their low energy structure.  The simplest examples have $\U(N)$ or $\SU(N)$ gauge group with matter in the adjoint and fundamental representation.  In \cite{Ashok:2002bi} we extended the analysis to $\SO(N)$ and $\Sp(N)$ gauge groups with adjoint matter, and in \cite{Kennaway:2003jt} we showed that a careful consideration of UV divergences requires the inclusion of a maximal number of fundamental matter fields in order to regulate those divergences.  We then studied in detail the structure of $\U(N)$ and $\SU(N)$ theories with adjoint and fundamental matter and developed simple, general formulae for the effective superpotentials, which reduce in special cases to previously known results. Other $\N=1$ theories have been treated in the literature including theories that do not arise from soft supersymmetry breaking of an $\N=2$ theory \cite{Landsteiner:2003ua}.

\section{Geometric engineering of gauge theories}
\label{sec:cy}
We begin by reviewing the construction from string theory of a softly broken $\N=2$ gauge theory with $SU$/$SO$/$Sp$ gauge group \cite{Cachazo:2002pr,Cachazo:2001jy,Edelstein:2001mw,Dijkgraaf:2002fc}. Consider type IIB string theory compactified on the non-compact $A_1$ fibration
\begin{equation}
u^2 + v^2 + w^2 + W'(x)^2 = 0,
\end{equation}
where $W(x)$ is a degree $n+1$ polynomial, which will later be related to the tree level superpotential for the adjoint chiral superfield $\Phi$.  This fibration has singularities at the critical points of $W(x)$.  In the neighborhood of those singularities, we can introduce the coordinate $x'=W'(x)$. Then it is easy to see that the singularities are all conifold singularities.

This generalized conifold can be de-singularized in two ways: it can be resolved or it can be deformed.  The resolution is given by the surface
\begin{equation}\label{resgenconi}
\left(\begin{array}{cc} u+iv & w+iW'(x) \\ -w+iW'(x) & u-iv \end{array}\right)
\left(\begin{array}{c} \lambda_1 \\ \lambda_2 \end{array}\right)=0
\end{equation}
in $\BC^4\times\BP^1$.  In this geometry each singular point is replaced by a $\BP^1$.  These $\BP^1$'s are disjoint, holomorphic, have the same volume and are homologically equivalent.  The latter property can be seen by making use of the fibration structure away from $W'(x)=0$.  This $A_1$ fibration over the $x$ plane induces a fibration of non-holomorphic $S^2$'s over the $x$ plane.  This $S^2$ cannot shrink to zero size as one approaches a critical point of $W$ in the $x$ plane, but it becomes the holomorphic $\BP^1$ of the resolution.

We can now construct a softly broken $\N=2$ $U(N)$ gauge theory with tree level superpotential $W(x)$ by wrapping $N$ D5-branes around the $S^2$.  The adjoint chiral superfield $\Phi$ parameterizes the normal deformations of the D-branes, and since these deformations are obstructed in the Calabi-Yau geometry there is a superpotential for $\Phi$, which is identified with the function $W(x)$ that describes the nontrivial $A_{1}$ fibration of the generalized conifold \cite{Brunner:1999jq,Kachru:2000ih}.  This is an UV definition of the theory; in fact it describes quantum gravity coupled to the gauge theory, because the excitation spectrum also includes the closed strings that propagate away from the D-branes, and which give rise to gravitons in the particle spectrum.  The bulk modes and massive open string modes can be decoupled by taking the 't Hooft limit $N \rightarrow \infty$, $g_{s} \rightarrow 0$, $\lambda = g_{s} N = {\hbox{\it const.}}$, which leaves only the lowest open string modes, the gauge and matter fields.

A classical supersymmetric vacuum of the gauge theory is obtained by minimizing the volume of the D5-branes. This amounts to distributing a collection of $N_i$ D5-branes over the $n$ minimal-volume holomorphic $\P^1$'s at the critical points of $W$.  The $U(N)$ gauge symmetry is then spontaneously broken to $U(N_1)\times\cdots\times U(N_{n-1})$.  $\SU(N)$ gauge group can be treated by decoupling the overall $\U(1) \subset \U(N)$ trace, which is a free theory.

If we flow this ultraviolet theory to the infrared (low energies), there will be a confinement transition.  In string theory this is described by a ``geometric transition'' in which the resolved conifold geometry with wrapped D5-branes is replaced by a deformed conifold
geometry~\cite{Vafa:2000wi}
\begin{equation}
\label{eq:genconifold}
u^2 + v^2 + w^2 + W'(x)^2 - f(x) = 0,
\end{equation}
where $f(x)$ is a polynomial of degree $n-1$.  For a reasonably small $f(x)$, each critical point of $W'(x)$ is replaced by two simple zeros of $W'(x)^2 - f(x)$.  This means that each $\BP^1_i$ is replaced by a 3-sphere $A_i$ with 3-form RR-flux $H$ through it, equal to the amount of D5-brane charge on the $\BP^1_i$.  After the geometric transition there are no more D-branes, so there are only closed strings in the spectrum.

The coefficients in $f(x)$ are normalizable modes that are localized close to the tip of the conifold.  The coefficients in $f(x)$ are determined by the periods
\begin{equation}
S_i=\frac{1}{2\pi i}\int_{A_i}\Omega.
\end{equation}
These periods $S_i$ are to be identified with the gaugino bilinear superfields of the gauge theory.  There are non-compact 3-cycles $B_i$ that are dual to the $A_i$.  The periods of the B-cycles are
\begin{equation}
\pder{\F_0}{S_i}=\int_{B_i}\Omega,
\end{equation}
where $\F_0$ is the prepotential of the Calabi-Yau geometry.  One needs to introduce a cutoff in order to make these periods finite; we will discuss the physical meaning of this cutoff in section \ref{sec:uv}.

The flux through the cycles $A_i$ is determined in terms of the RR-charges of the D-brane configuration
\begin{equation}
\begin{split}
N_i&=\int_{A_i}H,
\end{split}
\end{equation}
and the flux through the cycles $B_i$ is given in terms of the coupling constants
\begin{equation}
\tau_i=\int_{B_i}H.
\end{equation}
The effective superpotential $\Weff(S_{i})$  is then given by the  flux superpotential \cite{Taylor:1999ii,Becker:1996gj,Gukov:1999ya,Polchinski:1996sm}
\begin{equation}
\Weff(S_i)=\int H\wedge\Omega,
\end{equation}
Using the expressions for the periods and the fluxes, we get
\begin{equation}
\label{eq:weffclosed}
\Weff(S_i)=\sum_{i}\left(N_i\pder{\F_0}{S_i}+ \tau_iS_i \right). 
\end{equation}

\begin{figure}[t]
\begin{center}
\epsfig{file=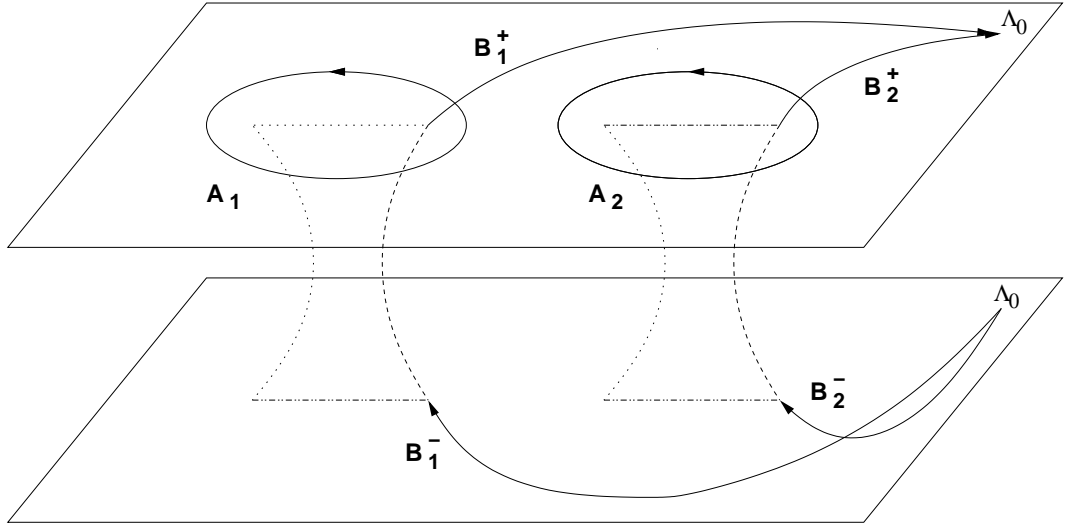}
\parbox{5.5in}{
\caption{The complex curve that results from projecting the Calabi-Yau to the base of the $S^{2}$ fibration.  It is a branched double cover of the complex plane, where the cuts are the projections of the $S^{3}$ cycles of the Calabi-Yau.  The A contours are compact cycles, and the B contours $B_{i} = B_{i}^{-} + B_{i}^{+}$ are non-compact and run from a point at infinity on the lower sheet, through the $i^{\text{th}}$ cut to the point at infinity on the upper sheet.  For later convenience the B contours have been regularized by a cutoff $\Lambda_{0}$.\label{fig:contour}}}
\end{center}
\end{figure}

In evaluating these period integrals, the $u$ and $v$ integrals can be performed trivially (the $A$ and $B$ cycles have the form of an $S^{2}$ fibration over lines in the complex plane, see figure \ref{fig:contour}), and the period integrals of the complex 3-dimensional Calabi-Yau (\ref{eq:genconifold})  can be reduced to the period integrals of a complex curve

\begin{equation}
\label{eq:spectralcurve}
y^{2} = W'(x)^{2} - f(x)
\end{equation}
The holomorphic 3-form $\Omega$, the periods of which define the effective superpotential, reduces to the meromorphic 1-form $y~dx$ on the curve.  The function $f(x)$, and therefore the curve itself, is fixed by a requirement of extremality, in a sense that will be made precise.  This curve is central to the construction of the gauge theory effective superpotentials, and we will rederive and study it from several points of view in the following chapters.

In order to study $SO$ or $Sp$ gauge theory, we can consider an orientifold of the previous geometry\footnote{Orientifolds were discussed in the A-model   in~\cite{Sinha:2000ap,Edelstein:2001mw,Acharya:2002ag,Fuji:2002vv}, while the discussion of~\cite{Gomis:2001xw} is more closely related to the B-model which is our interest here.}.  Since we started with a type IIB theory on a Calabi-Yau, we have to combine the worldsheet orientation reversal with a holomorphic involution of the \CY\ (an anti-holomorphic involution would be appropriate for the IIA theory).  Furthermore we want to fix one of the $\BP^1$'s and act freely on the rest of the Calabi Yau geometry.  This can be done if $W(x)$ is an even polynomial of order $2n$.  In terms of the fibration structure of the \CY, this means that the critical points of $W'(x)$ come in pairs $(-x_i, x_i)$ and one critical point is fixed at $x_0=0$. Then
\begin{equation}\label{holoinv}
(u,v,w,x,\lambda_1,\lambda_2)\mapsto(-u,-v,-w,-x,\lambda_1,\lambda_2)
\end{equation}
is a holomorphic involution of the geometry (\ref{resgenconi}), which leaves only the $\BP^1$ at $u=v=w=x=0$ fixed.  In the string theory this means that there is an O5-plane wrapping this $\P^1$ in the \CY\ geometry.  

There are essentially two choices of O5-plane with which we can wrap the fixed $\P^1$.  They are distinguished by a different choice of worldsheet action and carry RR 5-form charge of $\pm 1$ (the RR charge of an O$p^\pm$-plane is $\pm 2^{p-5}$ in conventions where we count the charge of $N/2$ D-branes but not their $N/2$ images). The orientifold contribution to  the RR charge of objects wrapping the $\P^1$ will cause a shift in the coefficient $N_0$ in the flux-generated superpotential on the deformed \CY\ geometry, as explained below.

Now we can construct a softly broken $\N=2$ $\SO(N)$ or $\Sp(N/2)$ gauge theory with tree level superpotential $W(x)$ by wrapping $N$ D5-branes around the $S^2$ and then performing the orientifold.  The gauge symmetry is again broken $\SO(N) \mapsto \SO(N_0) \times U(N_1) \times \cdots \times U(N_{n-1})$ or $\Sp(N/2)
\mapsto \Sp(N_0/2) \times U(N_1) \times \cdots \times U(N_{n-1})$ respectively, with $N=N_0+2N_1+\cdots+2N_{n-1}$.

At low energies the geometric transition again produces the deformed conifold
geometry~\cite{Vafa:2000wi}
\begin{equation}
u^2 + v^2 + w^2 + W'(x)^2 - f(x) = 0,
\end{equation}
where $f(x)$ is now an even polynomial of degree $2n-2$.  Such a polynomial represents the most general normalizable deformation of the singular conifold that still respects the holomorphic involution (\ref{holoinv}).  The orientifold acts on one 3-sphere $A_0$ as the antipodal map, while the other 3-spheres are mapped to each other in pairs $A_i$ and $A_{-i}$. Note that  there is no orientifold fixed plane anymore.

The 3-form RR-flux $H$ through each 3-sphere $A_i$ is  equal to the amount of D5-brane and O5-plane charge on the $\BP^1_i$ before the transition. 
\begin{equation}
\begin{split}
N_0\pm 2&=\int_{A_0}H,\\
N_i&=\int_{A_i}H, ~~~i\ne 0,
\end{split}
\end{equation}
and the flux through the cycles $B_i$ is again given in terms of the coupling constants
\begin{equation}
\tau_i=\int_{B_i}H.
\end{equation}
Since there is no orientifold fixed plane, there are no contributions  to the effective superpotential for the gaugino condensate from unoriented closed  strings~\cite{Acharya:2002ag}.  In the flux superpotential
\begin{equation}
\Weff(S_i)=\int H\wedge\Omega,
\end{equation}
the integral is now taken only over half of the covering space of the orientifold.  Using the expressions for the periods and the fluxes and taking into account the orientifold projection, we get
\begin{equation}
\label{eq:weffclosed.un}
\Weff(S_i)=\left(\frac{N_0}{2}\pm 1\right)\pder{\F_0}{S_0} + \sum_{i>0}N_i\pder{\F_0}{S_i}+ \half\tau_0S_0+\sum_{i>0}\tau_iS_i.
\end{equation}

This result could also have been computed on the open string side before the transition.  On the open string side there is no flux through any 3-cycles, so there is no contribution to the superpotential due to closed oriented strings.  But there are two kinds of other contributions to the effective superpotential: the open string contributions (disk diagrams) and the contributions due to closed unoriented strings at the orientifold fixed plane ($\RP^2$ diagrams). The contribution due to the open strings is the equal to one half that of the theory without the orientifold, {\it i.e.}, it is
\begin{equation}
\Weff^{O}(S_i)=\frac{N_0}{2}\pder{\F_0}{S_0}
+\sum_{i>0}N_i\pder{\F_0}{S_i}+
\half\tau_0S_0+\sum_{i>0}\tau_iS_i.
\end{equation}
The contribution due to the unoriented closed strings then must be
\begin{equation}
\Weff^{U}(S_i)=\Weff(S_i)-W^{O}_{eff}(S_i)=\pm\pder{\F_0}{S_0}.
\end{equation}
We will confirm this result in a matrix model computation in chapter \ref{ch:mm}.

\subsection{Computing the superpotential}
\label{sec:n1curve}

Consider pure  \ntwo\ Yang-Mills theory broken to \nOne\ via a tree-level superpotential of the form:
\begin{equation}
\Wtree  ~\equiv~ \sum_{p=1}^{n+1} \,  \frac{g_p}{p}  \, {\rm Tr}  \big(\Phi^p\big)
~\equiv~ \sum_{p=1}^{n+1} \,    g_p   \, u_p \,.
\end{equation}
The effective superpotential $\Weff(S)$ may be computed in terms of periods of the differential form (``resolvent''): 
\begin{eqnarray}
\omega(x) &=& \frac{1}{2} \left(W'(x) - \sqrt{(W'(x))^2 ~+~ f_{n-1}(x) } \right)\, dx \nonumber \\
&\equiv& \frac{1}{2}(W'(x) - y(x)) dx
\end{eqnarray}
which is single-valued on the genus $n-1$ Riemann surface $y^2 = W'(x)^2 + f_{n-1}(x)$ (the ``$\N=1$ curve'') that we encountered in the previous section.  In section \ref{sec:sw} we will rederive this curve by factorizing the Seiberg-Witten curve of the associated \ntwo\ theory obtained when $\Wtree=0$, and discarding the repeated roots of the curve that correspond to condensed monopoles.

The compact $A$-periods of the curve yield the gaugino bilinear superfields, $S_i$, while the non-compact $B$-periods, $\Pi_i$ yield the derivatives of the free energy $\frac{\del \F}{\del S_i}$.  Choose the branches of the square root so that on the first sheet $\omega(x)$ vanishes in the classical limit $f_{n-1} \rightarrow 0$; therefore on the second sheet $\omega(x) \rightarrow W'(x)$.

In this chapter we will focus on the maximally-confining phase of the theory (the vacua with classically unbroken gauge group $\U(N)$), for which the resolvent degenerates:
\begin{equation}
\label{confW}
y(x)  ~=~ \sqrt{(W'(x))^2 ~+~ f_{n-1}(x) } \, dx ~=~ G_{n-1}(x)  \, \sqrt{(x-c)^2 ~-~
\mu^2} \, dx\,, 
\end{equation}
for some polynomial, $G_{n-1}(x)$ of degree $(n-1)$.   For $\U(N)$ theories, it is convenient to use the freedom to shift $x$ so as to set $c=0$; this is not allowed for $\SU(N)$, for which the center of the cut is not a free parameter, but the $\SU(N)$ results may be obtained from the $\U(N)$ at the end of the calculation by decoupling the overall $\U(1)$ trace (we will come back to this point later).  The gaugino bilinear is then given by:
\begin{equation}
\label{Scomp}
S ~=~\frac{1}{2 \pi \imath} \oint_A \omega(x) =  \pm \frac{1}{4 \pi \imath} \oint_A y(x) ~=~\pm \frac{1}{2 \pi \imath} \int_{-\mu}^{\mu} \,
G_{n-1}(x)  \, \sqrt{x^2 ~-~ \mu^2} \, dx 
\end{equation}
where the sign depends on the orientation of the contour.  The $B$-period is given by integrating along a contour from infinity on the second sheet, through the cut to infinity on the first sheet, see figure \ref{fig:contour}.  The logarithmic divergence of this integral needs to be regularized, and this is usually done by a introducing a UV cut-off: 
\begin{equation}
\label{Picomp}
\Pi_B  ~=~ \int_B \omega ~=~ \int_{x_- = \Lambda_0}^{x_+  = \Lambda_0}  \omega   
= - \int_\mu^{\Lambda_0} G_{n-1}(x)  \, \sqrt{x^2 ~-~ \mu^2} \, dx  \,,
\end{equation}
where $x_-$ and $x_+$ denote the values of $x$ on the lower and upper sheets respectively.  
The effective superpotential is then given by:
\begin{equation}
\label{eq:weff}
W_{\rm eff}   ~=~ N \, \Pi_B + N W(\Lambda_0) ~+~ \tau \, S
\end{equation}
where $\tau$ is the bare gauge coupling, and the second term is added to cancel the contribution from the upper limit of the integral in $\Pi_B$.  As we saw in section \ref{sec:susy.ym}, the effect of the $\tau$ term is to combine with the log-divergent piece of $\Pi_B$ to give the (finite) dynamical scale $\Lambda$ of the theory.

In computing the effective superpotential by this method, the approach initially taken in the literature was to send $\Lambda_0 \rightarrow \infty$, causing its effects to decouple from the theory.  However, we will obtain physical insight into the nature of the computation by keeping the cut-off finite.  We will henceforth take the cut-off $\Lambda_{0}$ to be large but finite, and investigate the effects on the low-energy gauge theory; this amounts to keeping the $O(1/\Lambda_0)$ terms in $\Pi_B$ and subsequent calculations.

\subsection{Example: $\U(2)$}
\label{sec:periodu2}

Before analyzing the general case, consider the simplest example of $\U(2)$ with a tree-level mass: $W = \frac{1}{2} m \tr \Phi^2$.  The effective 1-form is

\begin{equation}
y(x) = m \sqrt{x^2-\mu^2}
\end{equation}
which is single-valued on a two-sheeted Riemann surface with a cut
between $x = \pm \mu$.  The gaugino bilinear is given by the A-period:

\begin{equation}
 S = \frac{1}{4 \pi \imath} \oint_A y(x) dx = \frac{1}{2 \pi \imath} \int_{-\mu}^{\mu} y(x) dx =  \frac{1}{4} m \mu^2
\end{equation}
and the B semi-period is
\begin{eqnarray}
\Pi_B &=&- \int_\mu^{\Lambda_0} y(x) dx \nonumber \\
&=&- \frac{m}{2} \left( \pm \Lambda_0^2 \sqrt{1 - \frac{\mu^2}{\Lambda_0^2}} + \mu^2 \log\left(\frac{\mu}{\Lambda_0\Big (1\pm\sqrt{1 - \frac{\mu^2}{\Lambda_0^2}}\Big)}\right)\right) \nonumber \\
&=& \mp \frac{m \Lambda_0^2}{2} \sqrt{1-\frac{4S}{m \Lambda_0^2}} - S \log\left(\frac{S}{\frac{m \Lambda_0^2}{2}\Big(1\pm\sqrt{1-\frac{4S}{m \Lambda_0^2}}\Big)-S}\right) \nonumber \\
\end{eqnarray}
where the integral is evaluated using hyperbolic functions and the two branches come from ${\hbox{sinh}}(x) = \pm \sqrt{{\hbox{cosh}^2(x)}-1}$ (this amounts to a choice of contour, i.e.~integrating to the point above $\Lambda_0$ on one of the two sheets).  As mentioned in the previous section, the role of $\tau$ in (\ref{eq:weff}) is to replace the $N \log(m \Lambda_0^2)$ term in $\Pi_B$ by the finite scale $N \log(\Lambda^3)$. This may be implemented in practice by setting $\tau = N \log(\frac{\Lambda^3}{m \Lambda_0^2})$ in (\ref{eq:weff}).

We find
\begin{equation}
\label{eq:u2plus}
W = N\left(S \Big(1- \log(\frac{S}{\Lambda^3}) \Big) - \frac{S^2}{m \Lambda_0^2} - \frac{2 S^3}{2 (m \Lambda_0^2)^2} - \frac{5 S^4}{3 (m \Lambda_0^2)^3} - \frac{14 S^5}{4 (m \Lambda_0^2)^4} - \ldots \right)
\end{equation}
Therefore in the limit $\Lambda_0 \rightarrow \infty$ (equivalently, keeping $\Lambda_0$ finite and considering energies $m << \Lambda_0$) the infinite correction series tends to zero and the effective superpotential (\ref{eq:u2plus}) reduces to the usual Veneziano-Yankielowicz superpotential.

The form of the series (\ref{eq:u2plus}) is the same as that obtained for $\U(2), N_f=4$, with $\Lambda_0$ identified with the quark mass.  The known formula for $W(S)$ with tree-level superpotential $W = \frac{1}{2} m \tr \Phi^2 + \sum_{i=1}^{N_f} \mu \tilde{Q}_i Q^i + \tilde{Q}_i \Phi^i_j Q^i$ is \cite{Argurio:2002xv,Brandhuber:2003va}

\begin{eqnarray}
\label{eq:weff.mass}
W(S) = N_c S (1- \log(\frac{S}{m \Lambda_0^2}) ) - N_f S \log(\frac{\mu}{\Lambda_0})\quad\quad\quad\quad\quad\quad&&  \nonumber \\
- N_f S \left(\frac{1}{2} + \frac{\sqrt{1 - 4 \alpha S} - 1}{4 \alpha S} - \log(\frac{1+\sqrt{1 - 4 \alpha S}}{2})\right)&&
\end{eqnarray}
with $\alpha = 1/(m \mu^2)$ (we will derive this expression in section \ref{sec:uv}).  Setting $N_c=2, N_f=4, \mu = \Lambda_0$ and performing  the  series expansion, we recover the   expression in (\ref{eq:u2plus}).  We will show in Section \ref{sec:uv} that this feature remains true for general $N_c$ and $W(\Phi)$, and the corrections obtained by keeping the cut-off dependence in the period integral indeed have the physical interpretation of $N_f = 2 N_c$ massive quark superfields, which serve to regularize the divergences of the calculation.

Choosing the other branch of $\Pi_B$ we obtain the negative of (\ref{eq:u2plus}).  This branch describes a Higgs branch \cite{Brandhuber:2003va}, where the gauge symmetry is broken by giving a vev to the scalar component of the quark superfields (an arbitrary Higgs vacuum can be obtained by writing $W = \tau S + \sum_{i=1}^{N} \Pi_B$ and choosing the branch of $\Pi_B$ termwise, i.e.~for each period integral we choose whether to integrate along a contour on the first or second sheet).

If instead of $\U(N)$ gauge theory we considered $\SU(N)$, the foregoing discussion would be modified by the need to ensure ``quantum tracelessness'' of the vacuum, i.e.~that $\langle u_1 \rangle = 0$.  This may be achieved by taking the tree-level superpotential $W = \frac{1}{2} m \tr \Phi^2 + \lambda \tr \Phi$ and proceeding with the above analysis, treating $\lambda$ as a Lagrange multiplier to enforce $\langle u_1 \rangle = \langle \tr \Phi \rangle = 0$.  Instead of repeating the calculation for $\SU(2)$, we will defer until later when we consider the general $\U(N)$ and $\SU(N)$ cases.

\subsection{Evaluation of the period integral for general W}
\label{sec:generalper}

The period integrals, (\ref{Scomp}) and (\ref{Picomp}), are elementary but one can obtain a simple closed form in terms of $\Wtree$.  This can be evaluated and gives a combinatorial formula for the moduli $u_k$ which can be compared to other techniques.  Make the change of variables\footnote{We again assume that $x$ has been centered on the cut.}:
\begin{equation}
\label{newvar}
x ~=~ \frac{1}{2} \,\mu \,  ( \xi ~+~ \xi^{-1}) \,,
\end{equation}
and define series expansions:
\begin{eqnarray}
 \label{Wser}
W \big(\frac{1}{2} \,\mu \,  ( \xi ~+~ \xi^{-1}) \big) &=&  b_0 ~+~
 \sum_{k= 1}^{n+1} ~ b_k \, ( \xi^k ~+~ \xi^{-k}) \,, \\
W'\big(\frac{1}{2} \,\mu \,  ( \xi ~+~ \xi^{-1})\big)  &=& c_0 ~+~ \sum_{k=1}^n ~
c_k \, ( \xi^k ~+~ \xi^{-k})
 \label{Wprimeser}
 \end{eqnarray}
Note that the series take this form because of the symmetry of (\ref{newvar}) under $\xi \to \xi^{-1}$.  Under this change of variables the integrand may be written:
\begin{eqnarray}
\label{magicident}
\frac{1}{2} \,\mu\, ( \xi ~-~ \xi^{-1})  \,   G_{n-1}\Big(\frac{1}{2} \,\mu\, ( \xi ~+~ \xi^{-1}) \Big)  
&=& G_{n-1}(x)  \, \sqrt{x^2 ~-~ \mu^2}    \,, \\
&=& W'(x) \, \sqrt{1 ~+~ \frac{f_{n-1}(x) }{ (W'(x))^2 }}  \\ 
&=&W'(x)  ~+~\CO(\xi^{-1}) \,.
\end{eqnarray}
The left-hand side is manifestly odd under $\xi \to 1/\xi$, while the right-hand side shows that all the non-negative powers in the $\xi$-expansion are given by (\ref{Wprimeser}).  It therefore follows that under the change of variables, one has 
\begin{equation}
\label{integrandseries}
\sqrt{(W'(x))^2 ~+~  f_{n-1}(x)} ~=~  G_{n-1}(x)  \, \sqrt{(x^2 -\mu^2)}
~=~ \sum_{k=1}^n ~ c_k \, ( \xi^k ~-~ \xi^{-k})  \,. 
\end{equation}
Note in particular that the left-hand side of (\ref{magicident}) is manifestly odd under $\xi \to 1/\xi$, therefore $c_0 =0$ in (\ref{Wprimeser}).

Define  $[\dots]_-$ to mean: discard all the  non-negative powers of $\xi$ in $[\dots]$. We  may then write the last equation as:
\begin{equation}
\label{niceform}
\sqrt{(W'(x))^2 ~+~  f_{n-1}(x)} ~=~ W'\Big(\frac{1}{2} \,\mu\, (\xi + \xi^{-1})
\Big) ~-~ 2\, \Big[ W'\Big(\frac{1}{2} \,\mu\, ( \xi +  \xi^{-1}) \Big)\Big]_- \,.
\end{equation}
One can now easily perform the integrals (\ref{Scomp}) and (\ref{Picomp}).  The former is simply given by taking $\xi =e^{\imath \theta}$ for $0 \le \theta \le \pi$, and it picks out the $\xi$-residue:
\begin{equation}
\label{Sres} 
S~=~   \frac{\mu}{2}   ~ c_1 
\end{equation}
To perform the second integral first note that:
\begin{eqnarray}
&&\frac{d }{d \xi}\, \Big[ W \big(\frac{1}{2} \,\mu\, 
( \xi +  \xi^{-1}) \big) \Big]_-   \nonumber \\
&&\quad\quad\quad\quad~=~   ~-~ 
\frac{1}{2} \,\mu\,  c_1\,   \xi^{-1}~+~ \frac{1}{2}  \,\mu\, (1 - \xi^{-2}) \,  
\Big[ W' \big(  \frac{1}{2} \,\mu\, (\xi +    \xi^{-1})  \big) \Big]_-
\end{eqnarray}
and therefore:
\begin{eqnarray}
\int  \sqrt{(W'(x))^2 ~+~  f_{n-1}(x)}\, dx  & = &
- \mu\,  c_1\,  \log(\xi) ~+~ W(x) - 2 \Big[ W(x) \Big]_-  
\label{indefinta}
\\ & = & - \mu\, 
c_1\, \log(\xi) ~+~b_0~ \nonumber \\
&&\quad\quad+~  \sum_{k=1}^n ~ b_k \, ( \xi^k ~-~ \xi^{-k}) \,,
\label{indefintb}
\end{eqnarray}
where $x = \frac{1}{2} \,\mu\, (\xi + \xi^{-1})$. To obtain $\Pi$, we must evaluate this between $\xi =1$ and $\xi = \xi_0$, where
\begin{equation}
\label{xizerodefn}
\xi_0 \equiv \xi(\Lambda_0) ~=~ \frac{\Lambda_0}{\mu} \, \bigg(1 ~+~ \sqrt{1 - \Big(\frac{\mu}{\Lambda_0}\Big)^2} \, \bigg)  \,.
\end{equation}
This yields:
\begin{equation}
\label{Pians}
\Pi ~=~    b_0 ~+~    \mu\,  c_1\,  \log(\xi_0) ~-~  \Big( W(\Lambda_0)   - 2 \big[ W(x) \big]_- \Big|_{\xi =\xi_0} \Big) \,,
\end{equation}
where the  definite integral has been evaluated using (\ref{indefinta}) at $\xi = \Lambda_0$
and using (\ref{indefintb}) at $\xi = 1$.

In the limit of large $\Lambda_0$ the last term in (\ref{Pians}) vanishes since it only involves negative powers of $\xi_0 \sim \Lambda_0^{-1}$. Taking this limit, and using (\ref{Sres})  one obtains:
\begin{equation}
\label{eq:periodpuregauge}
\Pi ~=~   b_0 ~+~  2\, S \,\log\Big(\frac{2 \Lambda_0}{\mu} \Big) ~-~ W(\Lambda_0) \,.
\end{equation}
Therefore
\begin{eqnarray}
\label{eq:glueballper}
\Weff(S) &=& N b_0 + 2 N S \log\Big(\frac{2 \Lambda}{\mu} \Big) \nonumber \\
\end{eqnarray}
We will show in section \ref{sec:extremize} that for general $\Wtree(\Phi)$, (\ref{eq:glueballper}) can be extremized with respect to $S$ by taking $\mu = 2 \Lambda$, and we find the previously known result \cite{Cachazo:2001jy}

\begin{equation}
\label{eq:glueballlow}
W_{\hbox{\it low}}(g_k, \Lambda)= N_c \sum_{p=1}^{\lfloor \frac{n+1}{2} \rfloor} \frac{g_{2p}}{2p} \comb{2p}{p} \Lambda^{2p}
\end{equation}
where we have evaluated the coefficients $b_0$ in the series expansion (\ref{Wser}). 

In the previous section we discussed the geometric engineering of this gauge theory from string theory, which involved D-branes wrapped on cycles of a \CY.  From the string theory perspective it is tempting to also interpret the cut-off of the period contour in terms of branes.  That is, it is really only physically natural to terminate the period integral on another brane.  Since D-branes carry gauge fields, having a stack of $M$ branes at the point $\Lambda_0$ would mean that one started with a larger (product) gauge group and that the original $\SU(N)$ theory is actually coupled to $M$ bi-fundamental matter multiplets with a (gauged) $\SU(M)$ ``flavor'' group (see \cite{Hofman:2002bi} for an analysis of this gauge theory).  However, when the second set of branes become non-compact, their associated gauge coupling tends to zero, and the $\SU(M)$ gauge factor becomes a global $\SU(M)$ flavor symmetry.  Thus, string theory suggests that keeping the UV cut-off terms should yield the superpotential associated with the coupling to fundamental matter multiplets.  This is indeed what we find in explicit calculations.

If one also recalls that the canonical form of the $B$-period integral, (\ref{Picomp}), involves an integral from the lower to the upper sheet of the Riemann surface, then this extra term  may be thought of arising  from $N_c$ branes  (or anti-branes) at each limit.   Thus one can also extract  the results for $N_f =N_c$ by regulating the upper and lower limits independently. We will develop and extend this observation in the next section.

\subsection{UV cut-off as regularization by $N_f=2N_c$ fundamental quarks}
\label{sec:uv}

As mentioned in section \ref{sec:n1curve}, the effective superpotential for the $N_f=0$ theory (in a maximally confining vacuum) is given by \cite{Cachazo:2001jy,Naculich:2002hr,Cachazo:2003yc}

\begin{equation}
\Weff \sim -2 N_c  \int_{\mu}^{\infty} \omega + \tau S
\end{equation}
where the integral is formally divergent and is usually cut off at a point $\Lambda_0$.  We will verify in section \ref{sec:mm.fund} that introducing $N_f$ fundamentals gives the (again formally divergent) contribution \cite{Dijkgraaf:2002dh,Argurio:2002xv}

\begin{equation}
\label{eq:wnf}
W_{N_f} \sim  \sum_{i=1}^{N_f} \int_{m_i}^{\infty} \omega
\end{equation}

However, when $N_f=2N_c$, the contours combine and the integration domains are now finite, so the divergence of the integrals have been regularized.  When all $m_i$ are equal we may write $m_i \equiv \Lambda_0$ and we can explicitly see the role of the $2N_c$ fundamental fields in implementing the cut-off of the $N_f=0$ integral: they act as regulators for the UV divergences of the calculation, by removing the short-distance divergences of the calculation.  Physically, the gauge theory with an adjoint chiral superfield and $N_f=2N_c$ fundamentals has vanishing beta function in the limit when all of the fields are effectively massless, i.e.~at energy scales much greater than their mass.  Thus, the theory has a nontrivial UV conformal fixed point, and is free from short-distance singularities.

In terms of the additional microscopic degrees of freedom we are forced to add, the tree level superpotential of the gauge theory is modified:

\begin{equation}
\label{eq:yukawa}
\Wtree(\Phi) \rightarrow \Wtree(\Phi) + \sum_{i=1}^{2N_c} \Lambda_0 \tilde Q^i Q^i + \tilde{Q}^i \Phi Q^i
\end{equation}
where $Q^i$ are the new ``quark'' superfields, and $\tilde{Q}^i$ are their conjugate antiquarks, and we have normalized the coefficient of the Yukawa interaction to 1 (the Yukawa coupling can be absorbed into the mass parameters $m_i$ by redefining the fields, since we are not interested in the kinetic terms).  In section \ref{sec:mm.fund} we show how the combinatorics of the Feynman diagrams involving the new quark fields combine to subtract the short-distance divergences of the theory without quarks.

As we have seen in the example of $\U(2)$, when $\Lambda_0$ is taken to be large but finite, it gives finite (but small) corrections to the expression for the effective superpotential $W(S)$.  Therefore, the vacuum expectation value for the gaugino bilinears $\langle S_i \rangle$ will be perturbed from that of the theory we started with (\nOne\ Yang-Mills theory with a massive adjoint and no fundamental matter).  In other words, in terms of the \nOne\ curve (\ref{eq:spectralcurve}), the presence of the cut-off at a finite distance from the cuts cause the size and center of the cuts to be perturbed.  Because of this deformation, it will turn out that this \nOne\ curve {\it cannot} be obtained by factorizing the SW curve of pure \ntwo\ Yang Mills.

Therefore, in regularizing the $N_f=0$ theory by imposing a finite cut-off on the divergent integral, we have gone off-shell (i.e.~the vacua of this theory do not solve the equations of motion of the $N_f=0$ theory).  Physically, this is because the presence of the cutoff is equivalent to introducing new physical degrees of freedom that contribute to the gaugino condensates.  This amounts to embedding the $N_f=0$ theory in a larger theory with $N_f=2N_c$ massive quark flavors; it is only in the limit of infinite quark mass (infinite cut-off) that the effects of the quarks on the vacuum structure of the theory decouple and we approach the on-shell vacua of the $N_f=0$ theory.

In practice we can think of the effective superpotential for $0\le N_f \le 2N_c$ fundamentals (as computed using the technique of \cite{Cachazo:2001jy} discussed here, also using the matrix model discussed in chapter \ref{ch:mm}), as always being generated by the UV-finite theory with $2N_c$ fundamental fields, with masses that are either kept finite or which are taken to infinity at the end of the calculation and decouple from the theory.  In other words, if we have $\tilde N_f$ fundamental fields of finite mass, then the remaining $2N_c-\tilde N_f$ are of mass $\Lambda_0 \gg m$.  Therefore:

\begin{eqnarray}
\Weff &\sim& -2 N_c \int_{\mu}^{\infty} \omega +  \sum_{i=1}^{\tilde N_f} \int_{m_i}^{\infty} \omega + (2N_c-\tilde N_f) \int_{\Lambda_0}^{\infty} \omega \nonumber \\
&=& -2 N_c \left(  \int_{\mu}^{\infty} \omega - \int_{\Lambda_0}^{\infty} \omega \right) - \sum_{i=1}^{\tilde N_f}\left( \int_{\Lambda_0}^{\infty} \omega -  \int_{m_i}^{\infty} \omega \right) \nonumber \\
&=&  -2N_c  \int_{\mu}^{\Lambda_0} \omega +  \sum_{i=1}^{\tilde N_f} \int_{m_i}^{\Lambda_0} \omega
\end{eqnarray}
and all integrals are finite.

We can then decouple the quarks of mass $\Lambda_0$ by taking $\Lambda_0 \rightarrow \infty$, and using the results of section \ref{sec:generalper} we find the following expression for $\Weff\ $:

\begin{eqnarray}
\label{eq:generalnf}
\Weff &=& N_c \left( b_0 + \mu c_1 \log(\frac{2 \Lambda_0}{\mu}) \right) + \nonumber \\
&& \hspace{0.5in} \sum_{i=1}^{N_f} \left(  \frac{\mu c_1}{2} \left( \log(\xi(m_i)) - \log(\frac{2 \Lambda_0}{\mu})\right)  + [W(\xi(m_i))]_- \right) + \tau S \nonumber \\
&=& N_c b_0 + \frac{ \mu c_1 }{2}\log\left(\frac{(2 \Lambda_0)^{2N_c-N_f} \prod_{i=1}^{N_f} \xi(m_i)}{\mu^{2N_c-N_f}}\right) + \sum_{i=1}^{N_f} [W(\xi(m_i))]_- + \tau S \nonumber \\
&=& N_c b_0 + \frac{ \mu c_1 }{2}\log\left(\frac{(2 \Lambda_0)^{2N_c-N_f} \prod_{i=1}^{N_f} m_i}{\mu^{2N_c}} \prod_{i=1}^{N_f}\left(1+\sqrt{1-(\frac{\mu}{m_i})^2}\right)\right) \nonumber \\
&&\hspace{0.5in}+ \sum_{i=1}^{N_f} [W(\xi(m_i))]_- + \tau S \nonumber \\
&=& N_c b_0 + \frac{\mu c_1}{2} \log\left(\frac{2^{2N_c-N_f} \tilde{\Lambda}^{2N_c}}{ \mu^{2N_c}} \prod_{i=1}^{N_f}\left(1+\sqrt{1-(\frac{\mu}{m_i})^2}\right)\right) \nonumber \\
&& \hspace{0.5in} + \sum_{i=1}^{N_f} [W(\xi(m_i))]_-
\end{eqnarray}
where we used the scale-matching relation $\tilde{\Lambda}^{2N_c} = \Lambda^{2N_c-N_f} \prod_i m_i$.  Using the definitions (\ref{Wser}) and writing explicit expressions for the coefficients $b_k$, this can be written as:

\begin{eqnarray}
\label{eq:generalnf2}
\Weff&=&N_c \sum_{i=1}^{\lfloor \frac{n+1}{2} \rfloor} \comb{2i}{i} \frac{g_{2i}}{2i} \left(\frac{\mu}{2}\right)^{2i} \nonumber \\
&&\hspace{0.5in} + S \log\left(\frac{2^{2N_c-N_f} \tilde{\Lambda}^{2N_c}}{ \mu^{2N_c}} \prod_{i=1}^{N_f}\left(1+\sqrt{1-(\frac{\mu}{m_i})^2}\right)\right) \nonumber \\
&& \hspace{0.5in}  + \sum_{i=1}^{N_f} \sum_{k=1}^{n+1} \frac{g_k}{k} \left(\frac{\mu}{2}\right)^k \sum_{l=\lfloor k/2 \rfloor + 1}^{k} \comb{k}{l} \xi(m_i) ^{k-2l}
\end{eqnarray}
An explicit general expression for $S = S(g_k, \mu)$ can be similarly obtained, but we will not need it here.

\subsection{Extremizing the superpotential}
\label{sec:extremize}

In order to find the physical vacua, we need to extremize (\ref{eq:generalnf2}) with respect to S.  This will fix $\mu$, the size of the cut, and give the vacuum superpotential in terms of physical quantities.  Varying with respect to $S$, this can be achieved by setting
\begin{equation}
\frac{\partial \mu}{\partial S} = 0
\end{equation}
\begin{equation}
\log\left(\frac{2^{2N_c-N_f} \tilde{\Lambda}^{2N_c}}{ \mu^{2N_c}} \prod_{i=1}^{N_f}\left(1+\sqrt{1-(\frac{\mu}{m_i})^2}\right)\right) = 0
\end{equation}
i.e.
\begin{equation}
\label{eq:extremize}
\mu^{2N_c} = (2 \tilde\Lambda)^{2N_c} \prod_{i=1}^{N_f}\left(\frac{1+\sqrt{1-(\frac{\mu}{m_i})^2}}{2}\right)
\end{equation}
Thus, the logarithmic term in (\ref{eq:generalnf2}) does not contribute in the vacuum, and the extremal superpotential is found by solving (\ref{eq:extremize}) to find $\langle \mu\rangle$.  Note that when $N_f=0$ the solution to (\ref{eq:extremize}) is given by taking $\mu = 2 \tilde \Lambda\equiv 2 \Lambda$, as claimed in section \ref{sec:generalper}.

When all quark masses are taken equal, $m_i \equiv m$, (\ref{eq:extremize}) can be written in the simplified form
\begin{equation}
\label{eq:extremize2}
(\mu^2)^{2N_c/N_f} - (4 \tilde\Lambda^2 \mu^2)^{N_c/N_f} + \frac{\mu^2}{4 m^2}(4 \tilde\Lambda^2)^{2N_c/N_f} =0
\end{equation}
Note that this condition is polynomial in $\mu^2$ when $N_c$ is a multiple of $N_f$.

\subsection{Examples}
\label{sec:examples}

We turn now to some other examples (in all cases the quark masses are set equal for simplicity).

\subsubsection{Quadratic tree-level superpotential}

The simplest tree-level superpotential of the form (\ref{eq:yukawa}) contains a mass term for the adjoint chiral superfield $\Phi$:

\begin{equation}
W(\Phi) = \frac{M}{2} \tr \Phi^2
\end{equation}
We consider arbitrary values of $N_c$ and $N_f$.  The gaugino bilinear takes the simple form

\begin{equation}
S = \frac{M}{4}{ \mu^2}
\end{equation}
and we can eliminate $\mu$ from $\Weff(m, M, \Lambda, \mu)$ to write the effective superpotential in terms of the physical parameters and gaugino bilinear:
\begin{eqnarray}
\Weff &=& N_c (S + S \log(\frac{M^{N_c} \tilde{\Lambda}^{2 N_c}}{S^{N_c}})) + N_f S \log(\frac{1+\sqrt{1-4S \alpha}}{2}) \nonumber \\
&& \hspace{0.5in}+ N_f S^2 \alpha \frac{1}{1-2 S \alpha + \sqrt{1 - 4 S \alpha}}\nonumber \\
&=& N_c S(1 + \log(\frac{M^{N_c} \tilde{\Lambda}^{2 N_c}}{S^{N_c}})) + N_f S \log(\frac{1+\sqrt{1-4S \alpha}}{2}) \nonumber \\
&&\hspace{0.5in}- N_fS(\frac{1}{2}+  \frac{ \sqrt{1 - 4 S \alpha}-1}{4 \alpha S}) \nonumber \\
\end{eqnarray}
where $\alpha = \frac{1}{M m^2}$.  This is the previously claimed result (\ref{eq:weff.mass}), first obtained by \cite{Argurio:2002xv}.  For the special case $N_c=2, N_f=1$ the extremization condition (\ref{eq:extremize2}) becomes

\begin{equation}
\mu^8 - (4 \tilde \Lambda^2 \mu^2)^2 + \frac{\mu^2}{4 m^2}(4 \tilde\Lambda^2)^4 =0
\end{equation}
\begin{equation}
\label{eq:scubic}
\Leftrightarrow S^4 - S^2 \bar \Lambda^6 + S \bar \Lambda^{12} \alpha =0
\end{equation}
where $\bar \Lambda^3 = M \tilde\Lambda^2$ is the scale of the theory below the mass $M$ of the adjoint.  Excluding the unphysical solution $S=0$ (which would correspond to a vacuum with unbroken chiral symmetry, and can be ruled out on general grounds \cite{Cachazo:2002ry}), there are three remaining solutions.  Taking the limit of infinite quark mass, $\alpha \rightarrow 0$, (\ref{eq:scubic}) degenerates further:

\begin{equation}
S^2(S^2-\bar \Lambda^6) = 0
\end{equation}
i.e.~two solutions $S=0$ are unphysical, and the two physical solutions are $S = \pm \bar \Lambda^3$.  At energies much lower than the mass $M$ of the adjoint field $\Phi$, the theory is described by $\N=1$ $\SU(2)$ Yang-Mills, and we indeed obtain the correct value of the gaugino condensates (\ref{eq:gauginocond}) of the Veneziano-Yankielowicz superpotential.

Keeping the mass of the fundamental fields finite gives a series of corrections to the pure $\N=1$ result:

\begin{eqnarray}
\langle S \rangle &=& \left\{ \begin{array}{ll}
{\bar \Lambda}^3 - \frac{1}{2} \alpha {\bar \Lambda}^6 - \frac{3}{8} \alpha^2 {\bar \Lambda}^9 - \frac{1}{2} \alpha^3 {\bar \Lambda}^{12} - \frac{105}{128} \alpha^4  {\bar\Lambda}^{15} + \ldots \\
{-\bar \Lambda}^3 - \frac{1}{2} \alpha {\bar \Lambda}^6 + \frac{3}{8} \alpha^2 {\bar \Lambda}^9 - \frac{1}{2} \alpha^3 {\bar \Lambda}^{12} + \frac{105}{128} \alpha^4  {\bar\Lambda}^{15} + \ldots \\
\end{array}\right. \\
W_{\mbox{\it low}} &=& \left\{ \begin{array}{ll}
2 {\bar \Lambda}^3 - \frac{1}{2} \alpha {\bar \Lambda}^6 - \frac{1}{4} \alpha^2 {\bar \Lambda}^9 - \frac{1}{4} \alpha^3 {\bar \Lambda}^{12} - \frac{21}{64} \alpha^4{\bar \Lambda}^{15} + \ldots \\
-2 {\bar \Lambda}^3 - \frac{1}{2} \alpha {\bar \Lambda}^6 + \frac{1}{4} \alpha^2 {\bar \Lambda}^9 - \frac{1}{4} \alpha^3 {\bar \Lambda}^{12} + \frac{21}{64} \alpha^4{\bar \Lambda}^{15} + \ldots \\
\end{array}\right.
\end{eqnarray}
This result agrees with that of \cite{Argurio:2002xv} (although they only explicitly considered one of the two vacua).  It shows clearly how the presence of the finite-mass quarks perturbs the vacua of the theory away from their $N_f=0$ values.

Equation (\ref{eq:scubic}) encodes the exact form of the effective superpotential of this theory.  In this case a closed-form expression for $\langle S\rangle$ and $\Wlow$ could also be obtained since the cubic branch of equation (\ref{eq:scubic}) may be solved explicitly; for higher-rank gauge groups the polynomial will be of degree $2N_c-1$ in S, and can always at least be evaluated as a series expansion to any desired order.

\subsubsection{Arbitrary tree-level superpotential with $N_f=N_c$}

In this example the extremization constraint (\ref{eq:extremize2}) becomes quadratic in $\mu^2$, and can be trivially solved for arbitrary tree-level superpotential $W(\Phi)$:

\begin{eqnarray}
\mu^4 - (4 \Lambda m - 4 \Lambda^2) \mu^2 = 0,
\end{eqnarray}
so $\mu^2 = 4 (\Lambda m - \Lambda^2)$ (the solution $\mu^2=0$ is again unphysical).  

When the cut in the $\N=1$ curve is centered away from the origin, centering the coordinate axes on the cut introduces a corresponding shift in the quark masses, $m \mapsto m+c$.  In the following section we will see that factorizing the $\U(N_c)$ Seiberg-Witten curve for $N_f=N_c$ fixes $c = \Lambda$, hence $\mu^2 = 4 \Lambda m$, and\ $\mu = 2 \tilde \Lambda$.  Moreover, the function $\xi(m+c)$ simplifies when evaluated at the extremal point:

\begin{equation}
\xi(m+\Lambda)|_{\mu = 2 \tilde \Lambda} = \frac{\tilde{\Lambda}}{\Lambda}
\end{equation}

Therefore, the expression (\ref{eq:generalnf2}) for the vacuum superpotential becomes:
\begin{eqnarray}
\label{eq:wnfnc}
\Weff &=& N_c \left[ \sum_{i=1}^{\lfloor \frac{n+1}{2} \rfloor} \frac{g_{2i}}{2i} \comb{2i}{i}  (\frac{\mu}{2})^{2i} + \sum_{i=1}^{n+1} \frac{g_i}{i} (\frac{\mu}{2})^i \sum_{k=\lfloor\frac{i}{2}\rfloor + 1}^{i}\comb{i}{k}\xi(m+\Lambda)^{i-2k} \right] \nonumber \\
&=&N_c \left[ \sum_{i=1}^{\lfloor \frac{n+1}{2} \rfloor} \frac{g_{2i}}{2i} \comb{2i}{i} \tilde\Lambda^{2i} + \sum_{i=1}^{n+1} \frac{g_i}{i} \sum_{k=\lfloor\frac{i}{2}\rfloor + 1}^{i}\comb{i}{k} \tilde\Lambda^{2(i-k)} \Lambda^{2k-i}\right]
\end{eqnarray}
Note that by contrast to the previous example, the effective superpotential now has the form of a finite series.

As we discuss in the next section, we should expect to recover this result by factorizing the Seiberg-Witten curve for \ntwo\ Yang-Mills with $N_f$ massive hypermultiplets.  In section \ref{sec:nffactor} section we will solve the factorization problem for general $N_f$ and verify the equivalence of the resulting vacuum superpotential for the case $N_f = N_c$.

Other examples can be treated similarly by solving the extremization condition (\ref{eq:extremize2}) to find the extremal size of the cut in the spectral curve, and substituting the result into (\ref{eq:generalnf2}).  These equations are exact, in that they receive no further quantum corrections, but in general they can only be solved as a series expansion.

%



\section{Seiberg-Witten curves and supersymmetric vacua}

\label{sec:sw}

In previous sections we studied the vacua of the \nOne\ gauge theory directly.  These results descend from the structure of the underlying \ntwo\ theory one obtains by setting $\Wtree=0$, and we turn our attention now to the \ntwo\ $\U(N)$ gauge theories with $N_f$ fundamental hypermultiplets.

As is well-known, the vacuum structure of \ntwo\ gauge theories are described by a fibration of a Riemann surface (the Seiberg-Witten curve) over the moduli space.  At points in the moduli space where the curve degenerates, physical degrees of freedom (monopoles, dyons or W-bosons) become massless.  

For example, the Seiberg-Witten curve of $\N=2$ $\U(N)$ or $\SU(N)$ pure gauge theory is the genus $N-1$ hyperelliptic curve
\begin{equation}
\label{eq:swcurve.un}
y^2 = P_N(x)^2 - 4 \Lambda^{2N}
\end{equation}
where $P_N(x) = x^N + \sum_{i=1}^{N} s_i x^{N-k}$, with $s_1=0$ for the $\SU(N)$ curve, and $\Lambda$ is the dynamically generated scale of the gauge theory.

Written in \nOne\ language, the effective superpotential for the \ntwo\ theory in the neighborhood of a point where $l$ monopoles simultaneously become massless is \cite{Seiberg:1994aj,Seiberg:1994rs}

\begin{equation}
W(M_m, \tilde{M}_m, u_p, \Lambda) = \sum_{m=1}^{l} \tilde{M}_m M_m a_{D,m}(u_p, \Lambda)
\end{equation}
where $M_m$ are the monopole hypermultiplets, $a_{D,m}$ are the periods of the Seiberg-Witten curve that determine the monopole masses, and $u_p$ are the gauge-invariant curve moduli

\begin{equation}
u_{p} = \frac{1}{p} \tr \Phi^p
\end{equation}
that parameterise the vacua of the \ntwo\ theory.  After breaking to \nOne\ by the addition of a tree-level superpotential, the Intriligator-Leigh-Seiberg linearity principle \cite{Intriligator:1994jr} implies that the exact superpotential becomes \cite{Cachazo:2001jy}

\begin{equation}
\label{eq:wmonopole}
W(M_m, \tilde{M}_m, u_p, \Lambda, g_p) = \sum_{m=1}^{l} \tilde{M}_m M_m a_{D,m}(u_p, \Lambda) + \sum g_p u_p
\end{equation}
The equation of motion for the monopole fields imposes that $a_{D,m} = 0$.  This is true iff the corresponding B-cycle of the Seiberg-Witten curve degenerates, therefore the vacua of the gauge theory are associated to a ``factorization locus'' in the moduli space of the Seiberg-Witten curve, where $l$ cycles of the Seiberg-Witten curve simultaneously pinch off to zero volume.    The equation of motion for the $u_p$ then implies that there is a nonzero monopole condensate in the confining \nOne\ vacua, i.e.~confinement of the \nOne\ theory is associated to monopole condensation.

The maximally-confining vacua correspond to the point in the \ntwo\ moduli space where all $N-1$ monopoles become massless, and the Seiberg-Witten curve degenerates completely to genus $0$.
 
After evaluating (\ref{eq:wmonopole}) at the factorization locus, the exact effective superpotential then becomes
\begin{equation}
\label{eq:wlow.fact}
W_{\hbox{\it low}}(g_p, u_p, \Lambda) = \sum g_p u_p|_{\{a_{D,m} = 0\}}
\end{equation}
Thus, evaluation of the effective superpotential is equivalent to solving the factorization of the spectral curve.  Once we know the moduli $\langle u_p \rangle$ at the factorization locus we can immediately read off the effective superpotential corresponding to any given $\Wtree$ using (\ref{eq:wlow.fact}).

The factorized Seiberg-Witten curve can be written as

\begin{equation}
y^{2} = G_{l}^{2}(x) F_{2(N-l)}(x)
\end{equation}
where the $l$ double roots of the factorization correspond to the collapsed cycles.  Since these collapsed cycles correspond to monopole fields that are frozen to a particular vacuum expectation value, they are no longer dynamical and the double roots can be dropped from the factorized curve, giving a ``reduced curve'' that describes the remaining low energy $\N=1$ dynamics.  This curve is to be identified with the \nOne\ curve $y^2 = W'(x)^2 - f_{n-1}(x)$ studied in section \ref{sec:n1curve}.  

For $\N=2$ $\U(N)$ or $\SU(N)$ pure gauge theory, the factorization of the curve is achieved as follows \cite{Douglas:1995nw}:
\begin{eqnarray}
\label{curvepure}
y^2 &=& P_N(x)^2 - 4 \Lambda^{2N}\nonumber \\
&=& 4 \Lambda^{2N}( T_N(x)^2 - 1)
\end{eqnarray}
where $T_N(x)$ are the Chebyshev polynomials of the first kind, defined by

\begin{eqnarray}
\label{eq:cheby}
T_N(x \equiv \Cos(\theta))&=& \Cos(N \theta) \nonumber \\
&=& \frac{N}{2} \sum_{r=0}^{\lfloor \frac{N}{2} \rfloor} \frac{(-1)^r}{N-r} \comb{N-r}{r}(2 x)^{N-2r}
\end{eqnarray}
which gives the expansion of $\Cos(N \theta)$ in terms of $\Cos(\theta)$.  In other words, by tuning the parameters $s_k$ of the curve (equivalently, the gauge-invariant moduli $u_k = \frac{1}{k} \tr \Phi^k$, which are related to the $s_k$ via $k u_k + k s_k + \sum_{i=1}^{k-1} i u_i s_{k-i} = 0$), we can obtain $P_N(x) =2  \Lambda^N T_N(\frac{x}{2 \Lambda})$, therefore 

\begin{eqnarray}
P_N(x)^2 - \Lambda^{2N} &=& \Lambda^N(\Cos^2(N \theta) - 1) = \Lambda^N(\Sin^2(N \theta)) \nonumber \\
&=& \Lambda^N \sqrt{1-\frac{x^2}{4\Lambda^2}} U_{N-1}(\frac{x}{2 \Lambda})^2
\end{eqnarray}
where $U_N(x)$ are the Chebyshev polynomials of the second kind, given by

\begin{equation}
U_N(x) = \sum_{r=0}^{\lfloor \frac{n}{2} \rfloor} (-1)^r \comb{n-r}{r} (2x)^{n-2r}
\end{equation}

From (\ref{eq:cheby}) one can read off the values of the $s_k$ in this vacuum.  To convert to $u_k$ we use  the product form

\begin{equation}
T_N(x) = 2^{N-1} \prod_{k=1}^N (x-\Cos( \frac{(2k-1) \pi}{2N}) \equiv 2^{N-1} \prod_{k=1}^N (x-x_k) 
\end{equation}
with

\begin{equation}
\label{eq:powersum}
u_k = \frac{1}{k} \sum_{i=1}^N x_i^k
\end{equation}
Expanding the power sum for $\SU(N)$ gives

\begin{equation}
u_k = \left\{ \begin{array}{ll}
0 & k \hbox{ odd} \\
\frac{1}{k} \comb{k}{k/2} \Lambda^{k} & k \hbox{ even}
\end{array} \right. 
\end{equation}
and therefore we have the effective superpotential 

\begin{eqnarray}
\label{eq:wpurelow}
W &=& \sum g_i \langle u_i \rangle \nonumber \\
&=& \sum \frac{g_{2k}}{2k} \comb{2k}{k} \Lambda^{2k}
\end{eqnarray}

The result for $\U(N)$ may be obtained from (\ref{eq:wpurelow}) by shifting $x \rightarrow x + u_1/N = x - \phi$ in (\ref{eq:powersum}) to account for the non-zero trace of $\Phi$, where the equality follows since we are in a maximally-confining $U(N)$ vacuum, for which classically $\langle \Phi \rangle = {\hbox{diag}}(\phi, \phi, \ldots, \phi)$.  Explicitly, for the maximally-confining $\U(N)$ vacua,

\begin{equation}
\label{ferrari:moduli}
u_p = \frac{N}{p} \sum_{q=0}^{\lfloor p/2 \rfloor} \comb{p}{2q} \comb{2q}{q} \Lambda^{2q} \phi^{p-2q} 
\end{equation}

If we wish, we can rewrite this expression in terms of the gaugino bilinear $S$, by performing a Legendre transformation with respect to the corresponding source $\log(\Lambda^{2N})$ (i.e.``integrating in S") \cite{Ferrari:2002jp}:

\begin{eqnarray}
\label{eq:glueballw}
W(\phi, g_p, S, \Lambda^2) &=& \sum_{p \ge 1} g_p u_p(\phi,\Lambda^2 = y) + S \log(\frac{\Lambda^{2N}}{y^N}) \nonumber \\
&=& N \sum_{p \ge 1} \frac{g_p}{p} \sum_{q=0}^{\lfloor p/2 \rfloor} \comb{p}{2q} \comb{2q}{q} y^q \phi^{p-2q} \nonumber \\
&&\quad+ S \log(\frac{\Lambda^{2N}}{y^N})
\end{eqnarray}
Of course, this form of $W$ does not contain any additional information, but it is useful for comparison with the other techniques we discuss.  For example, when we study the relationship between effective superpotentials and integrable systems in chapter \ref{ch:int} we will recover this expression from the Lax matrix of the affine Toda system.

\subsection{Factorization of the Seiberg-Witten curve for $N_f>0$}
\label{sec:nffactor}

The Seiberg-Witten curve for \ntwo\ gauge theory with $0 \le N_f  < 2 N_c$ fundamental hypermultiplets is \cite{D'Hoker:1997nv}

\begin{equation}
\label{eq:curvehypers}
y^2 = P_{N_c}(x)^2 - 4 \Lambda^{2N_c-N_f} \prod_{i=1}^{N_f} (x+m_i)
\end{equation}
where $m_i$ are the bare hypermultiplet masses.  When $N_f \ge N_c$ there is an ambiguity in the curve, and a polynomial of order $N_f - N_c$ in $x$ (multiplied by appropriate powers of $\Lambda$ to have well-defined scaling dimension $n$) may be added to $P_{N_c}(x)$ without changing the \ntwo\ prepotential.  For comparison to the results of section \ref{sec:examples}, we will mainly be interested in the case $N_f = N_c$  for which the ambiguity in $P_{N_c}(x)$ appears at constant order and is proportional to $\Lambda^N$.

The curve (\ref{eq:curvehypers}) can be scaled to recover the $N_f=0$ curve (\ref{curvepure}) by taking the limit

\begin{equation}
\Lambda \rightarrow 0,\hspace{0.3in} m_i \rightarrow \infty,\hspace{0.3in}  \Lambda^{2N_c-N_f} \prod m_i \equiv \tilde{\Lambda}^{2N_c}
\end{equation}
with $\tilde{\Lambda}$ finite.  Note that the latter identification is the scale-matching relation of the theories above and below the mass scale of the fundamentals.

We now show how the factorization using Chebyshev polynomials can be generalized to the hypermultiplet curve (\ref{eq:curvehypers}) (this problem has been studied indirectly using matrix models in \cite{Demasure:2002jb}).  Define the functions

\begin{eqnarray}
P_{N_c}(\theta) &=& \sum_{i=0}^{N_f}  \nu_i \Cos((N_c-i) \theta) \nonumber \\
Q_{N_c}(\theta) &=& \imath \sum_{i=0}^{N_f} \nu_i \Sin((N_c-i) \theta)
\end{eqnarray}
Then
\begin{eqnarray}
P_{N_c}^2 - Q_{N_c}^2 &=& \sum_i \nu_i^2 + 2 \sum_{i \neq j} \nu_i \nu_j ( \Cos(i \theta) \Cos(j \theta) + \Sin(i \theta) \Sin(j \theta) )\nonumber \\
&=& \sum_i \nu_i^2 + 2 \sum_{i \neq j} \nu_i \nu_j \Cos((i-j) \theta) \equiv R_{N_f}(\theta)
\end{eqnarray}
Therefore the equation
\begin{equation}
P_{N_c}^2 - R_{N_f} = Q_{N_c}^2
\end{equation}
gives the desired factorization of the Seiberg-Witten curve by setting $\Cos(\theta)= \frac{x}{2 \tilde \Lambda}$ for $\U(N)$, or  $\Cos(\theta)= \frac{x- \Lambda}{2 \tilde \Lambda}$ for $\SU(N)$, where the shift is needed to cancel the $x^{N-1}$ term in $P_N(x)$.  The parameters $\nu_i$ are related to the fundamental masses $m_i$, although the relations are polynomial in general.

This expression simplifies dramatically when $N_f = N_c, m_i \equiv m$, and we find

\begin{eqnarray}
P_N &=& \sum_{i=0}^N \comb{N}{i}\beta^{N-i} \Cos(i \theta) \nonumber \\
&=& (\beta + e^{\imath \theta})^N + (\beta + e^{-\imath \theta})^N \nonumber \\
Q_N &=& \imath \sum_{i=0}^N \comb{N}{i}\beta^{N-i} \Sin(i \theta) \nonumber \\
&=& (\beta + e^{\imath \theta})^N - (\beta + e^{-\imath \theta})^N
\end{eqnarray}
with $\beta = \Lambda/\tilde{\Lambda}$, where $\Lambda$ is the scale of the theory with flavors, and $\tilde{\Lambda}^2 = m \Lambda$ is the parameter defined above that corresponds to the dynamical scale of the theory in the limit where the fundamentals have been scaled out completely.  If we choose a limit where the fundamental masses become very large compared to the scale $\Lambda$, i.e.~such that $\beta$ becomes a small parameter, then the curve can be treated as a small deformation of the $N_f=0$ curve. 

After some algebra, we obtain the following expression for $P_N(x)$:

\begin{equation}
P_N(x) = 2 \Lambda^N  + \sum_{i=1}^N i \comb{N}{i} \Lambda^{N-i} \tilde \Lambda^i \sum_{r=0}^{\lfloor \frac{i}{2} \rfloor} \frac{(-1)^r}{i-r} \comb{i-r}{r} \frac{(x - \Delta)^{i-2r}}{\tilde \Lambda}
\end{equation}
where $\Delta = 0$ for $\U(N)$ and $\Delta=\Lambda$ for $\SU(N)$ to cancel the first subleading power of $x$.  We can resum this expression to extract the $s_k$ \cite{Moriarty}.  For $\U(N)$ we find

\begin{equation}
\label{eq:sk}
s_{N-j} = \Lambda^{N-j} \sum_{r=0}^{\lfloor \frac{i-j}{2} \rfloor} (j+2r)\comb{N}{j+2r} \frac{(-1)^r}{j+r} \comb{j+r}{r} \left(\frac{\tilde \Lambda}{\Lambda}\right)^{2r} + 2 \Lambda^N \delta_{j,0}
\end{equation}
and for $\SU(N)$ we find

\begin{eqnarray}
s_{N-j} &=& \Lambda^{N-j} \sum_{i=1}^N i \comb{N}{i}\sum_{r=0}^{\lfloor \frac{i-j}{2} \rfloor} \frac{(-1)^{i-j-r}}{i-r} \comb{i-r}{r} \comb{i-2r}{j}  \left(\frac{\tilde \Lambda}{\Lambda}\right)^{2r} \nonumber \\
&&+ 2 \Lambda^N \delta_{j,0}
\end{eqnarray}

We now compare to the results obtained in section \ref{sec:uv} based on period integrals of the \nOne\ curve.  Recall that for $N_f=N_c$ we obtained the expression (\ref{eq:wnfnc})

\begin{equation}
\label{eq:uk}
\Weff=N_c \left[ \sum_{i=1}^{\lfloor \frac{n+1}{2} \rfloor} \frac{g_{2i}}{2i} \comb{2i}{i} \tilde\Lambda^{2i} + \sum_{i=1}^{n+1} \frac{g_i}{i} \sum_{k=\lfloor\frac{i}{2}\rfloor + 1}^{i}\comb{i}{k} \tilde\Lambda^{2(i-k)} \Lambda^{2k-i}\right]
\end{equation}
From this expression can be read off the values of the gauge-invariant moduli $\langle u_k \rangle = \frac{\partial W}{\partial g_k}$.  Note that our result has the form of a finite series expansion in $\beta$, and in the limit $\beta=0$ we recover the superpotential (\ref{eq:glueballlow}) of the $N_f=0$ theory.  The $u_k$ are related to the curve parameters $s_k$ via the Newton formula\footnote{This footnote inserted to see if anyone notices it.}

\begin{equation}
k u_k + k s_k + \sum_{i=1}^{k-1} i u_i s_{k-i} = 0
\end{equation}

As in section \ref{sec:sw}, the $\SU(N)$ moduli $\tilde u_k$ may be obtained from the $\U(N)$ by shifting away the trace:

\begin{equation}
\label{eq:shiftedmoduli}
\tilde u_k = \sum_{i=1}^{N}(x_i - \frac{u_1}{N})^k
\end{equation}
Expanding the powers in (\ref{eq:shiftedmoduli}) one finds

\begin{equation}
\tilde u_k = \frac{1}{k} \left( \sum_{j=1}^k (\frac{-u_1}{N})^{k-j} j \comb{k}{j} u_j + N (\frac{-u_1}{N})^k\right)
\end{equation}

We have verified in a number of cases that the $u_k$ associated to the $s_k$ (\ref{eq:sk}) obtained from the factorized Seiberg-Witten curve agree with the values calculated from the superpotential (\ref{eq:uk}), up to a physically irrelevant sign $\Lambda \rightarrow - \Lambda$ (which can be absorbed into the conventions used to define the Seiberg-Witten curve (\ref{eq:curvehypers})) and the ambiguity in the top modulus $u_N$ at order $\Lambda^N$.

For example, the factorization for the first few $\U(N)$ curves is achieved by:
\\\\
$\U(2)$:
\begin{eqnarray}
P_2(x) &=& x^2 + 2x \Lambda - 2 \tLambda^2 + 2 \Lambda^2  \nonumber \\
u_1 &=& -2 \Lambda,\nonumber \\
u_2 &=& 2 \tLambda^2
\end{eqnarray}
$\U(3)$:
\begin{eqnarray}
P_3(x) &=& x^3 + 3 x^2 \Lambda + x(-3 \tLambda^2 + 3 \Lambda^2) - 6 \tLambda^2 \Lambda + 2 \Lambda^3 \nonumber \\
u_1 &=& -3 \Lambda,\nonumber \\
u_2 &=& 3(\tLambda^2 + \frac{1}{2} \Lambda^2),\nonumber \\
u_3 &=& 3(-\tLambda^2 \Lambda - \frac{2}{3} \Lambda^3)
\end{eqnarray}
$\U(4)$:
\begin{eqnarray}
P_4(x) &=& x^4 + 4 x^3 \Lambda + x^2(-4 \tLambda^2 + 6 \Lambda^2) + x(-12 \tLambda^2 \Lambda + 4 \Lambda^3) \nonumber \\
&&  + 2 \tLambda^4 - 12 \tLambda^2 \Lambda^2 + 2 \Lambda^4 \nonumber \\
u_1 &=& -4 \Lambda,\nonumber \\
u_2 &=& 4(\tLambda^2 + \frac{1}{2}\Lambda^2),\ \nonumber \\
u_3 &=& 4(-\tLambda^2 \Lambda^2 - \frac{1}{3} \Lambda^3),\nonumber \\
u_4 &=& 4(\frac{3}{2} \tLambda^4 + \tLambda^2 \Lambda^2)
\end{eqnarray}
which can be compared to the $u_k$ read off from (\ref{eq:uk}):
\begin{equation}
\begin{array}{c}
u_1 = N \Lambda,\ u_2 = N (\tLambda^2 + \frac{1}{2} \Lambda^2),\ u_3 = N(\tLambda^2 \Lambda + \frac{1}{3} \Lambda^3),\nonumber \\
u_4 = N(\frac{3}{2} \tLambda^4 + \tLambda^2 \Lambda^2 + \frac{1}{4} \Lambda^4)
\end{array}
\end{equation}

\section{$\SO$/$\Sp$ gauge groups}

In chapter \ref{ch:mm} we will describe our work \cite{Ashok:2002bi} on using matrix models to compute effective superpotentials for $\SO$ and $\Sp$ gauge theories.  Here we give a short discussion of field-theoretical aspects involving the Seiberg-Witten curve.  The Seiberg-Witten curves for $\N=2$ pure Yang-Mills theory with $\SO$/$\Sp$ gauge groups were found by \cite{Danielsson:1995is,Brandhuber:1995zp,Argyres:1996fw}.  For a rank-$r$ gauge theory, the spectral curve is a genus $r$ hyperelliptic curve, of the form
\begin{equation}
y^2 = P_{2r+2}(x, \{\phi_i\}),
\label{spectral1}
\end{equation}
where $P_{2r+2}$ is a polynomial of degree $2r+2$ in the $x$ that also depends on the moduli $\phi_i$.

The $\SO$ and $\Sp$ spectral curves can also be written as a genus $2r-1$ curve,
\begin{equation}
y^2 = P_{2r}(x^2, \{\phi_i\}),
\label{spectral2}
\end{equation}
which is therefore symmetric under the $\Z_2$ action $x \mapsto -x$ and is a double cover of the genus $N$ curve (\ref{spectral1}) via this map.  This is the form of the curve we will work with.  Because of the $\Z_{2}$ symmetry, each point is paired with its image; this will be important when we discuss the matrix models for $\SO$ and $\Sp$ gauge theories in section \ref{sec:mm.sosp}, since the matrix model eigenvalues live on the factorization of this curve, and therefore also come in pairs.

\ntwo\ \SUSY\ may again be broken to \nOne\ by an appropriate gauge-invariant superpotential term for $\Phi$.  Because the trace of odd powers of  matrices in the Lie algebra  of $\SO(N)$/$\Sp(N)$ vanishes, 
the superpotential deformation for  $\SO(N)$/$\Sp(N)$ only includes polynomial terms of even degree:
\begin{equation}
\Wtree(\Phi) = \sum_{k=1}^{n+1} \frac{g_k}{2k} \hbox{Tr}(\Phi^{2k}).
\label{wtree}
\end{equation}
A superpotential $\Wtree$ of order $2n+2$ breaks the gauge symmetry down to a direct product of $n+1$ subgroups, {\it e.g.}:
\begin{equation}
\SO(N) \rightarrow \SO(N_0) \times U(N_1) \times \ldots \times U(N_n),
\end{equation} 
where $N=N_0+2N_1+\cdots+2N_n$.

We saw in section \ref{sec:sw} that the supersymmetric vacua of the \nOne\ theory require $r-n$ mutually local monopoles to simultaneously become massless and condense, leading to confinement of the gauge theory.  Imposing this condition is therefore equivalent to the factorization~\cite{Cachazo:2002pr} 
\begin{equation}
y^2 = \prod_{i=1}^{r-n} (x^2 - p_i^2)^2 \prod_{j=1}^{2n}(x^2 - q_i^2),
\end{equation}
where $p_i \neq p_j, q_i \neq q_j$ for $i \neq j$.  On this locus we then obtain (after discarding the terms corresponding to the non-dynamical condensed monopoles) the reduced spectral curve
\begin{equation}
y^2 = \prod_{j=1}^{2n}(x^2 - q_i^2),
\label{n1curve}
\end{equation}
which has genus $2n-1$.  This curve parameterizes the \ntwo\ vacua that are not lifted by the deformation to \nOne\ (\ref{wtree}). Notice that
the curve is still invariant under $x \mapsto -x$: this implies that the branch points come in pairs: $(-q_i, q_i)$.  This reduced spectral curve is identified with the curve (\ref{eq:spectralcurve}) arising from string theory discussed in section \ref{sec:cy}.

The factorization problem was solved by \cite{Janik:2002nz} along the lines of the $\SU(N)$ discussion in section \ref{sec:sw}, however as discussed in section \ref{sec:n1curve} the effective superpotential of the $\N=1$ gauge theory can also be obtained from the periods of (\ref{n1curve}). It will take the form
\begin{equation}
\Weff = \sum_i \left( \hat{N_i} \Pi_i + \tau_i S_i \right),
\label{weff}
\end{equation}
where $4 \pi i S_i$ are the periods of the meromorphic 1-form $y\,dx$ around the A-cycles of the spectral curve, $\Pi_i$ the corresponding periods around the B-cycles, and $\hat{N_i}$ is
\begin{equation}
\hat{N_i} = \left\{\begin{array}{cc}N_i & \SU(N_i), \\
\frac{N_i}{2} - 1 & \SO(N_i), \\
N_i + 1 & \Sp(N_i).
\end{array}\right.
\end{equation}

In chapter \ref{ch:mm} we will see how the shift $N_i \mapsto \hat{N_i}$ emerges from a subleading correction to the effective superpotential, obtained using matrix integral techniques.


\chapter{Integrable Systems and $\N=1$ Vacua}
\label{ch:int}

In this chapter we investigate the relationship between \nOne\ superpotentials and integrable systems.  Integrable systems are known to underly the low energy structure of gauge theories with $\N=2$ supersymmetry (\cite{Martinec:1996by}, see \cite{D'Hoker:1999ft} for a  review), and this underlying structure again survives soft supersymmetry breaking to $\N=1$ to govern the effective superpotential.  The origin of these integrable structures in gauge theory is still incompletely understood.  In this chapter we obtain some new details about the correspondence, and it is an open problem to understand the results in a more general context.

In section \ref{sec:nffactor} we obtained simple combinatorial formulae for the moduli $u_{k}$ of the $\N=2$ Seiberg-Witten curve at the maximal factorization locus.  For $N_f=0$, these combinatorial formulae are encoded in the traces of powers of a particular matrix, namely the scalar component of the adjoint field $\Phi$, evaluated in the vacuum of interest:

\begin{equation}
\langle u_{k} \rangle = \frac{1}{k}  \tr \Phi^{k}
\end{equation}

The connection to integrable systems is via this matrix $\Phi$, which is identified with a Lax matrix for the associated integrable system.  The Lax matrix completely defines the dynamics of the integrable system.  The integrable system associated to pure $\N=2$ Yang-Mills is the periodic Toda chain \cite{Martinec:1996by}, and the known Lax matrix of this system can indeed be identified with $\langle \Phi \rangle$.  An algorithm was presented by \cite{Boels:2003fh} for computing the effective superpotential of $\N=1$ gauge theory with an adjoint chiral superfield, using this Lax matrix.

The integrable system associated to \ntwo\ SQCD (i.e.~$N_{f}\neq0$) was uncovered in \cite{Gorsky:1996hs,Gorsky:1996qp}, and is a particular spin chain system.  However the known Lax pair of this system is written in transfer matrix form as a chain of $2\times2$ matrices, for which the connection to $\N=1$ superpotentials is less direct since this form of the Lax matrix does not have an obvious physical meaning\footnote{See however the recent work \cite{Hollowood:2004dc}.}.  Therefore, it would be useful to find another Lax pair for this system that takes the form of a single matrix, similar to the $N_f=0$ case\footnote{Note that a given integrable system may have more than one Lax pair, and the matrices may even be of different rank, so we should not be discouraged from looking for a new Lax formulation.}.  We studied this problem in \cite{Kennaway:2003jt}, and in section \ref{sec:laxnf} we present the $N_c \times N_c$ matrix $\langle \Phi \rangle$ that encodes the $\langle u_k \rangle$ in the maximally-confining vacua, which is identified with a particular equilibrium value of the Lax matrix for the associated spin chain.

We begin this section by proving that the superpotential calculation of \cite{Boels:2003fh} using the integrable structure of the \ntwo\ gauge theory, yields the same result in the maximally-confining phase as (\ref{eq:glueballlow}), (\ref{eq:glueballw}) obtained from the $N_f=0$ period integral and factorization calculations \cite{Kennaway:2003jt}.

\section{The periodic Toda chain and $\N=1$, $N_{f}=0$ vacua}
\label{sec:intsys}

The integrable system associated to pure \ntwo\ Yang-Mills theory is the periodic Toda chain, which has Lax matrix:

\begin{equation}
\label{eq:laxmatrix}
L = \left( \begin{array}{ccccccc}
\phi_1 & y_1 & 0 & \ldots &0 & z \\
1 & \phi_2 & y_2 & 0 & \ldots & 0 \\
0 & \ddots & \ddots & \ddots & \ddots & 0 \\
0 & \ddots & \ddots & \ddots & \ddots & 0 \\
0 & \ddots & \ddots & \ddots & \ddots & y_{N-1} \\
y_N/z & 0 & \ldots & 0 & 1 & \phi_N
\end{array}\right)
\end{equation}
where $\phi_i$, $y_i$ are the dynamical position and momentum variables of the integrable system, whose precise definition will not be important for us (see \cite{D'Hoker:1999ft} for a review), and $z$ is a ``spectral parameter'', an auxilliary variable not associated to the physical system. The conserved quantities (Hamiltonians) of the Toda system $U_k = {\frac{1}{k}} \tr L^k$ are associated to the gauge-invariant polynomials $u_k = {\frac{1}{k}} \tr \Phi^k$ that parametrize the moduli space of the \ntwo\ gauge theory.  The spectral curve of the Lax system is defined by
\begin{equation}
{\hbox{det}}(x.I - L) \equiv P_N(x) + (-1)^N (z + \Lambda^{2N} z^{-1}) = 0
\end{equation}
where $P_N$ are the polynomials defined in section \ref{sec:sw}.  Under the change of coordinates
\begin{equation}
\label{yz}
y = 2z + (-1)^N P_N(x)
\end{equation}
the spectral curve becomes
\begin{equation}
\label{spectral}
y^2 = P_N(x)^2 - 4 \Lambda^{2N}
\end{equation}
which is the standard form of the Seiberg-Witten curve of \ntwo\ $\U(N)$ Yang-Mills theory.  Therefore, when we deform the \ntwo\ theory by turning on a tree-level superpotential
\begin{equation}
W = \sum_{i=1}^{n+1} g_i u_i
\end{equation}
the analogous quantity in the Toda system is the corresponding function of the conserved quantities $U_i$.  The essence of the proposal of \cite{Boels:2003fh} is that evaluating $W(L)$ gives the exact effective superpotential of the theory\footnote{When the superpotential $\Wtree$ contains terms of degree $N$ or higher, the spectral parameter $z$ that appears in the Lax matrix (\ref{eq:laxmatrix}) will appear in the $U_k$.  However, in the quantum \nOne\ gauge theory these moduli are ambiguous because the operators $\tr \Phi^k, k \ge N$ receive quantum corrections, and the resolution proposed in \cite{Boels:2003fh} was that all occurrences of $z$ in the Lax superpotential $W(L)$ should be discarded at the end of the computation (alternatively they can be suppressed to arbitrarily high orders by embedding $\U(N) \subset \U(tN)$).}.  The factorization of the spectral curve at the points corresponding to $\N=1$ supersymmetric vacua translates in the integrable system to equilibrium configurations that are stationary under the Hamiltonian flows generated by the $U_k$ \cite{Hollowood:2003ds}.

We will now obtain the explicit form of $\WLax$ for a given $\Wtree$ and recover the result in section \ref{sec:sw}.  For this purpose the form of the Lax matrix (\ref{eq:laxmatrix}) is slightly awkward to work with, because the $z$ entries are not on the same footing as the other variables.  To rectify this, conjugate $L$ by ${\hbox{diag}}(1, z^{1/N}, z^{2/N}, \ldots, z^{N-1/N})$ to bring it into the form:
\begin{eqnarray}
\label{eq:todalaxy}
L &\sim& \left( \begin{array}{ccccccc}
\phi & \frac{y}{z^{1/N}} & 0 & \ldots &0 & z^{1/N} \\
z^{1/N} & \phi & \frac{y}{z^{1/N}} & 0 & \ldots & 0 \\
0 & \ddots & \ddots & \ddots & \ddots & 0 \\
0 & \ddots & \ddots & \ddots & \ddots & 0 \\
0 & \ddots & \ddots & \ddots & \ddots & \frac{y}{z^{1/N}} \\
\frac{y}{z^{1/N}} & 0 & \ldots & 0 & z^{1/N} & \phi
\end{array}\right) \nonumber \\
&=& \phi I + \frac{y}{z^{1/N}} S + z^{1/N} S^{-1}
\end{eqnarray}
where $S$ is the $N\times N$ shift matrix, satisfying $S^N = I$.

Therefore,
\begin{eqnarray}
\tr(L^p) &=& \tr\Big(\sum_{l=0}^p \phi^{p-l} \comb{p}{l} I \sum_{m=0}^l \left(\frac{y}{z^{1/N}}\right)^m z^{-m/N} S^{2m-l} \comb{l}{m} \Big) \nonumber \\
&=& N \sum_{l=0}^{p} \phi^{p-l} \comb{p}{l} {\sum_{a=-\lfloor \frac{l}{2N} \rfloor}^{\lfloor \frac{l}{2N} \rfloor}} y^{(Na+l)/2} z^{-a} \comb{2l}{\frac{Na+l}{2}}
\end{eqnarray}
where in the second line we have used the fact that the terms can only appear on the diagonal if $2m-l = Na$, $a \in \Z$.  Suppressing powers of $z$ whenever they appear, we obtain

\begin{equation}
\WLax= N \sum_{p \ge 1} \frac{g_p}{p} \sum_{q=0}^{\lfloor p/2 \rfloor} \comb{p}{2q} \comb{2q}{q} \phi^{p-2q} y^q + S \log(\frac{\Lambda^{2N}}{y^N})
\end{equation}
which recovers the expressions (\ref{eq:glueballlow}), (\ref{eq:glueballw}) obtained using exact field theory techniques, and by evaluating period integrals.

\section{Results on a new Lax matrix for $N_f=N_c$}
\label{sec:laxnf}

The connection between \ntwo\ gauge theories and integrable systems can be summarized by identifying the matrix-valued field $\Phi$ of the quantum gauge theory with a Lax matrix for the integrable system.  Therefore, if we can evaluate $\langle \Phi \rangle$ in a given vacuum, we know the value of the Lax matrix in an equilibrium configuration of the integrable system.  Knowing the values of the moduli $\langle u_k \rangle$ in the particular \ntwo\ vacuum gives $N_c$ equations for the matrix $\langle \Phi \rangle$, which is enough in principle to determine $\langle \Phi \rangle$ up to gauge transformations.

In the previous section, we showed how evaluating the Toda Lax matrix in a particular equilibrium configuration (all position and momentum variables equal, i.e.~$\phi_i\ \equiv\ \phi,\\ y_i\ \equiv\ y\ \equiv\ \tLambda^2$) allows us to recover the $\langle u_k \rangle$ of the factorized Seiberg-Witten curve.  Conversely, given the $\langle u_k \rangle$, we can reconstruct the Lax matrix of the periodic Toda chain: the $\langle u_k \rangle$ may be obtained from the single matrix\footnote{The entry with coefficient 2 exists because (\ref{eq:todalax}) does not contain the spectral parameter $z$ (which does not have a physical meaning in the gauge theory), so we can absorb the entry $\tLambda^2/z$ of (\ref{eq:laxmatrix}) into this entry.}

\begin{equation}
\label{eq:todalax}
\langle \Phi \rangle = \left( \begin{array}{cccccc}
\phi & \tLambda^2 & 0 & \ldots & 0 \\
1 & \phi & \tLambda^2 & \ldots & 0 \\
\vdots & \ddots & \ddots &  \ddots & \vdots \\
0 & \ldots & 1 &\phi & 2 \tLambda^2 \\
0 & \ldots & 0 & 1 & \phi
\end{array}\right)
\end{equation}

One can explicitly see from this expression how the classical value of $\Phi = \hbox{diag}(\phi_1, \ldots, \phi_N)$ is deformed by quantum effects, specifically the interaction with the background magnetic field of the condensed monopoles, which generates the off-diagonal terms (this can most easily be derived via compactification to 3 dimensions, where the four-dimensional monopoles reduce to 3-dimensional instantons \cite{deBoer:1997kr}).

We follow the same philosophy for the $N_f=N_c$ vacua studied in section \ref{sec:nffactor}, and identify the matrix $\Phi$ from which the expectation values of the moduli $\langle u_k \rangle$ in the maximally-confining vacua may again be obtained by taking the trace of powers (recall that in this case the moduli took the form of a finite series).  We therefore have a candidate for a Lax matrix of the associated integrable system, which in these examples are spin chains \cite{Gorsky:1996hs, Gorsky:1996qp}.

We find for $\SU(N_c)$, $N_f=N_c$ and all quark masses equal, that the moduli $\langle u_k \rangle$ of the maximally confining vacuum may be obtained from the matrix

\begin{equation}
\langle \Phi \rangle = \left( \begin{array}{ccccccc}
0 & \tLambda^2 & \Lambda \tLambda^2 &  \Lambda^2 \tLambda^2 & \ldots &  \Lambda^{N-2} \tLambda^2 &  N \Lambda^{N-1} \tLambda^2 \\
1 & 0 & \tLambda^2 & \Lambda \tLambda^2 & \ldots &\Lambda^{N-3} \tLambda^2  & (N-1) \Lambda^{N-2} \tLambda^2 \\
0 & 1 & 0 & \tLambda^2 &  \ldots & \Lambda^{N-4} \tLambda^2 &  (N-2) \Lambda^{N-3} \tLambda^2 \\
& \ddots & \ddots & \ddots & & \ddots & \vdots \\
0 & 0 & 0 & 1 & 0 & \tLambda^2 & 3 \Lambda  \tLambda^2 \\
0 & 0 & 0 & 0 & 1 & 0 & 2 \tLambda^2 \\
0 & 0 & 0 & 0 & 0 & 1 & 0 \\
\end{array} \right)
\end{equation}
Note that this reduces to the Toda Lax matrix (\ref{eq:todalax}) in the appropriate scaling limit $\Lambda \rightarrow 0$ (here $\phi = 0$ for the $\SU(N)$ vacua to ensure tracelessness).  It remains an open problem to generalize this matrix to a general vacuum and to better understand the relationship with the degrees of freedom of the spin chain system.

\chapter{The Combinatorial Structure of Supersymmetric Vacua}
\label{ch:mm}




In this chapter, we first introduce matrix integrals and review how they may be solved in a genus expansion.  The solution produces a ``spectral curve'', which is isomorphic to the $\N=1$ curve (\ref{eq:spectralcurve}) when the potential of the matrix model is identified with the tree-level superpotential for the adjoint chiral superfield $W(\Phi)$.  The matrix integral is a generating function for Feynman diagrams that are in one-to-one correspondence with planar $\Phi$ diagrams for the $\SU(N)$ gauge theory, although the diagrams of the four-dimensional gauge theory carry four-dimensional momenta in the loops whereas the matrix diagrams are zero-dimensional and carry no internal momenta.  Nonetheless, as we discussed in chapter \ref{ch:geom} this spectral curve produces the exact effective superpotential of the four-dimensional gauge theory in terms of integrals of the resolvent along contours of the curve.  The insight afforded by the matrix model is that it provides a remarkably simple perturbative expansion of this effective superpotential.

The relationship between matrix integrals and gauge theory superpotentials was first observed for \nOne\ $\SU(N)$ gauge theories with adjoint matter~\cite{Dijkgraaf:2002fc} and for the ${\cal N} = 1^*$ deformation of ${\cal N}=4$ $\SU(N)$ SYM~\cite{Dijkgraaf:2002dh,Dorey:2002tj,Dorey:2002jc}.  The conjecture was subsequently extended to a number of other cases including~\cite{Dijkgraaf:2002vw,Chekhov:2002tf,Dorey:2002pq,Ferrari:2002kq,Berenstein:2002sn,Fuji:2002wd,Gopakumar:2002wx,Demasure:2002sc,Argurio:2002xv,McGreevy:2002yg,Suzuki:2002gp,Bena:2002kw,Gorsky:2002uk,Naculich:2002hi,Tachikawa:2002wk,Dijkgraaf:2002wr,Klemm:2002pa,Dijkgraaf:2002yn,Feng:2002zb,Feng:2002yf,Argurio:2002hk,Naculich:2002hr}.

The underlying quantum field theoretical reason for the correspondence is that after summing the four-dimensional Feynman diagram contributions of the field $\Phi$ to the effective superpotential, all dependence on internal loop momenta cancels and one is left with only the zero-momentum planar $\Phi$ diagrams of the theory \cite{Dijkgraaf:2002xd, Aganagic:2003xq} (and for $\SO$/$Sp$ gauge groups, and theories with fundamental matter, the leading non-planar), which are generated by the associated matrix integral.  In section \ref{sec:mm} we review the technology of zero-dimensional matrix models, the computation of the matrix model free energy and the gauge theory effective superpotential.

In order to understand the combinatorial origin of the divergent period integral appearing in the matrix model calculation, we extend the matrix integral to include $M\times 1$ and $1 \times M$ vectors (corresponding to gauge theory with matter in the fundamental and antifundamental representations).  The generating function of planar diagrams with 1 boundary recovers the previous expression (\ref{eq:wnf}) for the quark contributions to the superpotential in terms of period integrals of the spectral curve.  As in the geometrical analysis of chapter \ref{ch:geom}, adding $N_f=2N_c$ vectors to the matrix potential causes the divergences of the period integral to cancel, and we see explicitly the role of the ``quark'' vectors in regularizing the matrix integral computation of the effective superpotential.  Thus, the cancellation of this divergence is understood perturbatively in the gauge theory as coming from the contribution to the effective superpotential of planar Feynman diagrams with disk topology, in the limit where the quarks that propagate around the boundary of the diagram are very heavy.  This result was published in \cite{Kennaway:2003jt}.

Finally, we extend the analysis to gauge theories with $SO$ and $Sp$ gauge groups, which was first published in \cite{Ashok:2002bi}.  This amounts to including in the counting the non-orientable Ribbon diagrams with $\RP^{2}$ topology.  This requires an adaptation of the technique of higher-genus loop equations \cite{Ambjorn:1993gw,Akemann:1996zr}.  We show that the $\RP^{2}$ contribution to the resolvent -- and hence to the matrix model free energy -- has a simple form and is related to the genus-0 result.


\section{Matrix integrals and zero-dimensional matrix models}
\label{sec:mm}

Consider the matrix integral

\begin{equation}
\label{eq:matrixint}
Z = Z_{0} \int d \MM e^{-\frac{1}{g_s}\str W(\MM)}
\end{equation}
where $\MM \in \G$ is an $M\times M$ matrix, $W(\MM)$ is a polynomial, and $Z_{0}$ is a normalization factor.  The integral can be rewritten in terms of the eigenvalues of $\MM$ as

\begin{equation}
\label{eq:z}
Z = \int \prod_{i=1}^{M} d \lambda_{i} J(\{\lambda_{i}\}) e^{-\frac{1}{g_s}\sum_{i=1}^{M} W(\lambda_{i})}
\end{equation}
where $J$ is a suitable Jacobian for the change of variables, and $Z_0$ is fixed by the normalization of (\ref{eq:z}).  For Hermitian matrices, the change to the (diagonal) eigenvalue basis involves conjugation by unitary matrices

\begin{equation}
\MM \rightarrow {\bf U}^{\dagger} {\bf D} {\bf U}
\end{equation}
and the change of basis produces an integral over the Haar measure on $\U(M)$, which gives the volume of $\U(M)/\U(1)^M$ and fixes $Z_{0}$.  The Jacobian is given by $J=\Delta^{2}(\{\lambda_{i}\})$, where $\Delta$ is the Van der Monde determinant

\begin{equation}
\Delta(\{\lambda_{i}\}) = \prod_{i < j} (\lambda_{j} - \lambda_{i})
\end{equation}
The action for the matrix eigenvalues is then
\begin{equation}
\label{eq:mm.action}
S(\lambda_{i}) = -\frac{1}{g_s}\sum_{i=1}^{M} W(\lambda_{i}) + \sum_{i=1}^{M} \sum_{i<j} \log(\lambda_{j} - \lambda_{i})
\end{equation}

We will mostly be interested in the 't Hooft limit
\begin{equation}
\label{eq:thooftlimit}
M \rightarrow \infty,\quad S \equiv g_{s} M ={\hbox{\it const.}}
\end{equation}
Recall from section \ref{sec:cy} that this is also the limit in which gravity decouples from the string theory.
We wish to evaluate the matrix integral in this limit as a perturbative expansion around a saddle point (classical vacuum). Such a classical vacuum is given by a distribution of the eigenvalues of $M$ among the critical points $\{x_i\}$ of the function $W(x)$. We denote the number of eigenvalues at the critical point $x_i$ by $M_i$ and define the corresponding 't Hooft couplings $S_{i} = g_{s} M_{i}$.


The free energy $\F$ of the matrix model is given by

\begin{equation}
\F = \log Z = \F^{\hbox{\it pert.}} + \F^{\hbox{\it non-pert.}}
\end{equation}
where $\F^{\hbox{\it pert.}}$ comes from evaluating the integral perturbatively (as usual, it is given by the sum of connected Feynman diagrams), and $\F^{\hbox{\it non-pert.}}$ is a non-perturbative contribution that will be determined later. 

\begin{figure}[t]
\begin{center}
\epsfig{file=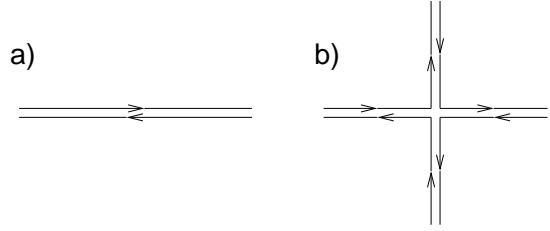}
\parbox{5.5in}{
\caption{Feynman rules for the Hermitian matrix model: a) propagator, and b) sample quartic vertex, giving the perturbative expansion in terms of ``ribbon graphs''.\label{fig:herm.rules}}}
\end{center}
\end{figure}

Consider the propagator for the Hermitian matrix model: it has a group theoretical factor \begin{equation}
\label{eq:su.propagator}
\langle \MM_{ij}\MM_{kl}\rangle\propto \delta_{il}\delta_{jk}.
\end{equation}
The propagator and interaction vertices may be represented in double-line notation, see Figure \ref{fig:herm.rules}.
We can now expand the free energy perturbatively around a given classical vacuum in terms of ribbon graphs, where the edges of the ribbon correspond to eigenvalues $\lambda_{i}$ of the matrix $\MM$.  
%
Thus, each closed loop of a ribbon graph edge contributes a factor of $M_i=S_{i}/g_{s}$, the number of eigenvalues on the $i^\text{th}$ critical point.  From the overall normalization of the action (\ref{eq:mm.action}), it is clear that each vertex of the diagram contributes a factor of $1/g_s$ and each propagator contributes $g_s$.  Thus the overall power of $g_s$ is

\begin{equation}
g_{s}^{p-v-l} = g_{s}^{-\chi} = g_{s}^{2g-2}
\end{equation}
where $p$ is the number of propagators, $v$ the number of vertices, $l$ the number of index loops and $\chi=2-2g$ is the Euler characteristic of the Riemann surface with minimal genus $g$ on which the diagram may be drawn.  Therefore the perturbative free energy has a topological expansion


\begin{equation}
\F(g_{s},S_{i}) = \F^{\hbox{\it np}}(S_{i}) + \sum_{g} {g_{s}}^{ 2 g-2} \F_{g}(S_{i})
\end{equation}
%
%
%
In the 't Hooft limit (\ref{eq:thooftlimit}) the planar (sphere-topology) diagrams dominate \footnote{Higher-genus contributions correspond to gravitational corrections from string theory.}.

Note that we are distinguishing the rank $M$ of the matrix from the rank $N_{c}$ of the four dimensional gauge theory we will soon make contact with.  As we have mentioned, it is only the planar diagrams of the four-dimensional gauge theory that contribute to the effective superpotential even for finite $N_{c}$, so the large $M$ limit taken in the matrix model is an auxilliary step designed to isolate the planar diagram contributions to the value of the matrix integral.  The explicit dependence on $M$ is hidden by rewriting $M=S/g_{s}$, and $S$ is identified with the gaugino bilinear superfield of the gauge theory.
%

A non-perturbative contribution to the free energy comes from the residual gauge invariance that exists when two or more eigenvalues coincide.  When $M_{i}$ eigenvalues are distributed in the $i^\text{th}$ critical point of the potential, the matrix integral is invariant under an additional $\prod_{i=1}^{n} \U(M_{i})$ gauge symmetry.  Thus, the path integral includes the orbit of the solution under this group, so the free energy of the matrix integral receives an additional contribution from the logarithm of the volume of these gauge factors \cite{Morozov:1995pb,Ooguri:2002gx}:

\begin{equation}
\F^{\hbox{\it np}} = \sum_{i} \log \mbox{vol}~\U(M_{i})
\end{equation}
This point was unclear in much of the literature, and the volume contribution was often confused with the normalization $Z_0$ of the matrix integral (\ref{eq:matrixint}).  However, this would not give the correct contribution in vacua with broken gauge symmetry, and the logarithm contributes with the opposite sign relative to the perturbative terms, which can be ruled out by an explicit evaluation of the free energy using the techniques we discuss below.

The asymptotic expansion of the volumes are worked out in Appendix \ref{app:mm}.  For $\U(M)$ we obtain

\begin{equation}
\log \mbox{vol}~\U(M) = -\frac{M^{2}}{2} \log M + \frac{1}{12} \log M + \frac{3}{4}M^{2} + \frac{1}{2} M^{2} \log 2 \pi + \CO(1)
\end{equation}
Changing variables using $M = S/g_{s}$ and extracting the leading term in $g_{s}$, we find that the leading-order non-perturbative contribution to the free energy is

\begin{equation}
\label{eq:f0.un}
\F_{0}^{\hbox{\it np}} = -\frac{S^{2}}{2} \log(S/g_{s}) + \frac{1}{2}S^{2} \log 2\pi + \frac{3}{4} S^{2}
\end{equation}
These non-perturbative matrix model contributions coincide with terms that describe non-perturbative physics in the gauge theory, namely the Veneziano-Yankielowicz superpotential, which is associated to the strongly-coupled gauge dynamics.  We will discuss this more later.  In principle there may be other non-perturbative contributions to the free energy.  In the limit of large $M$ we can evaluate the matrix integral by the method of steepest descent, which will allow us to compute the free energy directly.  It can then be verified in examples that the leading contribution to the free energy reproduces the perturbative and non-perturbative terms discussed above.

We will now discuss the solution of the matrix model in the eigenvalue basis.  From the eigenvalue action (\ref{eq:mm.action}), the equation of motion for a single eigenvalue $\lambda_{i}$ is

\begin{equation}
\label{eq:ev.eom}
2 \sum_{i \neq j} \frac{1}{\lambda_{i} - \lambda_{j}} = \frac{1}{g_{s}}W'(\lambda_{i})
\end{equation}
It can be solved by introducing the resolvent

\begin{equation}
\omega(z) = g_{s} \tr \frac{1}{\MM-z} = g_{s} \sum_{i} \frac{1}{\lambda_{i} - z}
\end{equation}
After multiplying by $1/(\lambda_{i}-z)$ and summing over $i$, equation (\ref{eq:ev.eom}) becomes

\begin{equation}
\label{eq:loopeq1}
\omega^{2}(z) - g_{s} \omega'(z) - W'(z) \omega(z) - \frac{1}{4} f(z) = 0
\end{equation}
where
\begin{equation}
\label{eq:fofx}
f(z)=\frac{4}{M} \sum_{i} \frac{W'(z) - W'(\lambda_{i})}{z-\lambda_{i}}
\end{equation}
Equation (\ref{eq:loopeq1}) is the {\it classical loop equation}.  In the 't Hooft limit (\ref{eq:thooftlimit}), the second term in (\ref{eq:loopeq1}) can be neglected, and performing the change of variables

\begin{equation}
\label{eq:yz}
y(z) = 2 \omega(z) - W'(z)
\end{equation}
it reduces to
\begin{equation}
\label{eq:spectral}
y^{2}(z) = W'(z)^{2} + f(z) = 0
\end{equation}
The coefficients in $f(z)$ are as yet undetermined.  Note that this curve has the same form as the curves discussed in the previous chapters, if the matrix potential $W(\MM)$ is identified with the tree-level superpotential for the adjoint chiral superfield $\Phi$.  The recovery of this curve from the matrix model is a signal that the matrix model is related to the four-dimensional gauge theory, since the same curve also emerged from string theory when we studied geometric engineering of the gauge theory in section \ref{sec:cy}.

Using (\ref{eq:yz}) and (\ref{eq:spectral}) the equation for the resolvent yields a formal solution~\cite{DiFrancesco:1995nw} 
\begin{equation}
\omega(x)=\frac{1}{2} \left(W'(x)-\sqrt{W'(x)^2+f(x)}\right)
\end{equation}
where the branch of the square root is fixed by the requirement that the resolvent have asymptotic falloff $\omega \sim S/x$, which vanishes in the classical limit $S \rightarrow 0$. The resolvent is thus expressed in terms of the $n$ unknown coefficients that appear in the polynomial $f(x)$ defined in (\ref{eq:fofx}). From the form of the solution, it is clear that the resolvent has square root branch cuts around the critical points of the matrix potential $W(\MM)$.  Physically, the eigenvalues sitting at the critical points feel a Coulomb repulsion from the logarithmic term in the eigenvalue action (\ref{eq:mm.action}), and spread out from their classical values to form the cuts.

In the large $M$ limit the distribution of matrix eigenvalues 
\begin{equation}
\rho(\lambda) = \sum_{i} \delta(\lambda-\lambda_{i})
\end{equation}
becomes continuous.  In terms of $\rho(\lambda)$ the resolvent can be rewritten as 
\begin{equation}
\omega(x)=g_{s} \int_{-\infty}^\infty\frac{\rho(\lambda)d\lambda}{\lambda-x}
\end{equation}
which implies that
\begin{equation}
\label{eq:discont}
\rho(\lambda)=\frac{1}{2\pi i g_{s}}(\omega(\lambda+i0)-\omega(\lambda-i0))=
\frac{1}{4\pi i g_{s}}(y(\lambda+i0)-y(\lambda-i0)).
\end{equation}
i.e.~the eigenvalue density is given by the discontinuity of the resolvent across its branch cuts.
The 't Hooft parameters associated to the number of eigenvalues in the $i^\text{th}$ cut are then given by
\begin{equation}
S_i= g_{s} N_{i} = \frac{1}{2\pi i}\oint_{A_i}\omega(x)dx
\end{equation}

The function $y(x)$ contains the singular part of the resolvent.  It can also be written as
\begin{equation}\label{eq:forceeqn}
y(\lambda)=-g_s{\frac{\partial S}{\partial\lambda}},
\end{equation}
where $S$ is the action, the derivative of which gives the force acting on an eigenvalue. Now, if the number of eigenvalues on the $i^\text{th}$ cut is varied by taking an eigenvalue to infinity along the non compact $B_i$ contour of the Riemann surface (\ref{eq:spectralagain}), the change in the free energy $\F_0$ of the matrix model is given by the line integral of the force along this contour: 
\begin{equation}
\label{eq:mm.bper}
\frac{\partial{\cal F}_0}{\partial S_i}=\int_{B_i^{+}}y(x)~dx = \int_{B_{i}} \omega(x)~dx - W(\Lambda_{0}).
\end{equation}
This is a differential equation that determines ${\cal F}_0$, the leading (genus 0) contribution to the free energy of the matrix model.  By equation (\ref{eq:forceeqn}) it is apparent that (\ref{eq:mm.bper}) gives the action for an eigenvalue to tunnel from the cut to infinity.

Note that the contour integral (\ref{eq:mm.bper}) is again logarithmically divergent, as we saw when we encountered the same integral in the context of factorized Seiberg-Witten curves.  As we will see in the following section, it can again be understood in the matrix model by introducing $2N_{c}$ vectors into the matrix potential, whose planar combinatorics cut off the integration contour and render the integral finite.  Thus, the matrix model provides a simple intuitive interpretation of the spectral curve and the related period integrals, in terms of the dynamics of matrix eigenvalues.

The fact that we have recovered the same curve from the matrix model suggests that the matrix model is related to the string theory (recall that the curve was obtained from the Calabi-Yau compactification manifold) and to the four-dimensional gauge theory that it engineers.  Indeed, this is the case, as first discussed in the seminal work \cite{Dijkgraaf:2002fc}: the action of B-type topological strings on the Calabi-Yau spaces of section \ref{sec:cy} reduces to the matrix models, and at the same time computes the gauge theory effective superpotentials.  

Since the spectral curve (\ref{eq:spectral}) and meromorphic 1-form $y~dx$ are the same as those obtained from the Calabi-Yau geometry discussed in section \ref{sec:cy}, the genus 0 free energy $\F_{0}$ of the matrix model is identified with the prepotential of the Calabi-Yau.  In other words, the large $M$ solution of the Hermitian 1-matrix model (\ref{eq:matrixint}) computes the prepotential of Type IIB string theory on the associated \CYM~\footnote{This is also true for other types of matrix integral, although there are relatively few matrix integrals that can be solved exactly.}.  Therefore as discussed in section \ref{sec:cy} the superpotential of the gauge theory is given in the matrix model by (\ref{eq:weffclosed})
\begin{equation}
\label{eq:weffclosed2}
\Weff(S_i)=\sum_{i}\left(N_i\pder{\F_0}{S_i}+\tau_iS_i \right). 
\end{equation}
%

Using the result (\ref{eq:f0.un}) for the non-perturbative contribution to the matrix model free energy we find

\begin{equation}
W = \sum_{i} N_{i} S_{i}(1-\log(\frac{S_{i}}{\Lambda^{3}})) + W^{\hbox{\it pert}}
\end{equation}
where the additional logarithms in (\ref{eq:f0.un}) have been absorbed into the definition of the cutoff\footnote{Strictly speaking we have not yet argued for the need to introduce a cutoff into the matrix integral; we will justify this point later.}.  Thus, the residual gauge symmetry of the matrix model vacua precisely gives rise to the Veneziano-Yankielowicz superpotential, which is associated to the strong-coupling gauge dynamics of the four-dimensional gauge theory.

Moreover, the planar diagrams of the matrix model, which contribute to the perturbative expansion of the free energy, are in 1-1 correspondence with planar Feynman diagrams of the gauge theory.  After summing these gauge theory Feynman diagrams, all dependence on four-dimensional loop momenta cancels, and their contribution is effectively zero-dimensional \cite{Dijkgraaf:2002xd}.  Thus, the matrix model can be used to compute the effective superpotential of the gauge theory.

In section \ref{sec:periodu2} we already evaluated the contour integral (\ref{eq:mm.bper}) for the simplest (Gaussian) matrix model (i.e.~gauge theory tree level superpotential $W(\Phi) = \frac{1}{2} m \tr \Phi^{2}$), and recovered the Veneziano-Yankielowicz superpotential together with correction terms that are understood as coming from the regulator superfields.  Equation (\ref{eq:mm.bper}) can be evaluated for arbitrary 1-cut matrix models (all eigenvalues in one critical point), using the techniques presented in chapter \ref{ch:geom}.  For multi-cut matrix models the calculations become more difficult, but still tractable in some cases.

As we have discussed above, the matrix model provides a simple alternative to evaluating the contour integrals: for a given vacuum distribution of eigenvalues we can simply enumerate planar Feynman diagrams of the matrix model to obtain the perturbative free energy to the desired order, and add to it the non-perturbative volume contribution from the residual gauge symmetries of the vacuum.  Summing the perturbative expansion to all orders is equivalent to evaluating the integral (\ref{eq:mm.bper}).

Comparing to the four-dimensional $\N=1$ gauge theory, we see explicitly the origins of the Veneziano-Yankielowicz superpotential for the strongly coupled gauge dynamics, as well as the perturbative corrections from planar $\Phi$ diagrams of the gauge theory.  This provides an elegant physical insight into the nature of the effective superpotentials calculated using the techniques of chapters \ref{ch:geom} and \ref{ch:int}, which involve many of the same calculations, but whose physical origins are less clear.




\section{Matrix models for adjoint and fundamental matter}
\label{sec:mm.fund}

The matrix model discussed in the previous section corresponds to $\N=1$ gauge theory with an adjoint chiral superfield.  As in the geometrical analysis of section \ref{sec:n1curve}, we encountered a logarithmic divergence in the contribution to the effective superpotential, which needed to be regulated by introducing a cutoff.  In the previous discussion we understood this cutoff as coming from $2N_c$ fundamental chiral superfields, which subtract the divergence at infinity of the integral and replace it by a cutoff equal to the mass of the fields.

The same analysis can be carried out in matrix language.  In other words, consider the matrix model with potential

\begin{equation}
\label{su2flavor}
\Wtree = W_\MM(\MM) + \sum_{i=1}^{M_{f}} \mu_{i} \tilde {\bf Q}_{i} {\bf Q}^{i} + g_{i} \tilde {\bf Q}_{i} \MM {\bf Q}^{i}
\end{equation}
where $\MM$ is an $M \times M$ Hermitian matrix, ${\bf Q}_{i}$ are $1 \times M$ vectors and $\tilde {\bf Q}_{i}$ are transpose $M \times 1$ vectors.  The ``Yukawa couplings'' can again be set to $g=1$ by a rescaling of $\bf Q$ and $\tilde{\bf Q}$.  This theory has been studied for $W_\MM(\MM) = \frac{1}{2} m \tr \MM^2$ in \cite{Argurio:2002xv, Brandhuber:2003va}.  We will derive the solution to the matrix model for a general $W_\MM$ using the combinatorics of planar diagrams, focusing on the contributions of the vectors ${\bf Q}$ to the free energy.\looseness=-1

As before, the matrix integral has a topological expansion, and the contributions at large-$M$ now come from planar diagrams with 0 and 1 quark boundary:

\begin{equation}
Z = \sum_{g,h}g_{s}^{2g+b-2} Z_{g,b}
\end{equation}
where $g$ is the genus and $b$ the number of boundaries, and we again recognise the Euler characteristic $\chi =  2 - 2g-b$.  Extending the result from the previous section, the superpotential is given by \cite{Dijkgraaf:2002dh,Argurio:2002xv}

\begin{equation}
\label{eq:weff.quarks}
W(S) = N_c \frac{\partial {\cal F}_{0,0}}{\partial S} + N_f {\cal F}_{0,1}
\end{equation}
Contributions to the first term come only from $\Phi$ self-interactions, so their combinatorics are the same as for the theory without quarks.  Diagrams with one external boundary can be counted by decomposing the counting problem into two parts: the combinatorics of the $\Phi$ diagrams on the interior of the disc, and the combinatorics of the boundary of the disc.

The first problem is equivalent to counting the planar $n$-point Green's functions $G_n(g_i)$ of the theory without quarks (i.e.~planar $\Phi$ diagrams -- possibly disconnected -- with $n$ external $\Phi$ legs).  This problem was solved in \cite{Brezin:1978sv}, as follows:

By definition,
\begin{equation}
\label{eq:greenfn}
G_n(g_i)=\langle \tr \Phi^n \rangle = \int_{a}^{b} d\lambda\,  y(\lambda)\,  \lambda^n
\end{equation}
where the second equality follows from the change of variables from the matrix integral to the eigenvalue basis and $a, b$ are the endpoints of the eigenvalue branch cut.  In other words, the sum of the planar Greens functions at each order are given by the corresponding moment of $y(\lambda)$.

The generating function for the Greens functions is
\begin{equation}
\label{genfn}
\phi(j) = \sum_{k=0}^{\infty} j^k G_k = \frac{1}{j} \omega(\frac{1}{j})
\end{equation}
where the second equality is given in terms of the resolvent
\begin{equation}
\omega(\lambda) = \frac{1}{2}(W'(\lambda) - \sqrt{W'(\lambda)^2 +
  f_{n-1}(\lambda)})
\end{equation}
by summing the geometric series in $\lambda$ coming from (\ref{eq:greenfn}), and converting the integral to a contour integral.  We also use the previous results that the eigenvalue density $\rho(\lambda)$ is equal to the discontinuity in $\frac{1}{2 \pi \imath} \omega(\lambda)$ across the branch cut (see (\ref{eq:discont})), and has asymptotic behavior $\omega(x) \sim S/x$ as $x \rightarrow \infty$.

To include the combinatorics of the boundary requires multiplying by $\frac{(k-1)!}{k!} = \frac{1}{k}$ at order $k$ in the expansion of $G$, to take into account the $(k-1)!$ distinct ways to connect a boundary quark with a leg of the internal Greens function\footnote{At first sight, it looks like an arbitrary connection of a boundary leg to an internal leg can make the overall graph non-planar, however we can always perform a corresponding crossing operation on the internal part of the diagram to undo this non-planarity.}, and the $\frac{1}{k!}$ coming from the expansion of $e^S$ to order $k$.  The factor $\frac{1}{k}$  can be incorporated into (\ref{genfn}) simply by integrating it:

\begin{eqnarray}
\Pi(j) &=& \int{\frac{1}{j^2} \omega(\frac{1}{j})} dj \nonumber \\
&=& - \int{\omega(x) dx} \nonumber \\
&=& -\frac{1}{2} \int(W'(x) - \sqrt{W'(x)^2 + f_{n-1}(x)} ) dx
\end{eqnarray}
where we have changed variables $x = \frac{1}{j}$ and used the definition of $\omega(x)$.  

The factors of $j$ count the number of external legs of the Greens function; therefore terms of order $j^k$ are associated to $k$ powers of the Yukawa coupling $g$, and $k$ quark propagators $\frac{1}{M}$ to connect up the $k$ external quarks on the boundary.  Therefore the one-boundary contribution to the matrix integral free energy is given by

\begin{eqnarray}
\label{eq:quarkgenfn}
\F_{0,1} &=& -\frac{1}{2} \int_M^{\infty}(W'(x) - \sqrt{W'(x)^2 + f_{n-1}(x)}) dx \nonumber \\
&=& - \int_M^{\infty}\omega(x) dx
\end{eqnarray}
and the contribution to the effective superpotential (\ref{eq:weff.quarks}) of the planar diagrams with 1 boundary precisely recovers the previously claimed result (\ref{eq:wnf}).  In the same way as in section \ref{sec:uv}, when there are $2M$ vectors the integral (\ref{eq:quarkgenfn}) combines with (\ref{eq:mm.bper}) to cancel the divergence at infinity, and cut off the integral at the value of the vector masses.\looseness=-1

\section{Matrix Models for $SO/Sp$ Gauge Theories}
\label{sec:mm.sosp}

In this section we extend the matrix integral techniques to analyze \nOne\ gauge theory with $\SO(N)$ and $\Sp(N)$ gauge groups and adjoint matter; which we first published in \cite{Ashok:2002bi}.  By a careful consideration of the planar and leading non-planar corrections to the large $M$ $SO(M)$ and $Sp(M)$ matrix models, we derive the matrix model free energy.  We do this both by applying the technology of higher-genus loop equations of~\cite{Ambjorn:1993gw,Akemann:1996zr} and by straightforward diagrammatics~(see {\it e.g.} \cite{Brezin:1978sv,Cicuta:1982fu}).  
%

 As for $\SU(M)$, we find that the loop equation for the resolvent of the matrix model (which is a Dyson-Schwinger equation for the matrix model correlation functions) describes a Riemann surface which is identified with a factorization of the spectral curve of the \ntwo\ gauge theory.  

In section~\ref{sec:corrections}, we discuss the application of the higher-genus loop equations to the computation of the $\RP^2$ contribution to the free energy.  The loop equations take the form of integral equations which give recursion relations between the contributions to the resolvent at each genus.  They suggest a very simple result for the $\RP^2$ contribution in terms of the sphere contribution.
%
%
We verify this relationship by explicitly enumerating ribbon diagrams with several types of vertex; as in the previous section, the combinatorics of these additional diagrams combine to reproduce the expected physical result. we find that the contribution to the free energy $\F_1$ from $\RP^2$ and $\F_0$ from $S^2$ are related by

\begin{equation}
\F_1 =\pm q \frac{\partial \F_0}{\partial S_0},
\end{equation}
where $S_0$ is half of the 't Hooft coupling for the $SO/Sp$ component of the matrix group. We determine the proportionality constant $q$ from the diagrammatics to be $q=\frac{g_s}{4}$.

Our results suggest a refinement of the proposal of Dijkgraaf and Vafa \cite{Dijkgraaf:2002dh} for the effective superpotential in the case of $\SO$ and $\Sp$ gauge groups. We find that
\begin{equation}
\label{eq:DVrefine}
\Weff = Q_{D5}\, \frac{\partial{\cal F}_0}{\partial S} + Q_{O5}\,\G_0+ \tau \, S,
\end{equation}
where $Q_{D5}$ is the total charge of D5-branes, $Q_{O5}$ is the total charge of O5-planes, $\F_0$ is the contribution to the matrix model free energy from diagrams with the topology of a sphere and $\G_0$ is proportional to $\F_1$, the contribution to the free energy from $\RP^2$ diagrams.  We use~\eqref{eq:DVrefine} to obtain results consistent with gauge theory expectations.  In particular, the subleading correction to the matrix model restores consistency with the requirement that there is a degeneracy of the massive vacua of the gauge theory given by $h$, the dual Coxeter number of the gauge group.

\subsection{The classical loop equation}
\label{sec:saddle}

We first consider the saddle point evaluation of the one matrix integral for $SO(M)$ or $Sp(M)$ matrices. Our discussion is analogous to that of section \ref{sec:mm} and consists of obtaining a loop equation for the resolvent. In the next section, we will formulate a systematic method for obtaining the $g_s$ corrections to the classical solution.
%

The partition function for the model with one matrix $\Phi$ in the adjoint representation of the Lie algebra of $G=SO(M)$ or $Sp(M)$ is
\begin{equation}
Z = Z_{0} \int d\Phi \, 
\exp \left( - \frac{1}{g_s} {\rm Tr}\, W(\Phi)\right).
\label{eq:onematrix}
\end{equation}
In Appendix~\ref{appvand}, we collect results that are useful for $SO/Sp$ groups, but here we shall discuss only the $SO(2M)$ group in detail.  

In the eigenvalue basis, the integral over an $SO(2M)$ matrix is given by  
\begin{equation}
Z = \int\prod_{i=1}^M \, d\lambda_{i}\,
\prod_{i<j} (\lambda_i^2-\lambda_j^2)^2 \,
e^{-\frac{2}{g_s}\sum_i W(\lambda_i)}.
\end{equation}
In terms of the number of eigenvalues $M_{i}$ in the neighbourhood of the critical point $x_i$, define the 't Hooft couplings

\begin{equation}
S_0 = g_s \frac{M_0}{2},~~S_i=g_s M_i.
\end{equation}
The effective action for the gas of eigenvalues is given by
\begin{equation}
S(\lambda)=-\sum_{i<j}\ln(\lambda_i^2-\lambda_j^2)^2+\frac{2}{g_s}
\sum_i W(\lambda_i).
\end{equation}
Note that $W$ is now a polynomial of order $2n$ with only even powers; this is because the trace of an antisymmetric matrix vanishes\footnote{In principle $W(\Phi)$ could also contain the Pfaffian, but we will omit this case.}.

This action gives rise to the classical equations of motion
\begin{equation}
\sum_{j\ne i}\frac{2\lambda_i}{\lambda_i^2-\lambda_j^2}
-\frac{1}{g_s}W'(\lambda_i) =0.
\end{equation}
Defining the resolvent
\begin{equation}
\omega_{0}(x)=g_s\tr\frac{1}{x-\Phi}=
g_s\sum_i\frac{2x}{x^2-\lambda_i^2},
\end{equation}
allows us to rewrite the equations of motion as: 
\begin{equation}
\label{eq:eomresolvent}
\omega_0(x)^2-g_s\left(\frac{\omega_0(x)}{x}-\omega_0'(x)\right)
+f(x)-2\omega_0(x)W'(x) =0,
\end{equation}
where
\begin{equation}\label{fofx}
f(x)=g_s\sum_i\frac{2\lambda_iW'(\lambda_i)-2xW'(x)}{\lambda_i^2-x^2}
\end{equation}
is a polynomial of order $2n-2$ with only even powers, {\it i.e.}, it
has $n$ coefficients.

In the small $g_s$ limit, \eqref{eq:eomresolvent} reduces to
\begin{equation}
\omega_0(x)^2+f(x)-2\omega_0(x)W'(x)=0,
\end{equation}
which may again be written in the form
\begin{equation}\label{eq:spectralagain}
y^2-W'(x)^2+f(x)=0,
\end{equation}
via the change of variables
\begin{equation}
y(x)=\omega_0(x)-W'(x).
\end{equation}
The force equation is now
\begin{equation}\label{forceeqn}
2y(\lambda)=-g_s\frac{\partial S}{\partial\lambda},
\end{equation}
where the factor of 2 comes from the fact that the force is acting on an eigenvalue and its image.



In terms of the eigenvalue density $\rho(\lambda)$,
\begin{equation}
\omega_0(x)=2\int_0^\infty\frac{x\rho(\lambda)d\lambda}{x^2-\lambda^2}=
\int_0^\infty\rho(\lambda)d\lambda\left(\frac{1}{x-\lambda}+
\frac{1}{x+\lambda}\right)=
\int_{-\infty}^\infty\frac{\rho(\lambda)d\lambda}{x-\lambda},
\end{equation}
the filling fractions are given by
\begin{equation}
\begin{split}
S_0&=\frac{1}{4\pi i}\int_{A_0}y(x)dx,\\
S_i&=\frac{1}{2\pi i}\int_{A_i}y(x)dx\quad,~i>0
\end{split}
\end{equation}
Note that we only integrate around half of the cycle $A_0$ because of the orientifold projection.


For $SO(2M+1)$ and $Sp(M)$, one can easily see that $\F_0$ and the Riemann surface are the same as in the case of $SO(2M)$.   These gauge groups are distinguished at subleading order in the $g_{s}$ expansion of the free energy; in the next section we will determine the leading contribution to the free energy from unoriented diagrams.

\subsection{$g_s$ corrections and loop equations}
\label{sec:corrections}



\begin{figure}[t]
\begin{center}
\epsfig{file=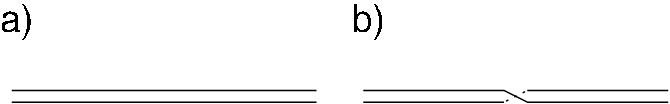}
\parbox{5.5in}{
\caption{Feynman rules for the $\SO$ and $\Sp$ matrix models: a) untwisted and b) twisted propagators\label{fig:twisted}}}
\end{center}
\end{figure}
We can now expand the free energy in terms of ribbon graphs as before.  The propagator of the $SO(M)$ matrix model is
\begin{equation}
\langle 
\Phi_{ij}\Phi_{kl}\rangle\sim\half(\delta_{ik}\delta_{jl}-\delta_{il}\delta_{jk}).
\end{equation}
Thus, the ribbon graphs now have the possibility of ``twisted propagators'' as well as the previous untwisted propagators (see figure \ref{fig:twisted}); an important point is that the twisted propagators comes with a relative minus sign.  The twisted propagators can give rise to non-orientable ribbon graphs, so the topological expansion includes a sum over diagrams that may be embedded in non-orientable Riemann surfaces.  As before, the overall power of $g_s$ associated to a ribbon diagram is

\begin{equation}
g_{s}^{-\chi} = g_{s}^{2g+c-2}
\end{equation}
where $g$ denotes the genus, and $c$ denotes the number of cross-caps of the Riemann surface on which the diagram is enscribed.



\subsection{The resolvent}

We shall now review the general technique of loop equations \cite{Ambjorn:1993gw,Akemann:1996zr}, which is an iterative procedure to calculate corrections to the partition function of the higher order in $g_s$. Central to this procedure is the loop operator, defined as
\begin{equation}
\frac{d}{dV}(x) = -\sum_{j=1}^{\infty}\frac{2j}{x^{2j+1}}\pder{}{g_{j}}.
\end{equation}
The resolvent, which is the generating functional for the single trace correlation functions of the matrix model, is defined as
\begin{equation}
\omega(x)=g_s\left\langle\tr\frac{1}{x-\Phi}\right\rangle = g_s\sum_{k=0}^{\infty}\frac{\langle\tr\Phi^{2k}\rangle}{x^{2k+1}}
\end{equation} 
Using the identity 
\begin{equation}
-(2k)\frac{d}{dg_{k}}\F=g_s\langle\tr\Phi^{2k}\rangle,
\end{equation}
the resolvent can expressed as
\begin{equation}\label{dFbydV}
\omega(x) = \frac{d}{dV}(x)\F + \frac{S}{x},
\end{equation}
where we used $S=\sum S_i=g_s M$.  We are using the variables $g_s$ and $S$ since we are working in the small $g_s$ limit with $S$ fixed. As mentioned before, the perturbative part of the free energy has an expansion in $g_s$ of the form
\begin{equation}
\F^{\hbox{\it pert.}} = \sum_{g,c}g_s^{2g+c-2}\F_{g,c}
\end{equation}
We will be interested in calculating the first two terms in this expansion, which are the contributions from diagrams with the topology of $S^2$ and $\RP^2$, although the analysis can in principle be extended to all orders to study gravitational corrections to the gauge theory superpotential. The resolvent has a similar expansion
\begin{equation}
\label{eq:resolventgenus}
\omega(x) = \sum_{g,c}g_s^{2g+c}\omega_{g,c}(x).
\end{equation}

The asymptotic behavior at infinity of the $\omega_{g,c}$ is clear from
the definition of $\omega(x)$
\begin{equation}\label{Wasympt}
\begin{split}
\omega_{0,0}(x)=&\frac{S}{x}+\CO(x^{-2}),\\
\omega_{g,c}(x)=& \CO(x^{-2}),\;\;\;\;2g+c>0.
\end{split}
\end{equation}
Using this fact and the existence of the genus expansion, we can write 
\begin{equation}\label{wfromf}
\begin{split}
\omega_{0,0}(x) =& \frac{d}{dV}(x)\F_{0,0}+ \frac{S}{x},\\
\omega_{g,c}(x) =& \frac{d}{dV}(x)\F_{g,c},\;\;\;\;2g+c>0.
\end{split}
\end{equation}
These equations determine the dependence of $\F_{g,c}$ on the coupling constants. There is still an additive constant that is undetermined, but this is physically meaningless. In the next section we will derive the loop equation, which will provide us with recursion relations to calculate $\omega_{g,c}$ as functions of the coupling constants $g_j$ appearing in the matrix potential $W(\MM) = \sum_j \frac{g_j}{j} \tr \MM^j$.  For the rest of the discussion, we denote $\omega_{0,0}$ by $\omega_0$ and $\omega_{0,1}$ by $\omega_1$.

\subsection{The loop equation}

In this section we will derive an important recursion relation between the different perturbative contributions $\omega_{g,c}$ to the resolvent. The loop equation can be derived by performing an infinitesimal reparametrization of the matrices $\Phi$ in the matrix integral and using the fact that the integral is trivially invariant under  reparametrization of $\Phi$. Let us reparametrize $\Phi$ by 
\begin{eqnarray}\label{reparam}
\Phi&=&\Phi'-\left(\frac{\epsilon}{x-\Phi'}\right)_\odd=
\Phi'-\epsilon\sum_{k=0}^\infty\frac{\Phi'^{2k+1}}{x^{2k+2}}\\
d\Phi&=&d\Phi'-\epsilon\sum_{k=0}^\infty
\sum_{l=0}^{2k}\frac{\Phi'^ld\Phi'\Phi'^{2k-l}}{x^{2k+2}}
\end{eqnarray}
where we only take the odd/even powers of $\Phi'$ in order to preserve the $SO/Sp$ Lie algebra. The Jacobian for this reparametrization, keeping only lowest powers of $\epsilon$, is then 
\begin{equation}
J(\Phi')=1-\frac{\epsilon}{2}\left(\tr\frac{1}{x-\Phi'}\right)^2
+\frac{\epsilon}{2x} \tr\frac{1}{x-\Phi'}.  
\label{eq:phijacobian}
\end{equation}
The action transforms as
\begin{equation}
\tr W(\Phi)=
\tr W\left(\Phi'-\left(\frac{\epsilon}{x-\Phi'}\right)_\odd\right)
= \tr W(\Phi')-\epsilon\, \tr\frac{W'(\Phi')}{x-\Phi'}.
\end{equation}
Inserting this into the matrix integral, the invariance under the small variation of $\Phi$ yields the identity
\begin{equation}
\begin{split}
& \half\int
d\Phi'\left[\left(\tr\frac{1}{x-\Phi'}\right)^2
-\frac{1}{x}\tr\frac{1}{x-\Phi'}\right]e^{-\frac{1}{g_s}\str
W(\Phi')} \\
& \hskip2cm =
\frac{1}{g_s}\int d\Phi'\tr\frac{W'(\Phi')}{x-\Phi'}
e^{-\frac{1}{g_s}\str W(\Phi')}.
\end{split}
\end{equation}
We can now make use of the identity
\begin{equation}
\frac{d}{dV}(x)\omega(x)=
\left\langle\left(\tr\frac{1}{x-\Phi}\right)^2\right\rangle-
\left\langle\tr\frac{1}{x-\Phi}\right\rangle^2
\end{equation}
(which is a rewriting of the steps leading to \ref{eq:loopeq1}) to get the loop equation
\begin{equation}
g_s\left\langle\tr\frac{W'(\Phi)}{x-\Phi}\right\rangle=
\half\omega(x)^2-\frac{g_s}{2x}\omega(x)
+\frac{g_s^2}{2}\frac{d}{dV}(x)\omega(x).
\end{equation}

We can rewrite the loop equation using
\begin{equation}
g_s\left\langle\tr\frac{W'(\Phi)}{x-\Phi}\right\rangle=
g_s\left\langle\sum_{i}\frac{W'(\lambda_{i})}{x-\lambda_{i}}\right\rangle
=\oint_{C}\frac{dx'}{2\pi i}\frac{W'(x')}{x-x'}\omega(x'),
\end{equation}
where $C$ is a contour that encloses all the eigenvalues of $\Phi$ but not $x$. In the large $M$ limit of the matrix model, we get a continuous eigenvalue distribution for $\Phi$ and all the eigenvalues are distributed over cuts on the real axis of the $x$-plane. The loop equation now reads
\begin{equation}\label{loopeqn}
\oint_{C}\frac{dx'}{2\pi i}\frac{W'(x')}{x-x'}\omega(x') = 
\half\omega(x)^2-\frac{g_s}{2x}\omega(x)
+\frac{g_s^2}{2}\frac{d}{dV}(x)\omega(x).
\end{equation}
We can now insert the $g_s$ expansions (\ref{eq:resolventgenus}) for the resolvent and iteratively solve for the $\omega_{g,c}$. The zeroth and first order equations are 
\begin{eqnarray}\label{loworderloop}
\oint_{C}\frac{dx'}{2\pi i}\frac{W'(x')}{x-x'}\omega_{0}(x') &=& 
\half\omega_{0}(x)^2, \\\label{loworderloop2}
\oint_{C}\frac{dx'}{2\pi i}\frac{W'(x')}{x-x'}\omega_{1}(x') &=& 
\omega_{0}(x)\omega_{1}(x)-\frac{1}{2x}\omega_0(x).
\end{eqnarray}
The resolvent that solves the loop equations must satisfy (\ref{Wasympt}), which imposes constraints on the end-points of the cuts in the $x$-plane. 

Equation \eqref{loworderloop2} is a linear inhomogenous integral equation for $\omega_1$. The homogeneous equation is solved by a derivative of $\omega_0$ with respect to any parameter which specifies the vacuum, {\it i.e.}, is independent of the coupling constants $g_j$. In our case there are only the parameters $S_i$, which specify the classical vacuum around which the matrix integral is expanded. 

\subsection{Solution to the loop equations}

We now solve the loop equations (\ref{loworderloop}) first for $\omega_0$ and then for $\omega_1$ in the case of a polynomial potential
\begin{equation}
W(\Phi)=\sum_{j=1}^{n}\frac{g_j}{2j}\Phi^{2j}.
\end{equation}

\subsubsection{Planar contributions}

In equation (\ref{loworderloop}), we deform the integration contour $C$ to encircle infinity, and rewrite it as 
\begin{equation}
\half\omega_0(x)^2 
=W'(x)\omega_0(x)
+\oint_{C_{\infty}}{x'}\frac{W'(x')\omega_{0}(x')}{x-x'}.
\end{equation}
Assuming that $\omega_0(x)$ has $k$ cuts in the complex $x$-plane, we make the ansatz  
\begin{equation}
\omega_0(x)=W'(x)-M(x)\sqrt{\prod_{i=1}^{2k}(x-x_i)},
\end{equation}
where $M(x)$ is an undetermined analytic function at the moment. Here the end points of the cuts, denoted by the $x_i$, are unknown and have to be determined. It is clear that if we have the maximum allowed number of cuts, $k=2n-1$, the function $M(x)$ is a constant. The loop equation determines $M$ in this case to be the coupling constant $g_n$.  For the $SO/Sp$ models the eigenvalues come in pairs, and the total number of ``independent'' cuts is $n$. There is one cut $[-x_0,x_0]$ centered around zero, and the other cuts come in pairs $[x_{2i-1},x_{2i}]$ and $[-x_{2i},-x_{2i-1}]$. Note that the cuts are simply the projections of the $S^{3}$ cycles of the \CY\ geometry that engineers this gauge theory, which we discussed in section \ref{sec:cy}.  

We now demand that the resolvent $\omega_0(x)$ falls off at infinity as $S/x$ (and hence vanishes in the classical limit $S\rightarrow 0$), and thus obtain $n$ constraints 
\begin{equation}\label{falloffconstr}
\delta_{k,n}=
\half\oint_{C}{x'}\frac{x'^{2k-1}W'(x')}{
\sqrt{\prod_{i=0}^{2(n-1)}(x'^2-x^2_i)}}
,\;k=1,2,\cdots,n.
\end{equation}
The most general solution to these $n$ constraints (\ref{falloffconstr}) is given by  
\begin{equation}
g_n^2\prod_{i=0}^{2(n-1)}(x^2-x^2_i)=W'(x)^2-f(x),
\end{equation}
where $f(x)$ is the most general even polynomial of order $2n-2$, 
\begin{equation}
f(x)=\sum_{l=0}^{n-1}b_l x^{2l}.
\end{equation}

Note that we have now recovered the solution to the classical loop equation that we obtained in section~\ref{sec:saddle}. We now repeat the procedure outlined there and define the Riemann surface $\Sigma$ given by
\begin{equation}
y^2=W'(x)^2-f(x).
\end{equation}
The filling fractions $S_i$ then become period integrals of the meromorphic 1-form $y\,dx$ over the 1-cycle $A_i$ of $\Sigma$ that encircles the $i^\text{th}$ branch cut
\begin{equation}
S_i=\oint_{A_i}\frac{y\,dx}{2\pi i}.
\end{equation}
We can then argue that the change in the free energy due to an eigenvalue tunneling to infinity from the  $i^\text{th}$ cut is
\begin{equation}\label{bcycleint}
\pder{\F_0}{S_i}=\int_{B_i}y\,dx.
\end{equation}
This again requires the introduction of a cutoff, which can be understood in terms of the combinatorics of diagrams with 1 boundary and the topology of a disc or Moebius strip.

\subsubsection{$\RP^2$ contributions}
\label{sec:rptwocontr}

Once we have the form of the solution for $\omega_0(x)$, we can substitute it in the loop equation, which is now a linear inhomogenous integral equation for $\omega_1(x)$, 
\begin{equation}
\oint_{C}{x'}\frac{W'(x')\omega_{1}(x')}{x-x'}
=\omega_{0}(x)\omega_{1}(x)-\frac{1}{2x}\omega_0(x).
\end{equation}

We can get a natural ansatz for $\omega_1$ from the string theory expectation that $\F_1$ should be a derivative with respect to $S_0$ of $\F_0$, 
\begin{equation}\label{foneansatz}
\F_1=q\pder{\F_0}{S_0},
\end{equation}
where $q$ is some constant which has to be determined. Inserting this into \eqref{wfromf}, we get 
\begin{equation}
\begin{split}
\omega_1(x)=&\oder{}{V}(x)\F_1= -q\sum_j\frac{2j}{x^{2j+1}}\pder{}{g_j}\pder{\F_0}{S_0} \\
=&q\pder{}{S_0}\left(\omega_0(x)-\frac{S}{x}\right)\\
=&q\pder{\omega_0}{S_0}-\frac{2q}{x}.
\end{split}
\end{equation}
It is easy to see that $q\pder{\omega_0}{S_0}$ solves the homogeneous part of the loop equation. The inhomogenous part of the loop equation is solved by $-\frac{2q}{x}$ if $q=-\frac{1}{4}$.  

More generally, in the case of multi-cut solutions, we could have added any solution to the homogeneous loop equations. This amounts to taking 
\begin{equation}
\F_1=\sum_iq_i\pder{\F_0}{S_i},
\end{equation}
such that $\sum q_i=-\frac{1}{4}$. However, corrections of the form $\pder{\F_0}{S_i}$ for $i>0$ should not be generated since these cuts represent $U(N_i)$ gauge physics for which there should be no $\RP^{2}$ contribution. We will give a short perturbative discussion of this in the next section.  


\subsection{Counting Feynman diagrams with $S^2$ and $\RP^2$ topology}

For a perturbative check of the relation
\begin{equation}
\F_1 = \pm q \frac{\partial {\cal F}_0}{\partial S_0}
\label{g0}
\end{equation}
we can enumerate ``ribbon'' graphs in the genus expansion of the matrix model.  Recall that the genus expansion is ordered by diagram topology, with diagrams of genus $g$ and $c$ cross-caps contributing at order $g_s^{-\chi} = g_s^{-2+2g+c}$.  The coefficient $q$ is related to the relative contribution of the planar (genus 0) diagrams which dominate at large $M$ and the leading $\frac{1}{M}$ correction coming from diagrams with topology $\RP^2$.

It is known that $\SO(2M)$ and $\Sp(M)$ matrix models are related by analytic continuation $M \mapsto -M$ (for the analogous gauge theory results see \cite{Mkrtchian:1981bb,Cvitanovic:1982bq, Cicuta:1982fu}). Therefore, at even orders in the  genus expansion, the contribution to the matrix model free energy is the same for both theories, while at odd orders the $\Sp(M)$ diagrams contribute to the free energy with an additional minus sign relative to $\SO(2M)$.  This fact determines the sign in \eqref{g0}.
Recall that
\begin{equation}
\chi = v - p + l
\label{eq:chi}
\end{equation}
where $v$ is the number of vertices in the ribbon graph, $p$ is the number of propagators and $l$ the number of boundary loops.   The Feynman rules are summarized in appendix~\ref{app:feynman}.
\begin{figure}[tp]
\begin{center}
\epsfig{file=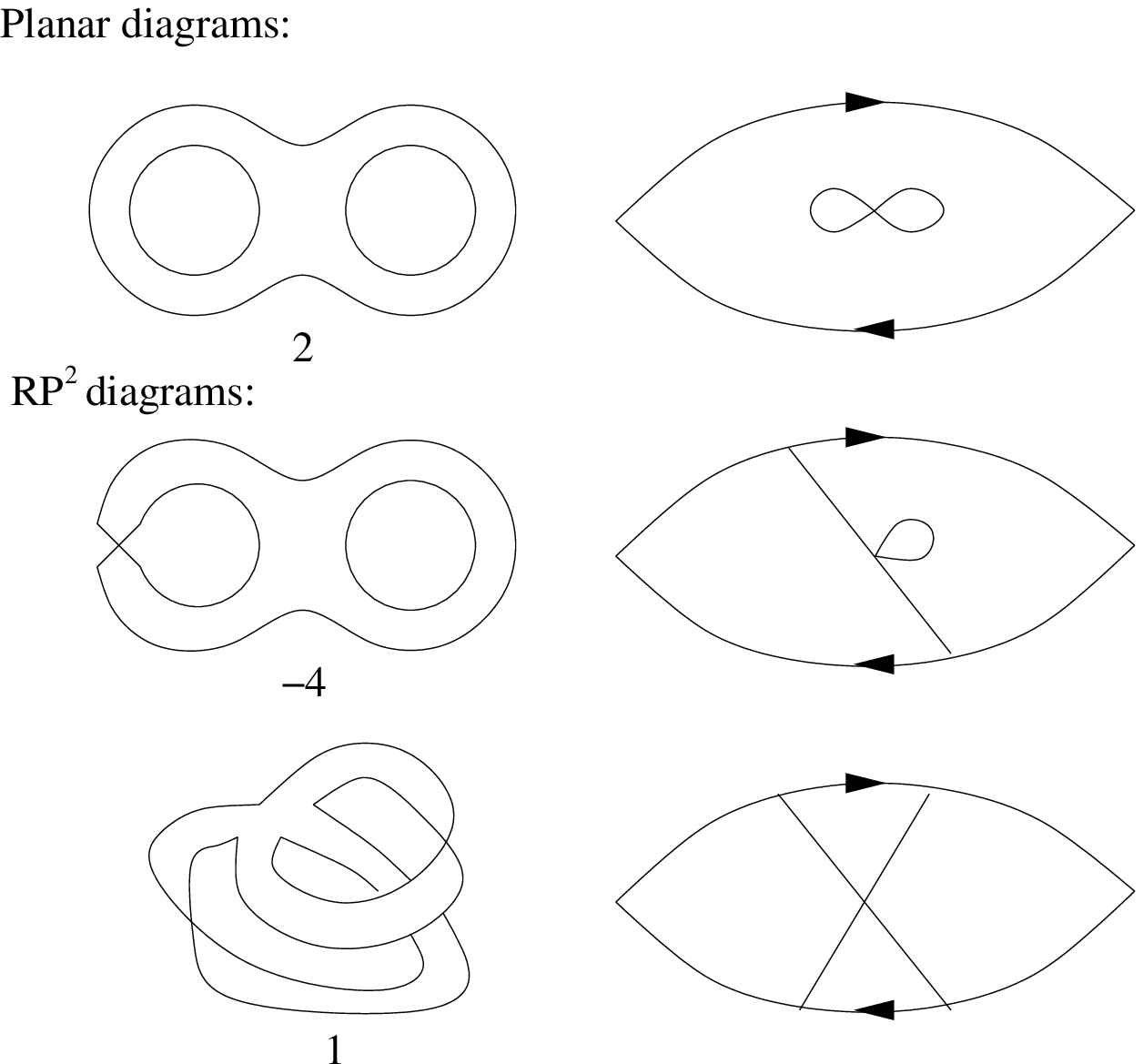}
\parbox{5.5in}{
\caption{$S^2$ and $\RP^2$ diagrams with one quartic vertex, written in terms of twisted and untwisted propagators and as diagrams on $\RP^2$ to show their   planarity.  Propagators that pass through the cross-cap become  twisted.\label{fig:diags}}}
\end{center}
\end{figure}
Let us evaluate the first-order quartic diagrams in fig. \ref{fig:diags}. The planar diagram has the value 
\begin{equation}
2\times\frac{1}{1!}\frac{g_2}{4g_s}\left(\frac{g_s}{2m}\right)^2M^3
\end{equation}
whereas the $\RP^2$ diagram with one twisted propagator contributes
\begin{equation}
-4\times\frac{1}{1!}\frac{g_2}{4g_s}\left(\frac{g_s}{2m}\right)^2M^2
\end{equation}
and the $\RP^2$ diagram with both propagators twisted contributes
\begin{equation}
1\times\frac{1}{1!}\frac{g_2}{4g_s}\left(\frac{g_s}{2m}\right)^2M^2.
\end{equation}
Using the fact, that $S=\frac{g_s}{2}M$, this shows that
\begin{equation}
\F_1=-\frac{1}{4}\pder{\F_0}{S_0}
\label{eq:derrel}  
\end{equation}
at the first order.   We have enumerated the Feynman diagrams to several higher orders and higher vertices and confirmed this relationship in those cases\footnote{This combinatorial result was previously unknown to mathematicians.} (see Appendix \ref{app:mm} for some examples).

In order to describe a multi-cut matrix model (corresponding to vacua with classically broken gauge group), we would need to use ghosts \cite{Dijkgraaf:2002pp} to expand around the classical vacuum.  In this prescription, one can think of the matrix model as several matrix models, which are coupled by bifundamental ghosts. Only one of those matrix models is actually an $SO(M_0)/Sp(M_0/2)$ matrix model, the other matrix models are $U(M_i)$ matrix models. The ghosts do not have twisted propagators, so the leading contribution from the $SO(M_0)/Sp(M_0/2)$ matrix model is again the same as for a single cut model. The loop equations still hold for the multi-cut model and the calculation can be extended to all orders.

\subsection{Computation of effective superpotentials}
\label{sec:superpot}

In this section we combine the results of the previous sections to compute the effective superpotential of the dual gauge theories. We will find that it is necessary to refine the formula originally conjectured by \cite{Dijkgraaf:2002dh} for the unoriented string contribution to the effective superpotential.
%
Recall that in a vacuum with coincident eigenvalues, there is a non-perturbative contribution to the matrix model free energy coming from the logarithm of the volume of the residual gauge transformations that preserve this vacuum.  In appendix B, following \cite{Ooguri:2002gx}, we have included the large $M$ expansion of the logarithm of the volume of the $SO/Sp$ groups. We find that, for $SO(M)$ when $M$ is even, the non-perturbative contribution to the free energy is
\begin{equation}
\begin{split}
{\cal F}^{np}=& \frac{1}{g_s^2} \F^{np}_0 + \frac{1}{g_s} \F^{np}_1
+ \cdots\\
= &  
\frac{1}{g_s^2} \left[ S^2 \log\frac{2\pi S}{m} 
- S^2 \left(\frac{3}{2} + \log \pi\right) \right] \\
& + \frac{1}{g_s} \left[ -\frac{S}{2} \log \frac{2\pi S}{m} 
+ \frac{S}{2} (1+\log\pi - \log 4 ) \right] + \cdots,
\end{split}
\end{equation}
with a similar expression for $M$ odd or $G=Sp(M)$. We see that 
\begin{equation}
{\cal F}_{1}^{np}= 
\mp \frac{1}{4}\frac{\del {\cal F}_{0}^{np}}{\del S} 
\pm \frac{1}{2} \log 2,
\end{equation} 
where the first $-/+$ sign is for $SO/Sp$ respectively. This is almost the same relationship as we found for the perturbative contributions~\eqref{eq:derrel}, but it is spoiled by the $\log 2$ term.  This amounts to a factor of 2 discrepancy in the volume of the gauge group\footnote{This mismatch may be related to the choice of whether or not to work in the covering group.}.

It is the non-perturbative sector, specifically the coefficient of the $S^{2} \log S$ term, that determines the number of gauge theory vacua, which is a main consistency test of the translation between matrix model quantities and the effective superpotential of the gauge theory. The number of vacua of a supersymmetric gauge theory is equal to the dual Coxeter number $h$ of the gauge group~\cite{Witten:1982df,Witten:1998bs}.   Therefore the total superpotential should lead to the conclusion that $S^h$ is single-valued.  

Open string physics tells us that the sphere contribution to the effective superpotential should be proportional to $Q_{D5}$, the total charge of D5-branes, while the $\RP^2$ contribution should be proportional to $Q_{O5}$, the total charge of O5-planes. We can express this by refining the suggestion of~\cite{Dijkgraaf:2002dh}:
\begin{equation}
\Weff = Q_{D5}\, \frac{\partial{\cal F}_0}{\partial S} + Q_{O5}\,\G_0
+\tau \, S,
\label{eq:WeffMM}
\end{equation}
We assume that  $\G_0$ is proportional to the total $\RP^2$ free energy,
\begin{equation}
\G_0 = a \left( \F^{np}_1 + \F^{p}_1 \right).
\end{equation}

Proceeding with this result, we find that 
\begin{equation}
\Weff = 
\left( \frac{N}{2} \pm \frac{a}{4} \right) S \log S + \frac{1}{2}
\tau\, S
+ \cdots,
\end{equation}
where the $+/-$ is for $SO/Sp$ respectively.  Consistency with both the closed string result~\eqref{eq:weffclosed} and the gauge theory\footnote{Note that, after including $a=\mp 4$, the effective superpotential naively suggests that for gauge group $Sp(N/2)$, $S^{N+2}$ is single-valued, whereas $h=N/2 +1$. The resolution to this puzzle was explained in~\cite{Gomis:2001xw}. Namely the D1-string wrapped on $\BP^1$ has instanton number {\it two} in $Sp(N/2)$. Properly accounting for this reproduces the $\BZ_{2h}$ chiral symmetry of the dual gauge theory.}  requires that we must have $a=\mp 4$.  This was confirmed by~\cite{Ita:2002kx} who gave a perturbative argument along the lines of~\cite{Dijkgraaf:2002xd}; it was found to be related to the measure on the moduli space of Schwinger parameters, a quantity that is intrinsic to the gauge theory.  

Note that the first subleading (non-planar) contributions combine with the leading (planar) contributions to give a shift in the overall coefficient.  This combination with  the leading-order  contributions is quite similar to the role of planar diagrams with one boundary discussed in the previous section, which conspire to soften the UV divergence of the planar free energy.

\chapter{Conclusions}

In this thesis we have studied effective superpotentials for confining gauge theories with $\N=1$ supersymmetry, focusing on theories where an underlying $\N=2$ supersymmetry is softly broken by a tree-level superpotential.  String theory provides insight into the structure of the vacua of these quantum theories, and a set of geometrical tools for computing the effective superpotential exactly, even in strongly-coupled regimes.  

The techniques we have discussed revolve around the computation of period integrals of a meromorphic 1-form on a particular hyperelliptic curve.  We studied this curve and found that the regularization of the divergence at infinity of the contour integrals requires the introduction of additional fundamental matter superfields into the gauge theory, which cut off the domain of integration and render the calculation finite.  This is physically pleasing, since the $\N=2$ gauge theory with $N_{f} = 2N_{c}$ massive fundamental hypermultiplets has vanishing $\beta$-function at high energies, indicating that the theory has a nontrivial conformal fixed point and is free from short-distance singularities.

We evaluated the period integrals explicitly for the maximally confining vacua (completely degenerate curve), and derived an explicit expression for the superpotential of an $U(N_c)$ gauge theory with $0\le N_f <2N_c$ fundamental superfields of arbitrary non-zero mass, and arbitrary tree-level interactions of the adjoint superfield $\Phi$.  Extremizing this superpotential gives the exact vacuum superpotential and the gaugino condensate, and agrees with previous special cases discussed in the literature. 

The $\N=1$ curve may be obtained by factorizing the Seiberg-Witten curve of the underlying $\N=2$ theory.  For the $\N=2$ theory with fundamental hypermultiplets, we solved the factorization problem for the case when the curve factorizes completely, and used this solution to verify the combinatorial form of the effective superpotential in $\N=1$ theories with fundamental matter.

The $\N=2$ gauge theories are known to have an underlying integrable structure, and this partly survives the soft supersymmetry breaking to $\N=1$.  The existence of this integrable system is equivalent to the statement that the vacuum of the gauge theory is completely characterized by the vev of the adjoint (matrix-valued) chiral superfield $\Phi$; this matrix is identified with the Lax matrix of the integrable system, which completely characterizes the integrable dynamics.  We considered the theory with $N_{f} = N_{c}$ fundamental hypermultiplets, and found the value of the Lax matrix of the corresponding integrable system in the maximally confining vacuum.  This form of the Lax matrix was not previously known.

The geometrical techniques involving the spectral curve, while computationally powerful in obtaining exact results about the confining phase of supersymmetric gauge theories, do not have a clear origin within the gauge theory.  The bridge between the geometrical techniques and the physics of the supersymmetric gauge theory is provided by the matrix models.  The reason for this correspondence is that after summing the gauge theory diagrams order by order, the 4-dimensional loop momenta cancel and only the planar combinatorics survive.  

We discussed how the $\N=1$ curve -- and therefore the gauge theory effective superpotential -- emerges from the study of a particular class of matrix integral, considered as a zero-dimensional path integral for the eigenvalues of the matrix.  Specifically, the free energy of the matrix model receives perturbative and non-perturbative contributions.  The former come from the planar (and for $\SO$ and $\Sp$ gauge theories, or theories with fundamental matter, the leading non-planar) diagrams of the matrix integral, which are in 1-1 correspondence with Feynman diagrams of the gauge theory.  

The non-perturbative contributions to the matrix integral come from the residual gauge transformations that exist when two or more eigenvalues populate the same critical point of the potential.  These correspond in the four-dimensional gauge theory to classical vacua where a subgroup of the gauge group remains unbroken in the classical theory.  Expanding the volume of the gauge groups reproduces the Veneziano-Yankielowicz superpotential.  In the four-dimensional gauge theory this superpotential is generated by the strong-coupling dynamics of the gauge field, and the only existing derivations come from anomalous symmetry constraints.  However, since the gauge field does not appear in the matrix integral (in a sense, it is integrated out), there is no complication from strong coupling and the contribution may be read off from the asymptotic expansion of the volume of the unbroken gauge group.

The various techniques we have used to study the effective superpotentials may be characterized as geometrical, algebraic and combinatorial in nature.  Each of them involves the spectral curve, but highlights a different aspect of its structure.  This structure is in turn reflected in the structure of the vacua of the $\N=1$ gauge theory.

These techniques teach us about confinement and other non-perturbative phenomena in the $\N=1$ gauge theories; for example in theories with an adjoint chiral superfield (which contains a scalar field), confinement of the low-energy gauge theory is associated to condensation of the magnetic monopoles of the gauge theory, and moreover the exact value of the monopole condensates can be calculated.   Extremizing the gauge theory effective superpotentials gives exact non-perturbative results about the vacua of the theory, such as the values of the gaugino condensates associated to chiral symmetry breaking.  We are therefore able to obtain exact results about confining theories that are believed to have many similar properties to non-supersymmetric Yang-Mills theory and QCD, for which analytical results are lacking.

Though string theory is fully a theory of gravity and other fundamental forces, it is commonly the case that the effects of gravity can be consistently decoupled, and string theoretical techniques can be used to study the remaining low-energy supersymmetric particle interactions in isolation.  Thus, if supersymmetry is realized in nature at an experimentally accessible energy scale, then -- {\it whether or not string theory is the correct unified theory of quantum gravity and fundamental forces} -- string theory has provided tools that will be useful for understanding aspects of physics in supersymmetric regimes.

\bibliographystyle{alpha}   
\bibliography{thesis}
\appendix    	
\chapter[Appendix: Matrix Integral Measures and Determinants]{Matrix Integral Measures and Determinants}
\label{app:mm}

In this appendix we collect some results on the group measure and adjoint action which are needed to do computations in the matrix models.

\section{The group measure for general matrices}
\label{appvand}

We wish to compute the Jacobian for the transformation from certain matrices $\Phi$ to their eigenvalues. This can be derived by a group-theoretic argument. In terms of the Cartan generators $H_i$ and ladder operators $E_\alpha$, for the algebra of the group $G$, satisfying
\begin{equation}
[H^{i},E^{\alpha}] = \alpha^{i}E^{\alpha},
\end{equation}
we can diagonalize a matrix $\Phi$
\begin{equation}
\begin{split}
& \Phi = U^{\dagger}\Lambda U, \\
& \Lambda = \sum_{i}\lambda_{i}H^{i}.
\end{split} \label{eq:diagonalize}
\end{equation}
We will define parameters $t_\alpha$ so that
\begin{equation}
dU = \left[ \sum_\alpha\, dt_\alpha E_\alpha\right] U, 
~~~t_\alpha^*=-t_{-\alpha} .
\end{equation}

The infinitesimal variation of $\Phi$ can then be written as
\begin{equation}
\begin{split}
d\Phi = & U^{\dagger}
\left[d\Lambda + \sum_{\alpha}dt_{\alpha}\, [\Lambda,E^{\alpha}]\right]U
\\
= & U^{\dagger}
\left[d\Lambda + \sum_{\alpha}dt_{\alpha}\, 
\left( \sum_i\lambda_i\alpha^i\right)E^{\alpha}\right]U.
\end{split}
\end{equation}
We now calculate the metric on the Lie algebra
\begin{equation}\label{liemetric}
{\rm Tr}\, d\Phi \, d\Phi^\dagger
=  \sum_i d\lambda_i^2
+  \sum_{\alpha,\beta} dt_\alpha \,dt_\beta \,
\left(\sum_i\lambda_i \alpha_i \right)
\left(\sum_j\lambda_j \beta_j \right) \,
{\rm Tr}\,E_\alpha E_\beta.
\end{equation}
Using the identity
\begin{equation}
\textrm{Tr}_{r} E^{\alpha}E^{\beta} = C(r)\delta_{\alpha+\beta,0}
\end{equation}
where $C(r)$ is a representation dependent constant, we can simplify
the second term in equation \eqref{liemetric} to 
\begin{equation}
C(r)\sum_{\alpha} \left| \sum_i \alpha^i\lambda_i \right|^{2}
|dt_{\alpha}|^{2}
\end{equation}
Up to numerical factors, the Jacobian is
\begin{equation}
\Delta (\Lambda) = \prod_{\alpha > 0}
\left| \sum_{i}\alpha^i \lambda_i \right|^{2}.
\end{equation}

We list the expressions for the roots and the corresponding
determinants for the different classical groups in Table~\ref{tab:dets}. 

\medskip
\begin{table}[ht]
\begin{center}
\begin{tabular}{|l|c|}
\hline
$G$   & $J(\Lambda)$ \\
Roots &\\
\hline
$A_{N-1}$ & $\prod\limits_{i<j}(\lambda_i-\lambda_j)^2$ \\
$e_i-e_j$ ($i\ne j$) &  \\
\hline
$B_N$     & 
$\prod\limits_{i<j}(\lambda_i^2-\lambda_j^2)^2\prod\limits_i\lambda_i^2$ \\
$\pm e_i\pm e_j$ ($i\ne j$), $\pm e_i$ &\\
\hline
$C_N$     & 
$\prod\limits_{i<j}(\lambda_i^2-\lambda_j^2)^2\prod\limits_i\lambda_i^2$ \\
$\frac{1}{\sqrt{2}}\left(\pm e_i\pm e_j\right)$ ($i\ne j$), $\pm \sqrt{2}e_i$ &\\
\hline
$D_N$     &
$\prod\limits_{i<j}(\lambda_i^2-\lambda_j^2)^2$ \\
$\pm e_i\pm e_j$ ($i\ne j$) &\\
\hline
\end{tabular}
\caption{The roots and the formul\ae\ for the Jacobians associated to the classical groups.}
\label{tab:dets}
\end{center}
\end{table}

\section{Asymptotic expansion of the gauge group volumes}

We now compute the asymptotic expansion of the volume of the gauge groups, which normalizes the partition function of the matrix model and provides the nonperturbative contribution to the free energy.  The volumes are given by 
\cite{Ooguri:2002gx}:
\begin{equation}
\begin{split}
& \hbox{vol}(\SU(N)) = \frac{\sqrt{N} (2 \pi)^{\half N^{2} + \half N - 1}}{(N-1)!(N-2)! \cdots 2! 1!}, \\
& \hbox{vol}(\SO(2N+1)) = \frac{{2^{N+1}(2\pi)^{N^2 + N - {\frac{1}{4}}}}}{(2N-1)!(2N-3)!\ldots 3! 1!}, \\
&\hbox{vol}(\SO(2N)) = \frac{{\sqrt{2}(2\pi)^{N^2}}}{(2N-3)!(2N-5)!\ldots 3! 1! (N-1)!}, \\
&\hbox{vol}(\Sp(2N)) = \frac{{2^{-N} (2\pi)^{N^2 + N}}}{(2N-1)!(2N-3)!\ldots 3! 1!}.
\end{split}
\end{equation}

We are interested in the large $N$ asymptotic expansion of the
logarithm of the volumes in order to compute the non-perturbative
contribution to the free energy.
Following \cite{Ooguri:2002gx}, we introduce the Barnes function
\begin{equation}
G_2(z+1) = \Gamma(z) G_2(z),\ G_2(1) = 1.
\label{barnes}
\end{equation}
Using the doubling formula for $\Gamma(z)$,
\begin{equation}
\Gamma(2z) = 2^{2z-1} \pi^{-{\half}} \Gamma(z) \Gamma(z + {\half}),
\label{gammadouble}
\end{equation}
and \eqref{barnes},  can evaluate the
denominator of the volume factors
\begin{equation}
G_d(N) \equiv (2N-1)!\ldots 3! 1! = \frac{1}{(4 \pi)^{N/2}} 2^{N(N+1)} G_2(N+1)G_2(N+{\frac{3}{2}})
\end{equation}
Using the Binet integral formula
\begin{equation}
\log \Gamma(z) = (z - {\half}) \hbox{log} z - z + {\half} \hbox{ log\ }2 \pi + 2 \int_{0}^{\infty} \frac{{\hbox{tan}({\frac{t}{z}})}}{{e^{2 \pi t} - 1}} dt,
\label{binet}
\end{equation}
the asymptotic expansion of $G_2(n)$ is
\begin{equation}
\log G_2(N+1) = \frac{N^2}{2} \log N - {\frac{1}{12}} \log N - \frac{3}{4} N^2 + {\half} N \log 2\pi + O(1).
\end{equation}
By expanding log($N-a$) for large $N$,  we obtain
\begin{equation}
\begin{split}
\log G_d(N) = & N^2 \log N + N^2(-\frac{3}{2} + \log 2) \\
& + {\half} N \log N - {\frac{1}{24}} \log N  + \frac{N}{2}(\log 4
\pi - 1) + O(1). 
\end{split}
\end{equation}

Putting all of this together, we find that
\begin{equation}
\begin{split}
& \log \hbox{vol}(\SU(N))  \\
& \hskip2cm = -N^2 \log N + \frac{1}{12} \log N \\
& \hskip2.5cm + \frac{3}{4} N^{2} +  \frac{1}{2} N^2 \log 2 \pi+ O(1), \\
& \log \hbox{vol}(\SO(2N+1))  \\
& \hskip2cm = -N^2 \log N+  N^2(\frac{3}{2} + \log \pi)  \\ 
& \hskip2.5cm - {\half} N \log N + \frac{1}{24} \log N 
+ \frac{N}{2}(1 + \log 4 + \log \pi) + O(1), \\
& \log \hbox{vol}(\SO(2N)) \\
&\hskip2cm = -N^2 \log N + N^2(\frac{3}{2} + \log \pi) \\
& \hskip2.5cm+ {\half} N \log N + \frac{1}{24} \log N 
+ \frac{N}{2}(-1 + \log 4 - \log \pi) + O(1), \\
& \log \hbox{vol}(\Sp(2N)) \\
& \hskip2cm = -N^2 \log N + N^2(\frac{3}{2} + \log \pi) \\
& \hskip2.5cm - {\half} N \log N + \frac{1}{24} \log N 
+ \frac{N}{2}(1 - \log 4 + \log \pi) + O(1).
\end{split} \label{eq:asymptoticV}
\end{equation}

\section{Matrix model Feynman rules and enumeration of diagrams}
\label{app:feynman}

We want to perturbatively evaluate the matrix integral
\begin{equation}
\int d\Phi\, e^{\frac{1}{g_s}\tr W(\Phi)},
\end{equation}
where the potential $W$ is given by
\begin{equation}
W(\Phi)=\sum_{j=1}^\infty\frac{g_j}{2j}\Phi^{2j}
\end{equation}
and $\Phi$ is a real antisymmetric $M \times M$ matrix.  We can write
this as
\begin{equation}
\int d\Phi \, \exp\left[
{\frac{1}{g_s}\tr\left(\frac{m}{2}\Phi^2
+\sum_{j=2}^\infty\frac{g_j}{2j}\Phi^{2j}\right)} \right],
\end{equation}
where $m=g_1$.  Expanding the exponential leads to traces of integrals
of the form
\begin{equation}
\begin{split}
& \int d\Phi \, e^{\frac{1}{g_s}\tr\frac{m}{2}\Phi^2}\, 
\Phi_{m_1n_1}\cdots\Phi_{m_kn_k}= \\
& \hskip2cm 
\pder{}{J_{m_1n_1}}\cdots\pder{}{J_{m_kn_k}}
\left(\int d\Phi \, \exp\left[{\frac{1}{g_s}\tr\frac{m}{2}\Phi^2
-\half\tr J\Phi}\right]\right)_{J=0}.
\end{split}
\end{equation}
This integral can now be evaluated, leading to
\begin{equation}
\left(\sqrt{\frac{2\pi g_s}{m}}\right)^{\frac{M(M-1)}{2}}
\pder{}{J_{m_1n_1}}\cdots\pder{}{J_{m_kn_k}}\left(e^{-\frac{g_s}{8m}\tr J^2}\right)_{J=0}.
\end{equation}
Differentiating step by step gives rise to expressions like
\begin{equation}
\begin{split}
&\pder{}{J_{mn}}\left(\frac{g_s}{2m}J_{m_1n_1}\cdots\frac{g_s}{2m}J_{m_kn_k}e^{-\frac{g_s}{8m}\tr J^2}\right) \\
&\hskip2cm
=\frac{g_s}{2m}(\delta_{mm_1}\delta_{nn_1}-\delta_{mn_1}\delta_{nm_1})\frac{g_s}{2m}J_{m_2n_2}\cdots\frac{g_s}{2m}J_{m_kn_k}e^{-\frac{g_s}{8m}\tr
J^2} \\
&\hskip2.2cm+\cdots \\
&\hskip2.2cm+\frac{g_s}{2m}J_{m_1n_1}\cdots\frac{g_s}{2m}J_{m_{k-1}n_{k-1}}\frac{g_s}{2m}(\delta_{mm_k}\delta_{nn_k}-\delta_{mn_k}\delta_{nm_k})e^{-\frac{g_s}{8m}\tr J^2} \\
&\hskip2.2cm+\frac{g_s}{2m}J_{mn}\frac{g_s}{2m}J_{m_1n_1}\cdots\frac{g_s}{2m}J_{m_kn_k}e^{-\frac{g_s}{8m}\tr J^2}.
\end{split}
\end{equation}
The indices $m_i$ and $n_i$ are contracted in traces as given in the
interaction which can be interpreted as forming vertices.  The
combinatorics can then be interpreted diagrammatically; one
must connect all the legs of the vertices in all possible ways with
untwisted and twisted propagators.  Each twisted propagator contributes
a factor of $(-1)$.

The rules for evaluating a diagram are then:
\begin{itemize}
\item Each kind of vertex with multiplicity $V_j$ contributes a factor
  of $\frac{1}{V_j!}(\frac{g_j}{2jg_s})^{V_j}$.
\item Each propagator contributes a factor of $\frac{g_s}{2m}$.
\item Each twisted propagator contributes an additional factor of $(-1)$.
\item Each index loop contributes a factor of $M=\frac{2S}{g_s}$.
\end{itemize}
The combinatorial factor of a diagram can be computed by counting all
topologically equivalent ways in which the legs of the vertices can be
connected.  This has some subtleties, since some diagrams with twisted
propagators can actually be planar.  To handle this, we make use of
the technique described in \cite{Cicuta:1982fu} to draw unoriented
diagrams (see also \cite{Mulase:2002cr,Mulase:2002xx} for recent work
on non-orientable ribbon diagrams in the mathematical literature).

An $\RP^2$ can be drawn in the plane as a disc, where antipodal
points on the boundary are identified.  $\RP^2$ diagrams can then be
drawn on that disc with some propagators going through the cross-cap
at the boundary.  The propagators going through the cross-cap are
twisted propagators, whereas all the others are untwisted propagators.

We can now also draw a planar diagram on the $\RP^2$.  If it has more
than one vertex, we can push one or several vertices through the
cross-cap without destroying the planarity, but all the propagators
going through the cross-cap are now twisted propagators.  This
operation contributes a multiplicative factor of $2^{v-1}$ to the
number of planar diagrams at each order $v$.  See Figure
\ref{fig:diags} for the enumeration of diagrams with 1 quartic vertex.

Using the relation between $p$ and the number of vertices $v_i$ of valency $i$ 
\begin{equation}
p = \frac{1}{2} \sum_i i v_i
\label{eq:prop}
\end{equation}
the contribution of planar diagrams to the free energy of the $\SU(M)$
matrix model is given by
\begin{equation}
{\cal F}_{0} = \sum_{v=1}^{\infty} \frac{d^{(n)}_v}{v!} (\frac{g_n}{n g_s})^v (\frac{g_s}{m})^p M^l
= \sum_{v=1}^{\infty} \frac{d^{(n)}_v}{v!} (\frac{g_n}{n g_s})^v (\frac{g_s}{m})^{\half{n}v} M^{2-(1-\frac{n}{2})v},
\end{equation}
where the sum is over diagrams with $v$ vertices of valence $2n$,
$d^{(n)}_v$ is the number of planar diagrams at each order, and $l$ counts
the number of boundary loops of the ribbon graph.  The propagator for
$\SU(M)$ theories is twice that of the $\SO$/$\Sp$ theories.  In the
second line we have simplified using (\ref{eq:chi}) and
(\ref{eq:prop}).  The number of diagrams of topology $S^2$
(i.e.~planar diagrams) in $\SU(M)$ matrix theory with a quartic
potential is given by \cite{Brezin:1978sv}
\begin{equation}
d^{(4)}_{v} = \frac{(2v-1)! 12^v}{(v+2)!} 
= 2, 36, 1728, 145152, \ldots .
\end{equation}
We are not aware of explicit generating functions for other vertex
valences $2n$, but these diagrams can be enumerated by computer to the
desired order.

If we now include twisted propagators (i.e.~enumerate planar diagrams
in the $\SO$ or $\Sp$ matrix models), there is an extra contribution
to the set of planar diagrams coming from vertices that have been
``flipped'', converting untwisted to twisted propagators according to
the rule described above.

\begin{equation}
{\cal F}_{0} = \sum_{v=1}^{\infty} \frac{d^{(n)}_v}{v!} (\frac{g_n}{n g_s})^v (\frac{g_s}{2m})^p M^l
= \sum_{v=1}^{\infty} \frac{d^{(n)}_v}{v!} (\frac{g_n}{n g_s})^v (\frac{g_s}{2m})^{\half{n}v} M^{2-(1-\frac{n}{2})v},
\end{equation}
\begin{equation}
d^{(4)}_{v} = \frac{1}{2} \frac{(2v-1)! 24^v}{(v+2)!} 
= 2, 72, 6912, 1161216, \ldots .
\end{equation}
A similar expression exists for the $\RP^2$ free energy
\begin{equation}
{\cal F}_{1} = \sum_{v=1}^{\infty} \frac{\widetilde{d^{(n)}_v}}{v!} (\frac{g_n}{n g_s})^v (\frac{g_s}{2m})^p M^{l-1} 
= \sum_{v=1}^{\infty} \frac{\widetilde{d^{(n)}_v}}{v!} (\frac{g_n}{n g_s})^v (\frac{g_s}{2m})^{\half{n}v} M^{1-(1-\frac{n}{2})v} .
\end{equation}
Here the number of diagrams $\widetilde{d^{(n)}_v}$ is counted with a
minus sign for each twisted propagator\footnote{Gaussian Ensembles are
  matrix models that have been well-studied in the physics and
  mathematics literature.  The Gaussian Orthogonal and Gaussian
  Symplectic Ensembles also contain non-oriented ribbon diagrams with
  twisted propagators, however the propagator is $\langle T^a_b T^c_d
  \rangle \sim \delta_{ac} \delta_{bd} + \delta_{ad} \delta_{bc}$,
  {\it i.e.}, there is no relative minus sign between the two terms.  This
  corresponds to counting $\RP^2$ diagrams with a positive sign
  always.  Therefore the free energy of the Gaussian Ensembles differs
  from that of the Lie Algebra matrix models at sub-leading orders in
  the genus expansion.}.  The relevant planar and $\RP^2$ diagrams
were enumerated by computer up to 4 vertices with a quartic potential
$\Wtree \sim \Phi^4$, to 2 vertices with a sextic potential $\Wtree
\sim \Phi^6$, and for a single vertex with a potential of degree up to
16.  The results are summarized in Table \ref{tab:enum} and verify the
desired relation:

\begin{equation}
{\cal F}_1 = -\frac{1}{2} \frac{\partial {\cal F}_{0}}{\partial M}.
\end{equation}

\begin{table}[h]
Diagrams with quartic vertices:
\begin{center}
\begin{tabular}{|c|c|r|r|r|r|}
\hline
Gauge group & Topology & $v=1$ & $v=2$ & $v=3$ & $v=4$ \\ \hline \hline
$\SU$ & $S^2$ & $2 M^3$ & $36 M^4$ & $1728 M^5$ & $145152 M^6$ \\ \hline
$\SO$/$\Sp$ & $S^2$ & $2 M^3$ & $72 M^4$ & $6912 M^5$ & $1161216 M^6$ \\
$\SO$/$\Sp$ & $\RP^2$ & $-3 M^2$ & $-144 M^3$ & $-17280 M^4$ & $-3483648 M^5$ \\
\hline
\end{tabular}
\end{center}

Diagrams with sextic vertices:
\begin{center}
\begin{tabular}{|c|c|r|r|r|}
\hline
Gauge group & Topology & $v=1$ & $v=2$ \\ \hline \hline
$\SU$ & $S^2$ & $5 M^4$ & $600 M^5$ \\ \hline
$\SO$/$\Sp$ & $S^2$ & $5 M^4$ & $1200 M^6$ \\
$\SO$/$\Sp$ & $\RP^2$ & $-10 M^3$ & $-3600 M^5$ \\ \hline
\end{tabular}
\parbox{5.5in}{
\caption{\small{Contribution to the free energy of the $\SU$/$\SO$/$\Sp$ matrix models at planar and $\RP^2$ level, for quartic and sextic potentials.  The first few terms in the perturbative expansion are listed, corresponding to the number of diagrams with increasing number of vertices (equivalently loops).}}
\label{tab:enum}}
\end{center}
\end{table}

\nc{\li}[2]{\large\item{#1}\\\normalsize{#2}}

\chapter[Appendix: Emergency Proof Techniques for Physicists]{Emergency Proof Techniques for Physicists}
\begin{enumerate}
\li{Proof by Intimidation}{Best applied to Graduate Students.}
\li{Proof by Divine Revelation}{See \cite{Witten:private}.}
\li{Proof by Exhaustion}{Keep going until your entire audience has fallen asleep or lost interest.}
\li{Proof by Vigorous Gesticulation}{If you think about it, this one is pretty similar to (\ref{analog}).}
\li{Proof by Extrapolation}{Prove the result in a certain limit, then assume it holds true over the entire parameter space.}
\li{Proof by Physicality}{Any result you dislike is declared to be unphysical and cast out.}
\li{Proof by Approximation}{Keep approximating until the result becomes trivial, then go back and fill in some of the gaps.}
\li{Proof by Assertion}{``It is clear that...''}
\li{Proof by Dimensional Transmutation}{$c = \hbar = \alpha' = 1$}
\li{Proof by Bastardized Notation}{It's easier than getting the mathematics correct.}
\li{Proof by Obscure Citation}{For full details, see \cite{Moriarty}.}
\li{Proof by Omission}{``It can be shown that...''}
\li{Proof by Peer Pressure}{``It should be completely obvious to every reader that..''}
\li{Proof by Recursive Citation}{Instead of citing the proof, cite a paper which refers to the proof, and iterate.  Bonus points if you can introduce a cycle into the graph of references.}
\li{Proof by Conclusion}{The consequences of the result are so profound that it must be true.}
\li{Proof by Redefinition}{Derive a result which is manifestly true, then redefine the meaning of the symbols and continue to use the result.}
\large\item\label{analog}{Proof by Analogous Reasoning}\\\normalsize{Compare the situation to a different, but vastly simpler one for which the result is true, and argue that the general case should have similar properties.}
\li{Proof by Trivial Limit}{The result reduces to the correct one in a suitably nice limit.}
\li{Proof by Rational Approximation}{$2 = \pi = \imath = -1 = 1$}
\li{Proof by Opressive Citation}{Cite an unrelated 100-page paper on the assumption that no-one will search through the entire thing for the proof.}
\li{Proof by Complication}{Spend $98\%$ of the paper deriving impenetrable technical results, and tie together at least 10 different threads which miraculously produce your result on the last page.}
\li{Proof by Notational Gymnastics}{Change your notational conventions at least three times to distract any hopes of pursuit by the reader.}
\li{Proof by Example}{Show the $n=1$ case.}
\li{Proof by Negative Reasoning}{The opposite of the result is false.}
\li{Proof by Forward Citation}{``We intend to present a proof of this result in work which is currently under preparation.''}
\li{Proof by Intuitive Diagram}{Draw a pretty enough picture and you can prove anything.}
\li{Proof by Deception}{Watch the hand...}
\li{Proof by Numerology}{We get the same numbers from two unrelated computations, so they must have the same meaning.}
\li{Proof by Profanity}{Should probably only be used as a last resort.}
\li{Proof by Association}{Tie the desired result to an unrelated discussion of obviously true material.}
\li{Proof by Universal Convergence}{It gives the right answer, so we'll let the mathematicians figure out why.}
\end{enumerate}

\end{document}